\documentclass{vldb}
 \usepackage{balance} 
  \usepackage[T1]{fontenc}
\usepackage{times,amsmath,epsfig}
\usepackage{amsthm}
\usepackage{enumerate}  
\usepackage{amssymb}
 \usepackage[center]{subfigure}     
\usepackage{graphicx}  
\usepackage{enumerate}  
\usepackage{cite}   
\usepackage{multicol}     
\usepackage{multirow}     
\usepackage[linesnumbered,  ruled, vlined]{algorithm2e}         
\newtheorem{thm}{Theorem}        
\newtheorem{prop}[thm]{Proposition}     
\SetNlSkip{0.5em} 
\IncMargin{0.5em}
\SetCommentSty{textrm}

  \renewenvironment{thebibliography}[1]{%
    \begin{oldthebibliography}{#1}%
      \setlength{\parskip}{0ex}%
      \setlength{\itemsep}{0ex}%
  }%
  {%
    \end{oldthebibliography}%
  } 
 
\newtheoremstyle{mdefstyle}% name
  {3pt plus 1pt}%      Space above, empty = `usual value'
  {2pt plus 1pt }%      Space below
  {}% Body font
  {}%         Indent amount (empty = no indent, \parindent = para indent)
  {\bfseries}% Thm head font
  {.}%        Punctuation after thm head
  {.5em}%     Space after thm head: " " = normal interword space;
  {\thmnote{#3}}% Thm head spec 
 
\newtheoremstyle{exmpstyle}% name
  {3pt plus 0.5pt}%      Space above, empty = `usual value'
  {2pt plus 0.5pt }%      Space below
  {}% Body font
  {}%         Indent amount (empty = no indent, \parindent = para indent)
  {\bfseries}% Thm head font
  {.}%        Punctuation after thm head
  {.5em}%     Space after thm head: " " = normal interword space;
  {\thmname{#1}\thmnumber{ #2}}% Thm head spec 
  
\theoremstyle{exmpstyle}
\newtheorem{exmp}{Example}

\theoremstyle{mdefstyle}
\newtheorem*{mdef}{}% all text supplied in the note

\theoremstyle{remark}

\SetNlSty{textbf}{}{.}
\SetKwFor{ForEach}{for each}{do}{endfch}
\SetArgSty{<textsc}

\usepackage{comment}

\begin{document}

% ****************** TITLE ****************************************

\title{Adding Logical Operators to Tree Pattern Queries on Graph-Structured
Data} 
 
% ****************** AUTHORS **************************************

% i$.$e$.$ three 'name/affiliation blocks' be placed beneath the title.

\numberofauthors{3} %  in this sample file, there are a *total*
\author{
{\fontsize{12pt}{12pt}\selectfont Qiang Zeng{\small $^{1, 2}$}\hspace{8mm}
Xiaorui Jiang{\small $^{1, 2}$}\hspace{8mm} Hai Zhuge{\small $^{1}$}} %
% add some space between author names and affils
\vspace{1.6mm}\\
\fontsize{10pt}{10pt}\selectfont
$^{1}$Key Lab of Intelligent Information Processing,
 Institute of Computing Technology, Chinese Academy of Sciences\\
\fontsize{10pt}{10pt}\selectfont
$^{2}$Graduate University of  Chinese Academy of Sciences 
\\ \fontsize{10pt}{10pt}\selectfont 
\{zengqiang, xiaoruijiang\}@kg.ict.ac.cn\hspace{8mm}zhuge@ict.ac.cn\\
}
     
\maketitle

\begin{abstract}
As data are increasingly modeled as graphs for expressing complex relationships,
the tree pattern query on graph-structured data becomes an
 important type of  queries  in real-world applications.
 Most practical query languages,  
such as XQuery and SPARQL, support  logical expressions using
logical-AND/OR/NOT operators to define  structural constraints of tree
patterns. In this paper, (1) we propose generalized tree pattern queries
(GTPQs) over graph-structured data, which fully support propositional logic of
structural constraints. (2) We make a
thorough study of fundamental problems including satisfiability, containment and
minimization, and analyze the computational complexity and the decision
procedures of these problems. (3) We propose  a compact graph representation of intermediate
results and a pruning approach to reduce the size of intermediate results and
the number of join operations -- two factors that often impair the
efficiency of traditional algorithms for evaluating tree pattern queries. 
(4) We
present an efficient algorithm for evaluating GTPQs using 3-hop as the
underlying reachability index. (5) Experiments on both real-life and
synthetic data sets demonstrate the effectiveness and efficiency of 
our algorithm,  from several times to orders of magnitude  faster than
state-of-the-art algorithms in terms of evaluation time, even
for traditional tree pattern queries with only conjunctive operations. 
\end{abstract}

\section{Introduction}
\label{section1}
Graphs are among the most ubiquitous data models for many areas, such as social
networks, semantic web and biological networks. As the most common tool for
data transmissions,  XML  documents  are  desirably modeled as graphs rather
than trees to represent flexible data structures by incorporating the concept
of ID/IDREFs. 
%  RDF documents are also essentially graph-structured data.
  Semantic Web data are also modeled as graphs, e$.$g$.$ in RDF/RDFS.  On graph
data, tree pattern queries (TPQs) are one of important queries of practical
interest. In  query languages such as XQuery and SPARQL, many queries
can be regarded as TPQs over graphs. As most of them support
logical operations including conjunction ($\wedge$), disjunction ($\vee$) and
negation ($\neg$) in the query conditions, it is  necessary to study TPQs over
graphs with multiple logical predicates, as illustrated in the following example.

% the  
% simplified tree pattern queries with exclusively
% the conjunctive operator, which we call conjunctive tree pattern queries in this
% paper. Besides, they primarily aim to find matches for all query nodes.
% \begin{figure}[t] 
% \subfigure[The graph representation of  part of the scheme of a DBLP file]{
% \label{dblp} %% label for first subfigure
% \begin{minipage}[b]{0.5\textwidth}
% \centering
% \includegraphics[height=0.6in]{dblp2.eps}
% \end{minipage}}
% \subfigure[The tree representation of $Q_{1}$, $Q_{2}$, and $Q_{3}$ in
% Example 1]{ \label{dblpquery} %% label for second subfigure
% \begin{minipage}[b]{0.5\textwidth}
% \centering
% \includegraphics[height=0.8in]{dquery_tu.eps}
% \end{minipage}}
% % \vspace{-4em}      
% \caption{DBLP scheme and tree pattern queries. Document elements matching  the nodes in the
% query with star symbols are required to be returned and the single- and
% double-lined edges denote the parent-child and ancestor-descendant relationships
% between elements respectively.}
% \label{expdblp} %% label for entire figure  
% \end{figure} 

% \vspace{-4em}  

\begin{exmp} 
A DBLP XML document  separately stores inproceeding  records for papers and proceeding
records for volumes, linked by \textsf{crossref} elements indicating where
a paper is published \cite{xmark}. The underlying data structure is clearly a
graph. Consider the following three queries which ask for information of
publications for which a certain tree pattern of data holds.
% \vspace{-3.5pt}
{
\small
\begin{enumerate}[$Q_1$:] \setlength{\itemsep}{0pt}
  \item Retrieve the information about Alice's conference papers that are
  published from 2000 to 2010 and co-authored with Bob.
  \item Retrieve the information about the conference papers of either Alice or
  Bob published from 2000 to 2010.
  \item Retrieve the information about Alice's conference papers that are not
  co-authored with Bob and published from 2000 to 2010. 
\end{enumerate}
}
% \vspace{-2pt}

% \vspace{-2pt} 

\setlength{\textfloatsep}{2pt plus 0pt minus 0pt}
\begin{figure}[tb]  
\begin{center}
  \includegraphics[height=1in]{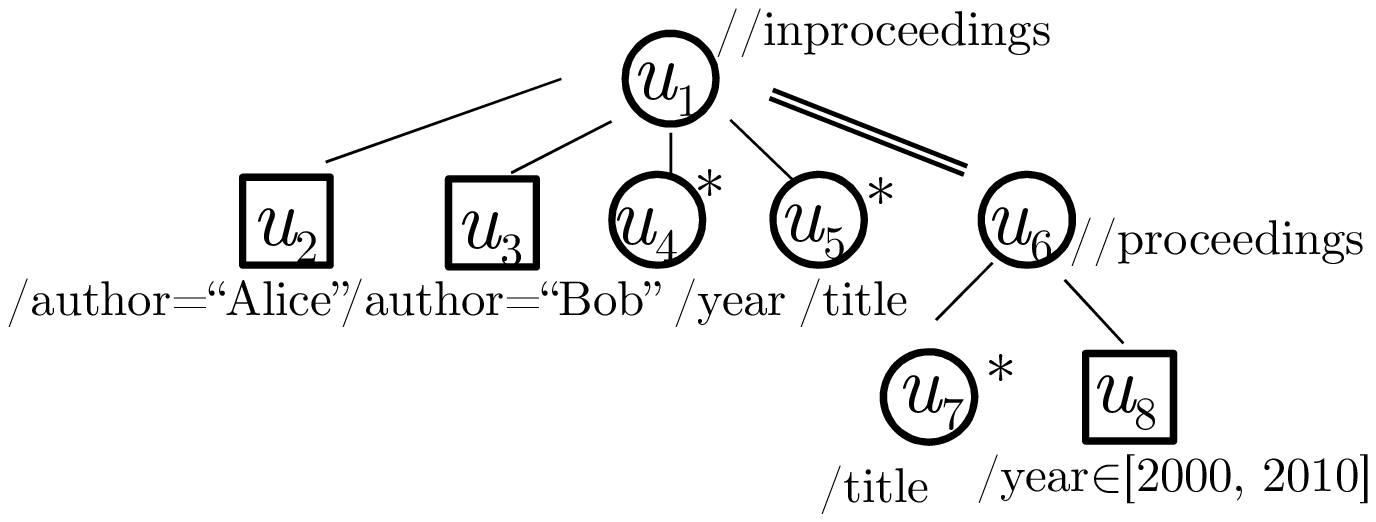}
  \caption{The tree representation of $Q_{1}$, $Q_{2}$, and $Q_{3}$ in
Example 1. Document elements
  matching the starred query nodes are required to be returned and the single-/double-lined
edges denote the parent-child/ancestor-descendant relationships between elements.}
  \label{expdblp} 
\end{center}
\end{figure}

They can be  expressed in XQuery and are essentially TPQs on
graph-structured data (see the Appendix), but $Q_2$ and $Q_3$ cannot be
expressed in traditional TPQs, which only contain conjunctive predicates. Indeed, they share the same tree representation as
 depicted in Fig$.$ \ref{expdblp}, but different structural predicates should be
 imposed on  the \textsf{inproceedings} element  $u_1$. For example, in
 $Q_1$, each embedding of the pattern should satisfy all paths specified in the query; but for $Q_2$, the two path conditions ``$u_1$--$u_2$'' and ``$u_1$--$u_3$''  are not required 
to be satisfied simultaneously. A 
predicate that specifies those edge constraints and  incorporates 
disjunction and negation needs to be attached to each query node in order to express $Q_2$
and $Q_3$. In general, (1) it is common in practice that logical expressions on query nodes 
needs to be imposed to specify complex relationships for not only attribute 
predicates (e$.$g$.$ $2000\leq \textrm{year}\leq 2010$) but also structural
constraints $\big($e$.$g$.$ ($u_1$--$u_2$ or $u_1$--$u_3$) in $Q_2$ and
$not(u_1$--$u_3)$ in $Q_3$$\big)$; (2) some of the nodes $\big($e$.$g$.$
$u_i (i\in\{1,2,3,6, 8\})$$\big)$ in the query pattern only serve as filters for
pruning unexpected results, which means that the
results of a TPQ should consist of matches for  a portion of the query
nodes only.
\qed   
\end{exmp}  

% \begin{comment}
% {\small
% \vspace{-3.5pt} 
% \begin{tabbing}
% let\qquad \=$\textdollar$dblp $:=$ doc(dblp.xml)\\
% for \>$\textdollar$paper in $\textdollar$dblp//inproceedings,
% \\\>$\textdollar$conf in $\textdollar$dblp//proceedings\\
% where \>$\textdollar$paper/author $=$ ``Alice'' and 
% $\textdollar$paper/author $=$ ``Bob'' and\\
% \>$\textdollar$paper/crossref $=$ $\textdollar$conf/@key and
%  data($\textdollar$conf/year) $\geq$ 2000 and\\ \>data($\textdollar$conf/year)
%  $\leq$ 2010\\ return\\
% \>if (exists($\textdollar$paper/year) and exists($\textdollar$conf/title))\\
% \>then
% \=$<$paper$>$\\
% \>\>$<$title$>\{\textdollar$paper/title$\}$$<$/title$>$\\
% \>\>$<$year$>\{\textdollar$paper/year$\}$$<$/year$>$\\
% \>\>$<$conf$>\{\textdollar$conf/title$\}$$<$/conf$>$\\
% \>\>$<$/paper$>$ 
% \end{tabbing}
% } 
%  \end{comment}

Although TPQs have been widely studied for many years, few of the proposed
processing algorithms can be used to efficiently evaluate such queries over
general graphs. They can neither support disjunction and negation on structural constraints nor be optimized
for the situation where output nodes take only a portion of query nodes (see
Related work for details).

 \begin{mdef}[Contributions \& Roadmap] This work makes the first effort to
 deal with TPQ over general graph-structured data with Boolean logic support.
 The contributions are summarized as follows.\\
(1) We introduce a new class of tree pattern queries over
graph-structured data, called generalized tree pattern queries (GTPQs) (Section \ref{section2}). In
a GTPQ, a node is not only associated with an attribute predicate,
which specifies the property conditions, but also a structural predicate
in terms of propositional logic with logic connectives including conjunction,
negation and disjunction to specify  structural conditions with
respect to its descendants. The query allows  a portion of the
query nodes to be output nodes. We also show that our formalization of query
is  advantageous over those in the literature on queries against
tree-structured data.\\  
(2) We investigate fundamental problems for GTPQs,
including satisfiability, containment, equivalence and minimization (Section \ref{section3}). We 
show that the satisfiability  of a special GTPQ with only conjunction and
disjunction is solvable in linear time, but the satisfiability and the other
three problems become computationally intractable when disjunction is incorporated.
We propose an exact algorithm to minimize GTPQs, which is supposed to be
sufficiently efficient, since the query sizes are typically small in practice.\\
(3) We propose a graph representation of intermediate results and a pruning approach to
 address notable problems in evaluating query patterns over graphs, develop an
 algorithm for GTPQs with ancestor-descendant edges and its extension to deal with
 parent-child edges (Section \ref{section4}). The algorithm  can
 largely filter nodes that cannot contribute to the final results, wisely  avoid
generating redundant intermediate results, and compactly represent the 
matches.\\ 
%  We show that the proposed approaches can be also used in other existing work.
(4) We implement our algorithm and conduct an experimental study using synthetic
and real-life data (Section \ref{section5}). We find that our evaluation
algorithm  performs significantly better than state-of-art algorithms even for  
conjunctive TPQs. It also has better scalability and  is robust for different
queries on different graphs. The experiments also  demonstrate the effectiveness
of the graph representation of results and the efficiency of the pruning
method. 
\end{mdef}  
\begin{mdef}[Related work]  
 There is a large body of research work on TPQs over
tree-structured data (see \cite{survey} for a survey). 
However, all studies heavily relied on the relatively simple structure of trees
and employed the node 
encoding schemes (including the interval \cite{holistic}, Dewey \cite{JTFast}
and sequence \cite{ViST} encodings) that are not applicable to graphs for
determining structural relationships.  Techniques  critical for their
efficiency, such as  stack encoding and  nodes skipping, can be only
applied to tree-structured data. For some sparse graph data whose
 structures  can be modeled by
disjoint trees connected by edges, such as many XML documents with ID/IDREFs,
although one can apply those existing algorithms for
tree-structured data to evaluate a query over such graphs by first decomposing
it to several TPQs over different trees and then merging
the results of distinct queries to form the final results, it is
inefficient due to large redundant intermediate results and costly merging
processes.
 
Some studies extended the traditional TPQs by incorporating additional
functions and restrictions. Chen et al. \cite{TPQ} included optional nodes to patterns  
and investigated efficient evaluation plans upon native XML database systems. The
generalized tree pattern is still against tree-structured data, which differs
from this work that studies TPQs over graph-structured data with logical predicates. 
Jiang et al. \cite{twigor} proposed new holistic algorithms based on a concept of OR-blocks
to process AND/OR-twigs, TPQs with OR-predicates. In the end of Section 2,
we shall show that (1) our query size can be always no larger than the
size of element nodes of AND/OR-twig for expressing a semantically identical query; (2)
constructing OR-blocks involves converting a propositional formula to
conjunctive normal form, thus taking exponential time in the worst case; (3)
the proposed algorithms  only support tree-struct-ured data as input.  
\cite{pathnot} studied  path queries with negation, while 
\cite{twignot} and   \cite{extend} added negation to TPQs. They 
cannot be applied to  GTPQs either, since they are
based on the classical holistic twig join algorithm  \cite{holistic} that only
works on tree-structured data. 

There has  been work on pattern queries for graph-structured
data. TwigStackD \cite{stackd} generalized the holistic algorithms, but it takes
considerable time and space without a pre-filtering process \cite{comm}. 
HGJoin \cite{hjoin} can evaluate general graph pattern queries
using OPT-tree-cover \cite{opt} as the underlying reachability indexing approach. 
It decomposes a pattern into a set of complete bipartite graphs and
 generates matches for them in order according to a plan. The time 
 cost of plan generation is always exponential since  it has
to produce a state graph with exponential nodes no matter for obtaining  an
optimal or suboptimal plan. Cheng et al. \cite{jointkde} proposed
\textit{R}-join/\textit{R}-semijoin processing for the graph pattern matching problem. It
relies on a cluster-based \textit{R}-join index whose size is typically prohibitively large, as
the index stores matches for every two labels derived from 2-hop indexing
\cite{2hop}. Unlike the plan generation of HGJoin, it adopts left-join to reduce the cost, but in the
worst case the time complexity is still exponential. Since both HGJoin and
\textit{R}-join/\textit{R}-semijoin use  structural joins similar to the earlier
work on tree-structured data, they typically have large intermediate results and need to perform large amounts of
expensive join operations. All these three algorithms also do not directly
support queries with negative/disjunctive predicates.  A
straightforward approach to apply them to the GTPQ processing is to decompose
the query into multiple conjunctive TPQs and perform the difference and merge operations 
on results of the decomposed queries. However,
the number of the resultant conjunctive TPQs may be exponential and large
intermediate results may need  to be generated and merged. 
% Also, data have to be  repeatedly scanned, thus incurring high I/O cost. 
 
A number of studies investigated various graph pattern matching problems
\cite{graphql, fan, distance}.  \cite{graphql} proposed a graph
query language GraphQL and studied graph-specific optimization techniques for
graph pattern matching that combines subgraph  isomorphism and
predicate evaluation. While the language is able to express queries with ancestor-descendant
edges and disjunctive predicates, the work focused on processing ¡°non-recursive¡±
and conjunctive graph pattern queries, where all edges of a query pattern
correspond to the parent-child edges of GTPQs, specifying the adjacent
relationship between desired matching nodes. \cite{fan} defined matching in terms of bounded simulation to
reduce its computation complexity. \cite{distance} studied distance pattern
matching, in which query edges are mapped to paths with a bounded length.
Queries of \cite{fan} and \cite{distance} do not
support negative/disjunctive predicates on edges and have quite different
semantics with ours. 
 
Most existing algorithms  are  to find  all
instances of patterns containing matches of all query nodes. In real-world
applications, however, the answer to the query often only require matches of
several but not all query nodes. Indeed, many query nodes only serve as filters
for imposing structural constraints on output nodes. Our framework can avoid
 generating redundant matches at
run time. 
 
Satisfiability, containment, equivalence and minimization are fundamental
problems for any query languages. The minimization of TPQs
over tree-structured data has been investigated in several papers. Amer-Yahia et
al. \cite{min1} proposed algorithms for the minimization with and without
integrity constraints. Ramanan \cite{min2} studied this problem for TPQs
defined by graph simulation. Chen et al. \cite{min3} used a richer class of
integrity constraints for query minimization of TPQs with an unique output node.
However,  we are not aware of previous work on minimization as well as the other
three problems for TPQs with logical predicates either over tree-structured
data or over graph-structured data. 
 \end{mdef}  
 \setlength{\textfloatsep}{3pt plus 2pt minus 2pt}
\section{Data model and generalized\\ tree pattern queries}\label{section2}
\begin{mdef}[Data graphs]
A data graph is a directed graph $G=(V, E, f)$, where (1) $V$ is a finite set of
nodes; (2) $E\subseteq V\times V$ is  finite set of edges, in which each pair
($v, v'$) denotes an edge from $v$ to $v'$; (3) $f$ is a function  on $V$
defining attribute values associated with nodes. For each
node $v\in V$, $f(v)$ is a tuple ($A_1=a_1, \ldots, A_n=a_n$), where the expression
$A_i=a_i (i\in[1, n])$ represents that $v$ has a attribute denoted by $A_i$ and
its value is a constant $a_i$.  For example, in a data
graph $G=(V, E, f)$ of a DBLP document, the node properties in $f$ may include 
tags, string values, typed values, and  attributes specified in the elements.

Abusing notions for trees and traditional tree pattern queries, we refer to a
node $v_2$ as a \emph{child} of a node $v_1$ (or $v_1$ as a parent of $v_2$) and
say they have a \emph{parent-child} (PC) relationship if there is an edge $(v_1,
v_2)$ in $E$, and refer to $v_2$ as a descendant of $v_1$ (or $v_1$ as an
\emph{ancestor} of $v_2$) and say they have an \emph{ancestor-descendant} (AD)
relationship if there is a nonempty path from $v_1$ to $v_2$ in $G$.
\end{mdef}
\begin{mdef}[Generalized tree pattern queries]
A generalized tree pattern query (GTPQ) $Q=(V_b, V_p, V_o, E_q, 
f_a, f_e, f_s)$, where:\\ 
(1) $V_b$ and $V_p$ are both a finite set of nodes, called \emph{backbone
nodes} and \emph{predicate nodes}, respectively. The complete set of query nodes
is denoted as $V_q$, i.e$.$, $V_q=V_b\cup V_p$.\\ (2) $V_o\subseteq V_b$. The
nodes in $V_o$ are called \emph{output nodes}.\\ (3) $E_q\subseteq \{(u_1, u_2)|u_1,
u_2\in V_b\}\cup \{(u_1, u_2)|u_1\in V_b\cup V_p, u_2\in V_p\}$, is a
finite set of edges. Here, $(V_q, E_q)$ is restricted to a directed tree .\\ 
(4) $f_a$ is a
function defined on $V_q$ such that for each node $u\in V_q$, $f_a(u)$ is an
\emph{attribute predicate} that  is a conjunction of atomic formulas of the form 
of ``$A$ op $a$'', in which $A$ is an attribute name, $a$ is a constant and  op is a 
comparison operator in $\{<, \leq, =, \neq, >, \geq\}$.\\
(5) $f_e$ is a function on $E_q$ to
specify the type of the edge. Each edge $(u_1, u_2)$ represents either
PC relationship or AD relationship. \\
(6) $f_s$ is a function defined on internal nodes. For each internal node $u\in
V_q$ with $k$ children being predicate nodes, $f_s(u)$, called a
\emph{structural predicate}, is a propositional formula  in $k$ variables $p_{u'_1},
\ldots, p_{u'_k}$,  each corresponding to a tree edge directing to a predicate
child of $u$. In particular, if $u$ has no predicate children, $f_s(u)=1$.
Each node $u$ is  associated with a distinct propositional
variable denoted by $p_u$.
\end{mdef}
\vspace{1pt}
% As indicated in the definition, (1) backbone nodes and the edges connecting them
% constitute a backbone tree structure of the whole  query tree; (2) all
% output nodes must be backbone nodes (reasons are explained later in the Remark);
% (3) negation and disjunction are \emph{only} 
% %[[]]
% included
% in the predicates of predicate nodes.
 
We call a GTPQ a union-conjunctive GTPQ if the structural predicates on all
query nodes are negation-free, and call it a conjunctive GTPQ if the  structural
predicates on all the query nodes only have conjunction connectives.

Before giving the semantics of GTPQs, we  add variables for non-root backbone
nodes to extend the structural predicate. For an internal node $u$ with $k'$
backbone children, denoted by $u_1,\ldots, u_{k'}$, the \emph{extended
structural predicate} $f_{ext}(u)=p_{u_1}\wedge \ldots\wedge p_{u_{k'}}\wedge
f_s(u)$.

\begin{exmp}
In Example \ref{expdblp}, $Q_{1}=(V_b, V_p, V_o, E_q, 
f_s, f_e, f_s)$  is a conjunctive GTPQ, in which (1)
 $V_b=\{u_1, u_4, u_5, u_6, u_7\}$, $V_p=\{u_2, u_3, u_8\}$, $V_o=\{u_4, u_5,
 u_7\}$; (2) the attribute predicate $f_a$ for a query node is a
 conjunction of comparisons among tags and typed values $\big($e$.$g$.$
 $f_a(u_2)=({\rm tag }= $ ``author'' $ \wedge$ value $= $   ``Bob'')$\big)$; (3)
 $f_s(u_1)=p_{u_2}\wedge p_{u_3}$, and $f_s(u_6)=p_{u_8}$.  The only difference
 between $Q_{2}$ and $Q_{1}$  is that in $Q_{2}$,  $f_s(u_1)=p_{u_2}\vee
 p_{u_3}$. In $Q_{3}$, $f_s(u_1)= p_{u_2}\wedge \neg p_{u_3}$. As an example of  extended
 structural predicates, for $Q_2$, $f_{ext}(u_1)=(p_{u_2}\vee
 p_{u_3})\wedge p_{u_4}\wedge p_{u_5}\wedge p_{u_6}$.
 \qed 
\end{exmp}
\begin{mdef}[Semantics]
Consider a data graph $G=(V, E, f)$ and a GTPQ $Q=(V_b, V_p, V_o, E_q, 
f_a, f_e, f_s)$. We say that a data node $v$ in $G$ \emph{downwardly matches} a
query node $u$ in $Q$, denoted by $v\models u$, if the following
conditions are satisfied:\\
(1)  $v$ satisfies the attribute\
predicate of $u$, denoted by $v\sim u$.  That is, for each formula ``$A$ op
$a$'' in $f_a(u)$, there is an element ($A=a'$) in $f(v)$ such that $a'$ op $a$.
$v$ is called a \emph{candidate matching node} of $u$. $mat(u)$ denotes the set
of candidate matching nodes of $u$, i$.$e$.$, $mat(u)=\{v|v\in V, v\sim u\}$.\\   
(2)	If $u$ is an internal node, the data node $v$ determines a truth assignment
to the variables of $f_{ext}(u)$ such that $f^v_{ext}(u)=1$, where
$f^v_{ext}(u)$ denotes the truth-value of $f_{ext}$ under the assignment. For
each variable $p_{u'}$, the truth-value $p^v_{u'}$ is assigned
as follows: for each PC (resp$.$ AD) child $u'$ of $u$, $p^v_{u'}=1$ if there exists 
a child (resp$.$ descendant) $v'$ of $v$ such that $v'\models u'$; otherwise,
$p^v_{u'}=0$.
\end{mdef}
Let $V_b=\{u_1, \ldots, u_m\}$. A $m$-ary tuple ($v_1, \ldots, v_m$) of nodes in
$G$ is said to be a match of $Q$ on $G$, if  the following conditions
hold: (1) for each $v_i (i\in [1, m])$, $v_i\models u_i$;
(2) for each edge $(u_i, u_j)\in E_q (i, j\in [1, m])$, if $u_j$ is a PC child
of  $u_i$, $v_j$ is a child of $v_i$; otherwise, $v_j$ is a descendant of $v_i$.
 
%[[]]
The answer $Q(G)$ to $Q$  is a set of results in the form of tuples,
where each tuple consists of the images of output nodes $V_o$ in a match of
$Q$. For each match, there is at least an assignment for all variables that 
makes the extended structural predicates of all internal backbone nodes and some
of internal predicate nodes evaluate to true, which we
%[[]] can
  call a certificate  of
the match. For a match and an assignment as a certificate of the match, an instance of 
$Q$ on $G$ is a tuple consisting of such nodes that
each of them matches a distinct query node whose corresponding propositional 
variable is true under the assignment. 
In particular,
 an instance of conjunctive GTPQ is exactly a match of the query.
\begin{figure}[t]
\centering  
\subfigure[Data graph $G$]{
\label{datagraph} %% label for first subfigure
% \begin{minipage}[b]{0.4\textwidth}
% \centering  
\includegraphics[height=1in]{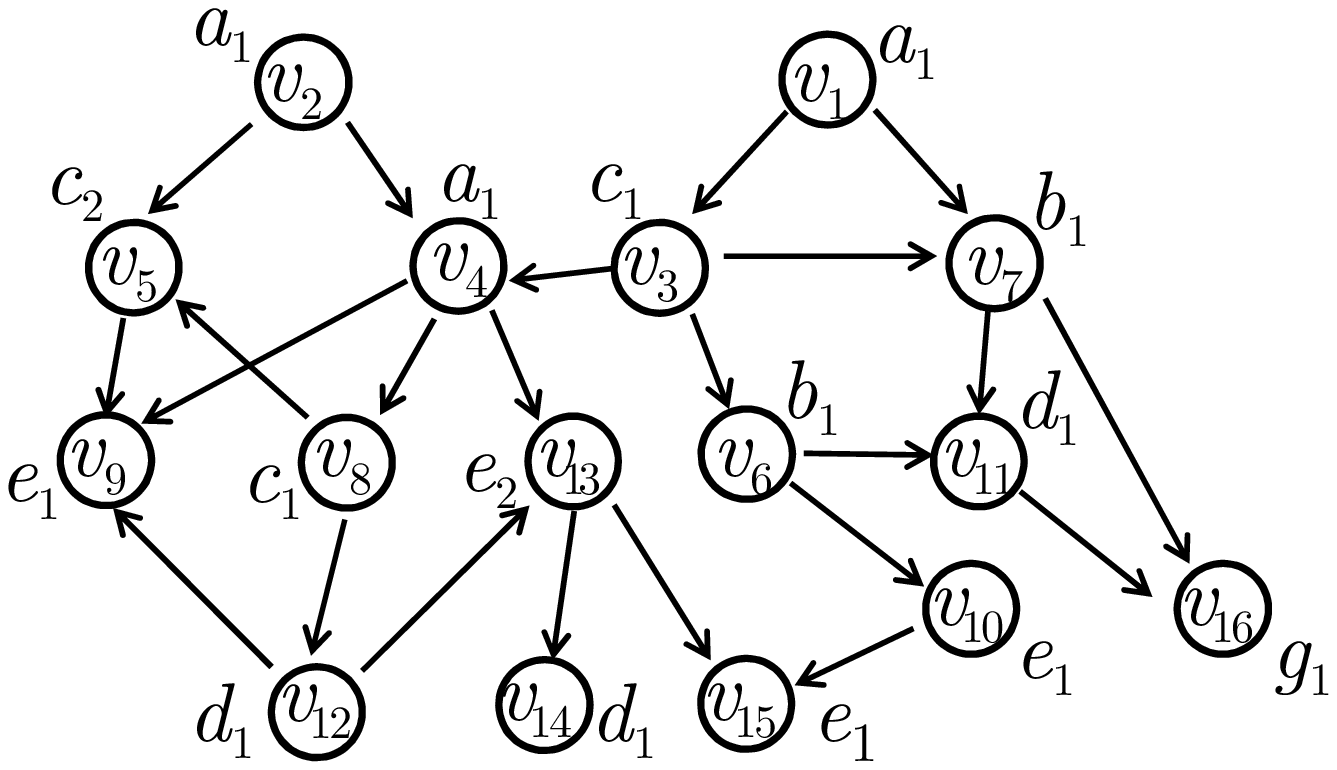}
% \end{minipage}  
}\\    
\subfigure[GTPQ $Q$ on $G$]{    
\label{gtpq} %% label for second subfigure   
% \begin{minipage}[b]{0.4\textwidth}g
% \centering   
\includegraphics[height=0.9in]{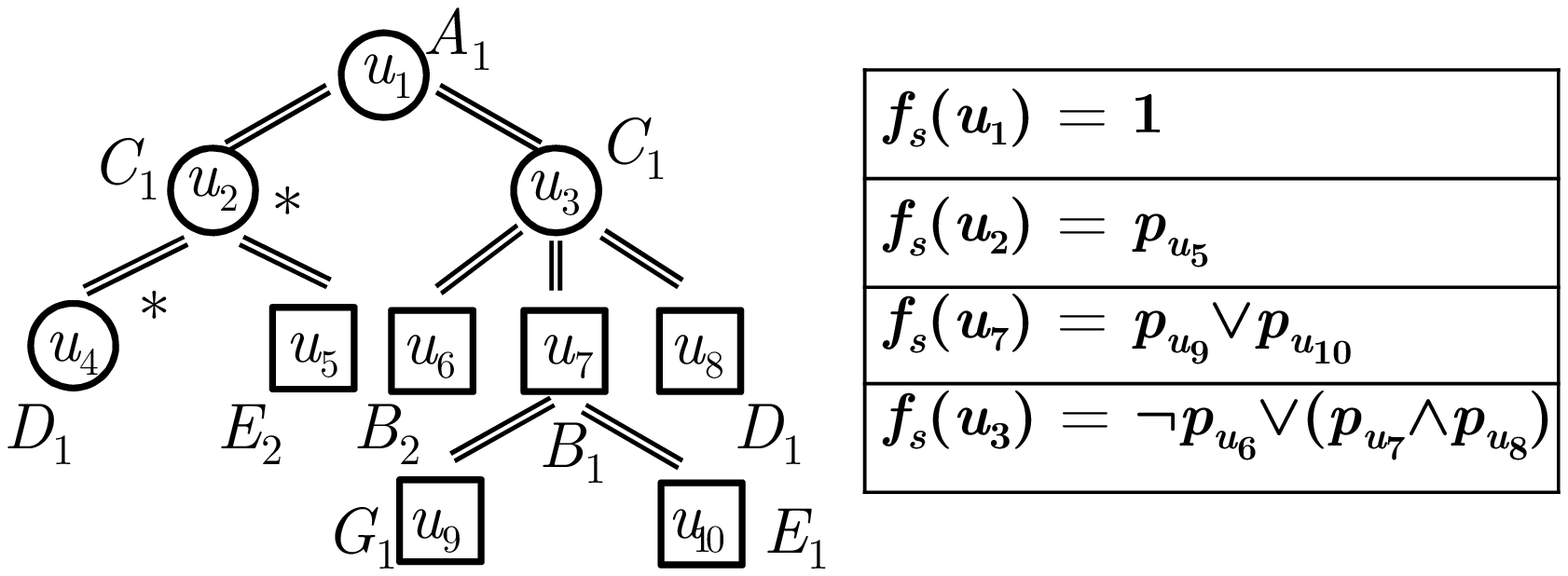} 
% \end{minipage}
} 
\caption{Example of a data graph and a GTPQ. We use a
rectangle to represent a predicate node and a circle  to represent a backbone
node. }
\label{exmp:des} %% label for entire figure
\end{figure} 
 
\begin{figure}[t]
\centering
\subfigure[B-twig query]{
\label{cmp:alltwig}
% \begin{minipage}[t]{0.2\textwidth}
\centering 
\includegraphics[height=1in]{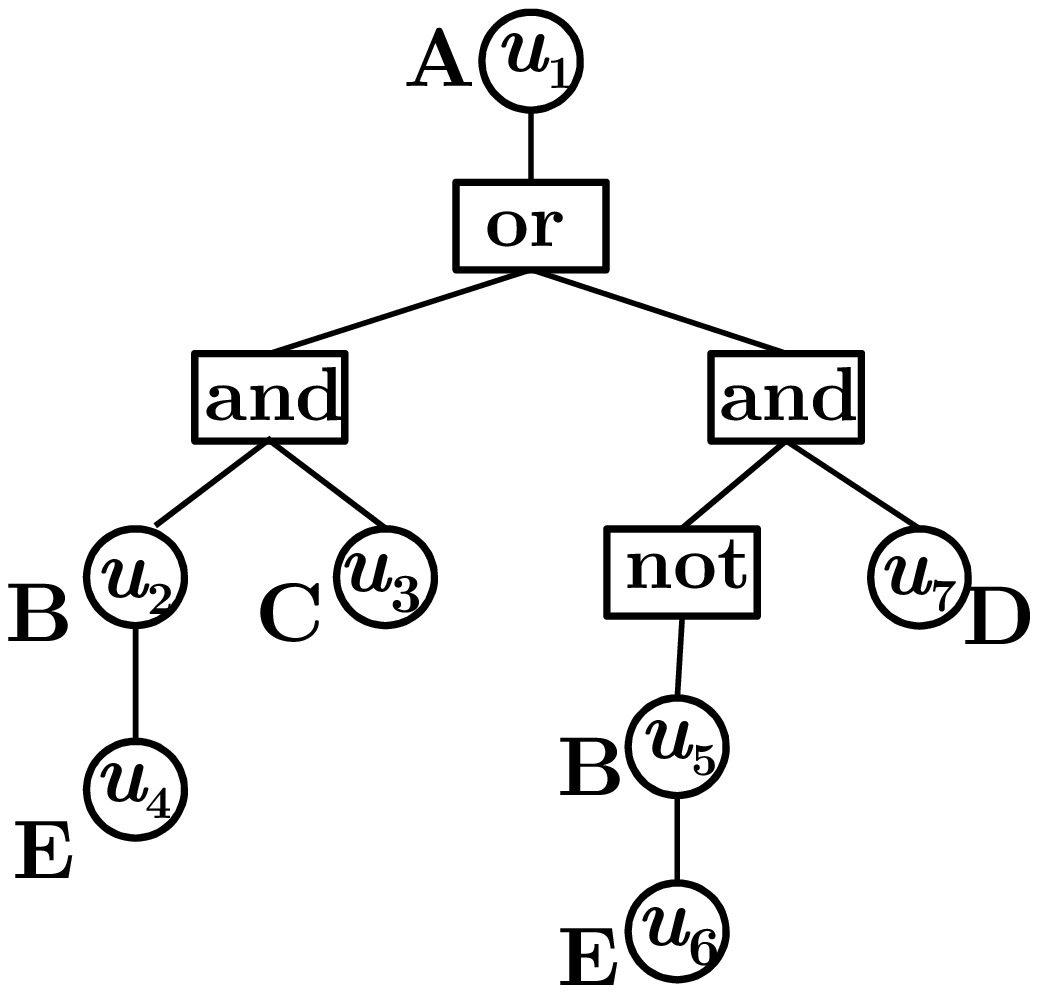}
% \end{minipage}
}   
\subfigure[GTPQ]{ 
\label{cmp:gtpq}
% \begin{minipage}[t]{0.2\textwidth}
\centering  
\includegraphics[height=1in]{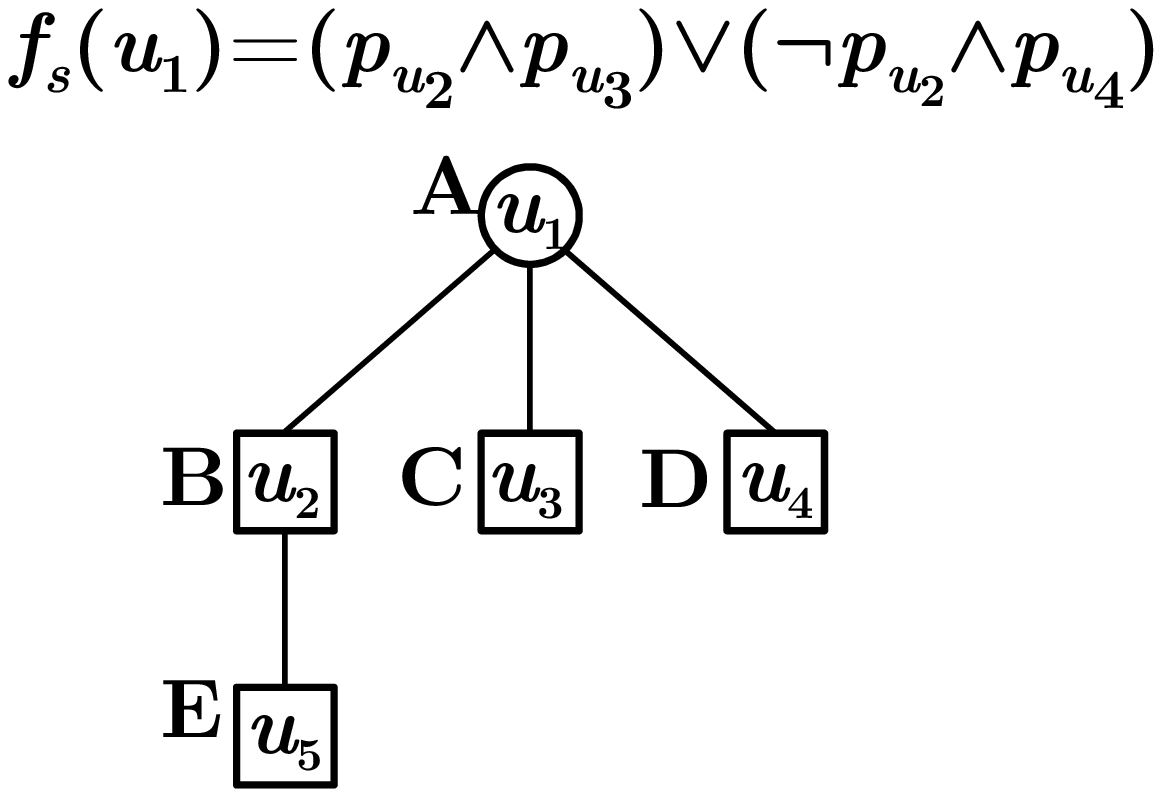}
% \end{minipage}
}
\caption{Comparison between   a B-twig query and  a GTPQ }
\label{cmp}
\end{figure} 
%   
   
% 
% \begin{figure*}
% \begin{minipage}[b]{1\textwidth}
% \centering
% \subfigure[A data graph $G$]{
% \label{datagraph} %% label for first subfigure
% \begin{minipage}[b]{0.5\textwidth}
% \includegraphics[width=1.8in]{data.eps} 
% \end{minipage}}  
% \subfigure[A GTPQ $Q$ on $G$]{
% \label{gtpq} %% label for second subfigure
% \begin{minipage}[b]{0.5\textwidth}
% \includegraphics[width=2.4in]{query.eps}
% \end{minipage}}
% \caption{An example of a data graph and a GTPQ. In this paper, we use a
% rectangle to represent a predicate node and a circle  to represent a backbone
% node. All output nodes are annotated by a star symbol.} 
% \label{exmp:des} %% label for entire figure
% \end{minipage}% 
%     
% 
% \begin{minipage}[b]{0.4\textwidth}
% \centering
% \subfigure[]{
% \label{cmp:alltwig} 
% \begin{minipage}[t]{0.2\textwidth}
% \includegraphics[height=1.3in]{duibit.eps}
% \end{minipage}} 
% \subfigure[]{ 
% \label{cmp:gtpq} 
% \begin{minipage}[t]{0.2\textwidth}
% \includegraphics[height=1in]{duibim.eps}
% \end{minipage}}
% \caption{Comparison between (a) a B-twig query and (b) a GTPQ. }
% \label{cmp}
% \end{minipage}%
% 
% 
% \end{figure*}

\begin{exmp}  
\label{exmp:gtpq}
For simplicity of presentation, a lower-case letter $x_i$  in all figures throughout
this paper denotes
$f(v)$ for a data node $v$ and a capital letter  $Y_j$ denotes $f_a(u)$ for  a
query node $u$ such that $v\sim u$ if $j\leq i$ and $X=Y$. 

Consider the data graph and the query shown in Fig$.$ \ref{exmp:des}. 
$v_{13}\sim u_5, v_{15}\not\sim u_5$. Accordingly, $mat(u_5)=\{v_{13}\},
mat(u_{10})=\{v_9,v_{10}, v_{13}, v_{15}\}$.  The answer $Q(G)=\{ (v_3,
v_{11}), (v_3, v_{12}), (v_3,\\ v_{14}), ($$v_8$$, $$v_{12}),(v_8$$,$$
v_{14})\}$. One of the query  matches leading to $(v_3, v_{11})$ is $(v_1,
v_3, v_3, v_{11})$, where elements are
%[[]]
sorted
 in the ascending order of the subscripts of corresponding query nodes. An
instance of
%[[
 this match is  $\{u_1:v_1, u_2:v_3, u_3:v_3, u_4:v_{11}, u_7:v_{6},
u_8:v_{11}, u_9:v_{15}\}$,  where `$u:v$' means $v$ is a match of $u$. Indeed,
$v_3\models u_3$,  because (1) $v_3\sim u_3$, and (2) $f^{v_3}_{ext}(u_3)=1$
since $v_6\models u_7$ and  $v_{11}\models u_8$. Also, $v_5\models u_3$, because $v_5$
cannot reach a node  matching $u_6$ and hence $p^{v_5}_{u_3}=0$, thereby
$f^{v_5}_{ext}(u_3)=1$.
\qed
\end{exmp}

% the reason why attribute predicate does not include not operators.

For simplicity of
semantics, we require a query to explicitly specify backbone nodes and predicate
nodes and restrict output nodes to backbone ones. The distinction between
the two types of nodes is that propositional variables associated with backbone
nodes are disallowed to be operands of negation and disjunction as
those associated with predicate nodes, which guarantees that each backbone node
has an image in a match of the query. Permitting negation and
disjunction on any query nodes leads to issues that are not computationally
desirable. If each query result is still required to have 
an image for each output node, the expressive power does not change; but to
determine whether a query is valid is effectively to  check whether the
variables associated with output nodes are always  true for all certificates of  matches, which is a co-NP-complete
problem.  Otherwise, the output 
structures become not fixed. They can either be specifically defined in the
query, or consist of exponential combinations of output nodes by default. Our algorithm described in Section
 \ref{section4} can be straightforwardly extended to process queries with
 multiple output structures (see the Appendix).

We now compare GTPQ with
%[[]]
 the works in \cite{twignot}
and \cite{alltwig}. \cite{twignot} deals with \emph{AND/OR-twig} against
tree-structured data.
\cite{alltwig} further extends \cite{twignot} to handle \emph{B-twig}, which
additionally introduces the logical-NOT operation into the query.  Both
represent a query by defining special types of nodes for operators, namely
 logical-AND  nodes, logical-OR nodes and logical-NOT nodes.
For each occurrence of a variable in a structural predicate of a GTPQ,
the corresponding AND/OR-twig or B-twig needs to use a distinct subtree to
express the structural constraints with respect to descendants as specified by the
variable, since in  AND/OR-twigs and B-twigs, the query nodes connected to
different operator nodes are considered as  distinct. The query size of
AND/OR-twigs or B-twigs hence may be much larger than the size of a GTPQ for
expressing  complex tree patterns. In Fig$.$ \ref{cmp}, the
 B-twig query has to use  two
 paths $u_2$--$u_4$ and $u_5$--$u_6$ to represent the constraints that can be
 imposed by a single path $u_2$--$u_5$ in the semantically equivalent GTPQ.   Moreover,
before evaluating the query, \cite{twignot} and \cite{alltwig} have to construct
 OR-blocks to normalize the twig. The normalization process
 is essentially a CNF conversion of propositional formulas. Since a CNF
 conversion can lead to an exponential explosion of the formula, the time cost
 of a conversion is  exponential in the size of original query, and the
 resulting  query size also becomes exponential in the worst case.  Therefore, our query representation is more powerful and compact than 
 the tree representation  of  \cite{twignot} and \cite{alltwig}.

\section{Fundamental Problems for Generalized Tree Pattern
Queries}\label{section3} In this section, we study the problems of
satisfiability, containment, equivalence, and minimization of GTPQs, which are
important for query analysis and optimization.

\subsection{Satisfiability}
A GTPQ $Q$ is \emph{satisfiable} if there is a data graph $G$ on which the
answer $Q(G)$ to $Q$  is nonempty. We first introduce some definitions
before showing how to determine the satisfiability and establishing the property
of the problem.

We say $u$ is an \emph{independently constraint} node if (1) the formula
$\big(f_s(u')[p_u/1]\oplus f_s(u')[p_u/0]\big)\wedge f_s(u) $ is satisfiable, 
in which $u'$ is the parent of $u$, $f_s(u')[p_u/x]$ is  the formula produced by
assigning $x$ to the variable $p_u$ $(x\in\{0, 1\})$,  and $\oplus$ is the
exclusive-or logical operator;  (2) all ancestors of $u$ are independently
constraint nodes. Intuitively, the variables  of independently constraint nodes
can independently affect the resulting  truth-value of the structural predicates of
their parents and ancestors. Backbone nodes are clearly independently
constraint nodes, if their structural predicates are satisfiable.

A \emph{transitive structural predicate} $f_{tr}(u)$ for a node $u$ is
constructed from $f_{ext}(u)$ in a bottom-up sweep as follows. (1) For each leaf node and each non-independently constraint node $u$ , the transitive structural predicate is the same as
 the extended structural predicate, i$.$e$.$
$f_{tr}(u)=f_{ext}(u)$.  (2)  For an internal node $u$ such that the transitive
structural predicates of all  children have been defined, $f_{tr}(u)$ is
produced by substituting $\big(p_{u'}\wedge f_{tr}(u')\big)$ for each variable
$p_{u'}$ of independently constraint node $u'$ in $f_s(u)$.

For two non-root nodes $u_1, u_2$ in $Q$,
we say that $u_2$ is \emph{similar} to $u_1$, denoted by $u_1\triangleleft
u_2$, if the following conditions hold. (1)  For each formula ``$A$ op $a_1$''
in $f_a(u_1)$, there is a formula ``$A$ op $a_2$'' in $f_a(u_2)$ such that (a)
if $\text{op }\in\{\leq, <\}$, $a_2\leq a_1$, (b) if $\text{op } \in \{\geq, >\}$, $a_2\geq a_1$, (c) if $\text{op } \in \{=,\neq\}$, $a_1=a_2$. We use $u_2\vdash u_1$ to
denote that $u_1$ and $u_2$ satisfy this condition. (2) For each PC (resp$.$ AD)
child $u'_1$ of $u_1$ such that $u'_1$ is an independently constraint node,
there is a PC child (resp$.$ a descendant) $u'_2$ of $u_2$  such that
$u'_1\triangleleft u'_2$.  (3) The formula  $f_{tr}(u_2)\to    
f_{tr}(u_1)[u_1\mapsto u_2]$ is a tautology,   where $f_{tr}(u_1)[u_1\mapsto
u_2]$ is a formula transformed  from $f_{tr}(u_1)$ by  replacing $p_{u'}$
with $p_{u''}$ for each pair ($u', u''$) such that (a)  $u'$  is a descendant of
$u_1$, (b) $u''$ is a descendant of $u_2$ and (c) $u'\trianglelefteq u''$.  We
say that $u_1$ is \emph{subsumed} by $u_2$, denoted by $u_1\trianglelefteq u_2$, if (1)
$u_1\triangleleft  u_2$, and (2) the parent of $u_1$ is the lowest common
ancestor $u_{lca}$ of $u_1$ and $u_2$, and (a) if $u_1$ is a PC child of $u_{lca}$, $u_2$ is also a PC child of
$u_{lca}$; (b) otherwise $u_2$ is a descendant of $u_{lca}$.

We finally define 
\emph{complete structural predicates} 
to characterize  the whole structural constraints of a GTPQ. For a node $u$,
the complete structural predicate $f_{cs}(u)$ is created from the corresponding transitive structural predicate $f_{tr}(u)$ by performing the following operations: 
(1) for each descendant $u'$ of  $u$,
if its attribute predicate is unsatisfiable,    
$f^{new}_{cs}(u)=f^{old}_{cs}(u)[p_{u'}/0]$, where $f^{old}_{cs}(u)$ is
the old formula before this transformation   and $f^{new}_{cs}(u)$ is the newly
generated formula;  (2) for every two nodes $u_1$ and $u_2$ in two
distinct subtrees of $u$ such that
$u_2\trianglelefteq u_1$, 
$f^{new}_{cs}(u)=f^{old}_{cs}(u)\wedge  \big(\neg p_{u_1}\vee (p_{u_2}\wedge
f_{ext}(p_{u_2})\big)$, where $f^{old}_{cs}(u)$ and $f^{new}_{cs}(u)$ have the
same meaning as above in (1).

Theorem \ref{thm:sat} shows that the satisfiability of a GTPQ is equivalent to
the satisfiability of the complete structural predicate of the root, if
given that the attribute predicate of the root is satisfiable. If
the query is a conjunctive or union-conjunctive GTPQ, the problem of satisfiability can 
be solved in linear time. When negation is added into the
query, the satisfiability becomes NP-complete.

\begin{thm} 
\label{thm:sat}
A GTPQ $Q$ is satisfiable if and only if for the root node $u$ of $Q$, $f_a(u)$
and $f_{cs}(u)$ are both satisfiable. \qed
\end{thm}

\begin{thm}
\mbox{}
\vspace{-2mm}
\begin{enumerate} [{\normalfont 1.}]  \setlength{\itemsep}{0pt}
\setlength{\topsep}{0pt}
  \item The satisfiability of a union-conjunctive GTPQ can be determined in
  linear time.
  \item The satisfiability of a GTPQ is NP-complete.\qed
\end{enumerate} 
\label{thm:sat:com}
\end{thm}

\begin{figure}[t] 
\subfigure[$Q_1$]{
\label{satisfy1} 
\begin{minipage}[t]{0.17\textwidth}
\centering
\includegraphics[height=1in]{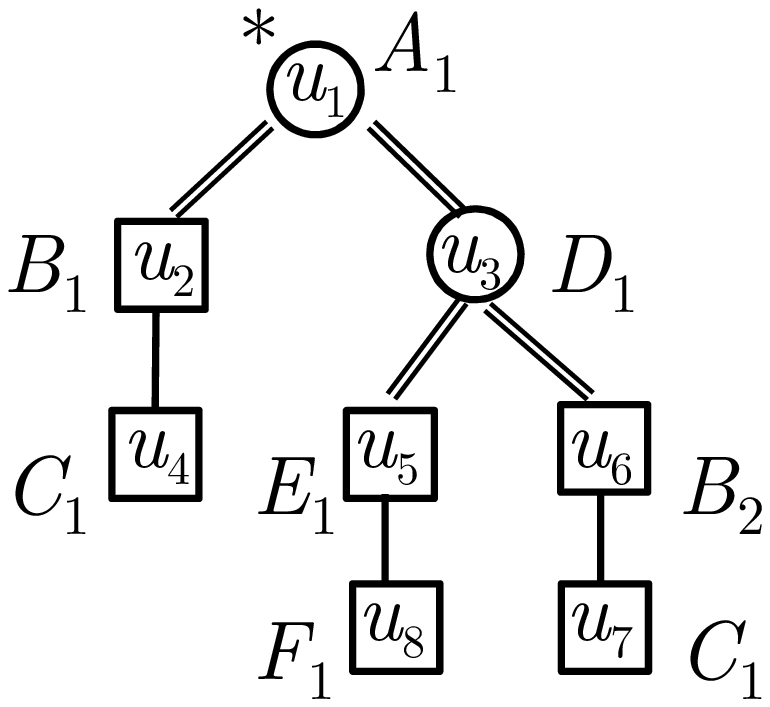}
\end{minipage}} 
\subfigure[$Q_2$]{ 
\label{satisfy2} 
\begin{minipage}[t]{0.15\textwidth}
\centering
\includegraphics[height=1in]{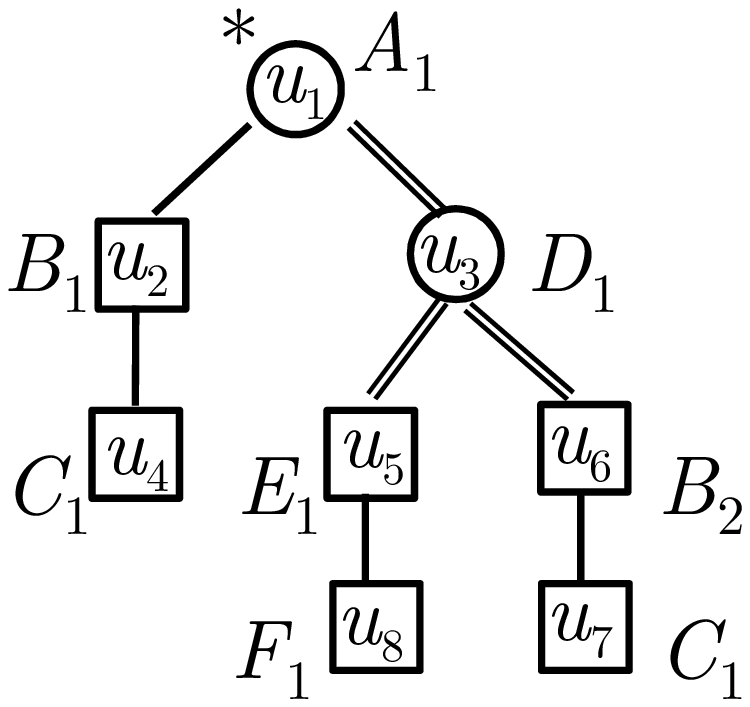}
\end{minipage}}
\subfigure[$Q_3$]{     
\label{mingtpq}   
\begin{minipage}[t]{0.12\textwidth} 
\centering
\includegraphics[height=1in]{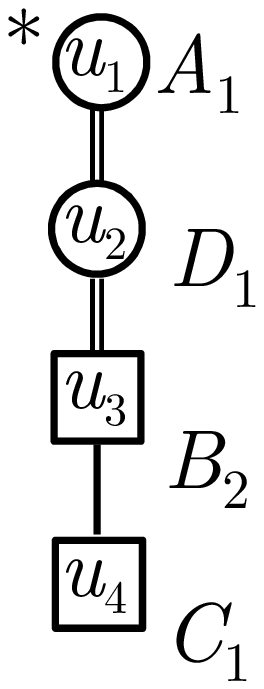}
\end{minipage}}
\caption{Examples for four fundamental problems of GTPQs}
\label{satisfy} 
\end{figure}

\begin{exmp}
Consider the query in Fig$.$ \ref{gtpq}. All query
nodes are independently constraint nodes. Replacing $p_{u_7}$ with $p_{u_7}\wedge(p_{u_9}\vee p_{u_{10}})$
in $f_{ext}(u_3)$, we have 
$f_{tr}(u_3)=\neg p_{u_6}\vee(p_{u_7}\wedge(p_{u_9}\vee p_{u_{10}})\wedge
p_{u_8})$.   Since there are no two nodes $u$ and $u'$ such that $u\trianglelefteq u'$,
 $f_{cs}(u_1)=f_{tr}(u_1)=p_{u_5}\wedge p_{u_4}
\wedge p_{p_5} \wedge p_{u_3} \wedge \big(\neg
p_{u_6}\vee(p_{u_7}\wedge(p_{u_9}\vee p_{u_{10}})\wedge p_{u_8})\big)$. Due to the satisfiability of $f_{cs}(u_1)$,
we see that the query is satisfiable. Indeed, we can get a nonempty answer by
posing $Q$ on $G$ in Fig$.$ \ref{gtpq} as shown in Example
\ref{exmp:gtpq}.
 
Let us turn to $Q_1$ and $Q_2$ depicted in Fig$.$ \ref{satisfy}. The following
table presents structural predicates of internal nodes for $Q_1$ and $Q_2$.

\begin{table}[h] 
\small
\centering
\begin{tabular}{|l|l|l|}  
\hline 
$f_s(u_1)=\neg p_{u_2}$ & $f_s(u_2)= p_{u_4}$ &
$f_s(u_5)=p_{u_8}$ \\
\hline
\multicolumn{2}{|c|}{$f_s(u_3)=(p_{u_5}\wedge p_{u_6})\vee (\neg p_{u_5}\wedge
p_{u_6})$} & $f_s(u_6)=p_{u_7}$\\ 
\hline 
\end{tabular} 
\end{table}

For both queries, $u_5$ and $u_8$ are two non-independently constraint nodes. 
In $Q_1$, we have $u_2\trianglelefteq u_6$, because
(1) $u_6\vdash u_2$, (2) $u_4\trianglelefteq u_7$, (3) 
$f_{tr}(u_6)\rightarrow f_{tr}(u_2)[u_2\mapsto u_6]= p_{u_7}\rightarrow
p_{u_7}$, which is a tautology, (4) $u_2$ is an AD child of $u_1$ which is an
ancestor of $u_6$. In contrast, for $Q_2$, $u_2\not\trianglelefteq u_6$, since
now $u_2$ is a PC child of $u_1$ but $u_6$ is not. Suppose attribute predicates
of all nodes are satisfiable. Then for $Q_2$, $f^2_{cs}(u_1)=\neg(p_{u_2}\wedge
p_{u_4})\wedge p_{u_3}\wedge\big((p_{u_5}\wedge p_{u_6}\wedge p_{u_7})\vee(\neg
p_{u_5}\wedge p_{u_6}\wedge p_{u_7})\big)$,
which is satisfiable; but for  $Q_1$,
$f^1_{cs}(u_1)=f^2_{cs}(u_1)\wedge\big(p_{u_6}\rightarrow (p_{u_2}\wedge
p_{u_4})\big)$, which is unsatisfiable.  Therefore, we know that $Q_2$ is
satisfiable and $Q_1$ not.
\label{exmp:sat}
\qed
\end{exmp}

\subsection{Containment and Equivalence} 
For two GTPQs $Q_1$ and $Q_2$, $Q_1$ is \emph{contained}
in $Q_2$, denoted by $Q_1\sqsubseteq Q_2$, if for any data graph $G$, $Q_1(G)\subseteq
Q_2(G)$. $Q_1$ and $Q_2$ is \emph{equivalent},
denoted by $Q_1\equiv Q_2$, if $Q_1(G)\subseteq Q_2(G)$ and
$Q_2(G)\subseteq Q_1(G)$.

\begin{mdef}[Homomorphism]
Given two GTPQs $Q_1$ with query nodes $V^1_q$ and
$Q_2$ with query nodes $V^2_q$, a homomorphism 
from $Q_1$ to $Q_2$ is a mapping $\lambda$ from $V^1_q$ to $V^2_q\cup\{\perp\}$
such that (1) the two sets of output nodes of  $Q_1$ and  $Q_2$ are bijective;   
(2) for any non-independently constraint node $u\in V^1_q$, $\lambda(u)=\perp$; (3) 
for any independently constraint node $u_1$ in $V^1_q$,
(a) for any PC (resp, AD) child node $u'_1$ of $u_1$ such that $u'_1$ is also an
independently constraint node, $\lambda (u'_1)$ is a PC child (resp, a
descendant) of $\lambda(u_1)$, and (b) $\lambda(u_1)\vdash u_1$; (4) the 
formula $f_{cs}(u^2_{root})\to f_{cs}(u^1_{root})[u^1_{root}\mapsto
\lambda(u^1_{root})]$   is a tautology, where $u^1_{root}$ is the root node of
$Q_1$  and $f_{cs}(u^1_{root})[u^1_{root}\mapsto \lambda(u^1_{root})]$ is a
formula  transformed  from $f_{cs}(u^1_{root})$ by replacing $p_{u'}$ with 
$p_{\lambda(u')}$ for each independently constraint node $u'\in V^1_q$.
\end{mdef}

Theorem \ref{thm:homo} yields a decision procedure for containment and
equivalence between two GTPQs. Theorem \ref{thm:homo:time} states the
intractability of the two problems of
containment and equivalence.
% \vspace{-1.5mm}
\begin{thm}
\label{thm:homo}
For two GTPQs $Q_1$ and $Q_2$, $Q_1\sqsubseteq Q_2$ iff there exists a
homomorphism from $Q_2$ to $Q_1$. \qed
\end{thm}
\vspace{-1.4mm}   
\begin{thm}
\label{thm:homo:time}
The containment checking for GTPQs is  co-NP-hard. \qed
\end{thm}
% \vspace{-1mm}
\begin{exmp}
\label{exmp:satisfy}
Recall the queries in Fig$.$ \ref{satisfy}. We now assume  $f_s(u_1)=p_{u_2}$
and others the same as in Example \ref{exmp:sat}. Let $Q_3$ be a
conjunctive GTPQ, and $u^j_i$ denote $u_i$ in $Q_j$ to distinguish
nodes in different queries. We have that $Q_2\sqsubseteq Q_3$, $Q_2\sqsubseteq Q_1$ and
$Q_1\equiv Q_3$. Indeed, there is a homomorphism $\lambda_{3,2}$ from $Q_3$ to $Q_2$, where $\lambda_{3,2}(u^3_1)=u^2_1, \lambda_{3,2}(u^3_2)=u^2_3, \lambda_{3,2}(u^3_3)=u^2_6, \lambda_{3,2}(u^3_4)=u^2_7$. There is
also $\lambda_{1,3}$ from $Q_1$ to $Q_3$, in which
$\lambda_{1,3}(u^1_i)=\perp (i=5,8), \lambda_{1,3}(u^1_j)=u^3_3 (j=2,6),
\lambda_{1,3}(u^1_k)=u^3_4(k=4,7), \lambda_{1,3}(u^1_1)\\=u^3_1,
\lambda_{1,3}(u^1_3)=u^3_2$. We can also derive $\lambda_{3,1}$  and $\lambda_{1,2}$.
\qed
\end{exmp}

\begin{algorithm}[t] 
% \begin{multicols}{2}
\small
\DontPrintSemicolon
\SetKwInput{KwData}{Input}\SetKwInput{KwResult}{Output}
\KwData{GTPQ $Q=(V_b, V_p, V_o, E_q,
 f_a, f_e, f_s)$ with the root  $u_r$.}
\KwResult{A minimum equivalent GTPQ $Q_m$ of $Q$.} 
% %\BlankLine
construct an equivalent query $Q_m$ from $Q$ by removing  subtrees
rooted at a node whose attribute predicate is unsatisfiable and assigning the
variables of the removed nodes to 0 for respective structural predicates
\label{alg:rmvattribute}
\;
 
check each structural predicate to determine for each node whether it is an
independently constraint node  and remove all non-independently constraint nodes
followed by assigning the variables of them
to 0 for respective structural predicates \label{alg:simp2} \;

% compute the transitive structural predicate $f_{tr}(u)$ for each node $u$ in
% $Q_m$ from bottom to up \label{alg:ts}\;
% % 
% % \ForEach {pair $(u_i, u_j)\in V^m_q\times V^m_q (i\neq j)$ \emph{s.t.}
% % $u_j\vdash u_i$}{
% % 	\lIf {$u_i\trianglelefteq u_j$}{
% % 		$S:=S\cup{(u_i, u_j)}$\;
% % 	}
% % }

compute the complete structural predicate $f_{cs}(u)$ for each node $u$  in
$Q_m$ in bottom-up order\label{alg:cs}\;
 \ForEach{$u\in V^m_q$ in bottom-up order do \label{alg:for}}{
	\If{$f_{cs}(u)$ is unsatisfiable}{
		$f_{s}\big(parent(u)\big):=f_{s}\big(parent(u)\big)[p_u/0]$ \;
		remove the whole subtree rooted at $u$ from $Q_m$ \label{alg:ifend}\;
	}
}

\ForEach{node $u\in V^m_q$ \label{alg:ifend}}{
	\If{the formula $f_{cs}(u_r)\rightarrow p_{u}$ is a tautology}{
		\ForEach{ $u'$ such that $u'\trianglelefteq u$}{
			$f_{s}\big(parent(u')\big):=f_{s}\big(parent(u')\big)[p_{u'}/1]$ \;
			\ForEach{ output node $u_o$ in the subtree rooted  at $u'$}{ 
% 				\ForEach{ $u''$ in the subtree rooted by $p_u$ such that $u_o\triangleleft
% 				u''$ } {
% 					\If{the subtree query pattern rooted by $u''$ and that rooted by $u_o$ are
% 					isomorphic}{
% 						remove $u_o$ from the output node list and add $u''$ into it\;
% 					}
% 			    }
				\If{there exists $u''$ such that $u_o\triangleleft u''$ and the subtree
				query pattern rooted at $u''$ and that rooted at $u_o$ are isomorphic}{
						remove $u_o$ from the set of output nodes and add $u''$ into it\;
					}
			 }
				remove nodes in the subtree rooted at $u'$ from $Q_m$ that are not
				ancestors of any output nodes and corresponding edges they connect\;
			
	}}\ElseIf{the formula $f_{cs}(u_r)\rightarrow \neg p_{u}$ is a tautology}{
		\ForEach{pair $(u, u')\in S$ }{
			$f_{s}\big(parent(u')\big):=f_{s}\big(parent(u')\big)[p_{u'}/0]$ \;
			remove the whole subtree rooted at $u'$ from $Q_m$\;
		}
	} 
 
}
 
\Return $Q_m$
\caption{minGTPQ\label{min}}
% \end{multicols}
\end{algorithm}

\subsection{Minimization}
 
Since the
efficiency of processing a query depends on the size of it, it is necessary
to identify and eliminate redundant nodes. For a GTPQ with query nodes $V_q$, we
define its size as $|Q|=|V_q|$. 

\begin{mdef}[Minimization] Given a
GTPQ $Q$, the minimization problem is to find  another GTPQ $Q_m$ such that (1)
$Q\equiv Q_m$, (2) $|Q_m|\leq|Q|$, and (3) there exists no other such $Q'$ with $|Q'|<|Q_m|$.
\end{mdef}

From Theorem \ref{thm:homo}, we have that for a GTPQ $Q$, there is a minimal
equivalent GTPQ of $Q$ whose query nodes are a subset of query nodes of $Q$. We
say two GTPQs $Q_1$ and $Q_2$ are isomorphic, if there is a homomorphism between
them that is a one-to-one mapping. The following proposition shows that the
minimal equivalent query of a GTPQ is unique up to isomorphism. 
% \vspace{-1.5mm}
\begin{prop} 
Let GTPQs $Q_1$ and $Q_2$ be minimal and equivalent. Then $Q_1$ and $Q_2$ are
isomorphic.\qed
\end{prop} 
% \vspace{-1.5mm}  
Algorithm \ref{min} shows how to minimize a GTPQ. We
% It  removes all nodes     
% whose attribute predicates (line ) or structural predicates (line ) are
% unsatisfiable and the subtrees rooted b y them. The propositional formulas of
% structural predicates then are
% simplified(line ) by removing redundant variables and accordingly the
% corresponding nodes are removed. Finally, the subtrees which  The
% correctness can be proved based on Theorem \ref{thm:homo}.
give an example to illustrate it.
 
\begin{exmp}
In
Fig$.$ \ref{satisfy}, the query $Q_3$ is a minimum equivalent query of $Q_1$
with structural predicates given in Example \ref{exmp:satisfy}.  (1)  Since we suppose all
attribute predicates are satisfiable, there are no nodes  to be removed in this
step, and $Q_m=Q_1$ (line 1). (2) All nodes except $u_5$ and $u_8$  are
independently constraint nodes, hence we remove $u_5$ and $u_8$ and assign 0
to  $p_{u_5}$  in $f_s(u_3)$, thereby having that $f_s(u_3)=p_{u_6}$ (line 2).
In this step, all propositional formulas of structural predicates are \emph{simplified} to
equivalent formulas with minimum variables. (3) There are
no nodes whose complete structural predicates are unsatisfiable, and so none is
removed (line 4--7).  (4)  The formula $f_{cs}(u_1)\rightarrow p_{u_6}$ is a
tautology and $u_2\trianglelefteq u_6$, so $u_2$ and its child $u_4$ is removed,
and we have $f_s(u_1)=1$,   thereby generating the query $Q_3$ (line 8--19).
This step is to remove subtrees which can be \emph{semantically subsumed} by others.
\qed
\end{exmp}

The
correctness can be proved based on Theorem \ref{thm:homo}. Since the algorithm
involves solving SAT problems, the worst-case time complexity is exponential in
the query size. In fact, Theorem \ref{thm:minnp} shows that the minimization
problem is NP-hard and hence it is difficult to find a polynomial-time algorithm.
 Nevertheless, because there are many high-performance algorithms for SAT and
 the query size is not much large in practice, it is still worth minimizing a GTPQ
 considering the benefits of efficiency of evaluation.

\begin{thm}
\label{thm:minnp}
\mbox{The minimization problem for GTPQs is NP-hard.\qed}
\end{thm}

\section{Evaluating Generalized Tree\\ Pattern queries}
\label{section4}
 
\subsection{Framework}  
\label{frame}
Recall that two major problems that impair the efficiency of algorithms for
processing TPQs over graphs are large intermediate results and
expensive join operations on them. In the following, we propose two new
techniques to address them.
 
\begin{mdef}[Graph representation of intermediate results]
To reduce the cost of storing intermediate results and avoid  merge-join
operations, we represent intermediate results as a graph rather than sets of
tuples. Each match for a path or a substructure of the  query pattern
  can be embedded into the tree pattern and hence naturally can
be represented as a tree. By grouping all the candidate matches by the
corresponding matched query nodes and adding an edge to connect a pair of data nodes whenever 
there's an edge between the corresponding pair of query nodes in the query pattern, 
we can represent the intermediate and final results as graphs.   In such a graph representation, 
each data node exists at most once, in contrast to the tuple representation in which a
data node may be in multiple tuples. Also, the AD or PC relationship between
two nodes is exactly represented by only one edge, while in the tuple form the
corresponding two nodes may be put as an element  in more than one tuple to
repeatedly and explicitly represent their relationship. Since the size of the intermediate
matches may be huge, even exponential in both the query size and the data size in the worst case,
 the graph representation is much more compact with at most quadratic space
cost. Moreover, to enumerate all resulting matches of a pattern query, we only need to
perform one single graph traversal on a presumably small graph  instead of 
multiple merge-join operations over large intermediate results. 

It is worth noting that such a way of representing intermediate results can be
also applied to algorithms for other graph pattern queries to boost their
evaluation.  For TPQs, it is particularly optimal because we can
enumerate matches directly from the graph. However, for graph pattern queries, additional matching
operations including joins may be unavoidable because it
 is difficult to locally determine which nodes should be traversed to
form a match. The additional matching operations are in essence an easier evaluation
of a pattern matching on a  smaller graph, such a technique can thus  still be
expected to speed up the whole processing.
\end{mdef}
\begin{mdef}[Reachability index enhanced effective pruning]
Since the number of data nodes to be processed 
significantly affects the efficiency of  pattern query evaluation, 
it is desirable to perform effective pruning to 
reduce the number of candidate matching nodes. In the literature, \cite{stackd}
and \cite{jointkde} have developed two pruning approaches for reachability
query pattern matching. TwigStackD \cite{stackd} proposed a pre-filtering
approach that can select nodes guaranteed to be in final matches. Since it has to 
perform two graph traversals on the data graph, it is likely unfeasible
 for large-scale real-world graphs. The work
 \cite{jointkde} on pattern queries over labeled graphs
proposed another pruning process, namely \textit{R}-semijoin, using a special
index called cluster-based \textit{R}-join index. It can filter nodes that cannot possibly
contribute to partial matches for an AD edge between two labeled query nodes. However, (1) the
selected nodes may be still redundant since the nodes only satisfy the reachability condition
imposed by one edge and the global structural satisfaction is not checked.
(2)  It is highly costly to construct and store the \textit{R}-join index for a large
data graph since the index essentially precomputes and stores all matches for pairwise
labels and the index size is quadratic  in the graph size. (3) It cannot be used
to perform pruning for queries that have expressive attribute predicates rather
than a fixed set of labels associated with nodes. Since
predicates of query nodes are often not fixed and predictable, the
index actually cannot be precomputed and this approach cannot be used.
 
We explore the potentials of existing reachability
index for effective pruning. It is interesting to note that most reachability indexing
 schemes follow a
 paradigm. They first utilize a relatively simple reachability index which often
 assigns two or three labels to
each node in order to cover the reachability of a substructure, called a cover, such as 
tree-cover in \cite{opt, dual}, path-tree in
\cite{path}, and chain-cover in \cite{chain2, th}. To cover the
remaining reachability information, each node keeps one or two lists where complete or
just  a portion of ancestors and descendants are stored.   When answering whether a node can reach another, the algorithms 
typically  use nodes stored in the lists as the intermediate to determine the
reachability. 
%  Specifically, they look up the index in the lists of one node to 
% determine the reachability on the cover with the target node or its list
% nodes using the reachability labels on the cover. 

When it comes to answer a number of reachability queries between two sets of
nodes, the following two observations are helpful: (1) the lists of different nodes often share a number of nodes, 
(2) the nodes in different lists have rich reachability information.
If we merge the 
lists of a set of nodes by eliminating the duplicates  and those whose 
reachability information can be derived from others, the merged list
``subsumes'' all the reachability information in the original lists of 
the node set but the size will not be much larger,  and possibly even much
smaller, than the list size of any individual node. Using the merged list, 
reachability patterns are likely to be  evaluated more efficiently.

For example, considering a  reachability pattern $u_A$---$u_B$, we want to
filter data nodes in $mat(u_A)$ that cannot reach any nodes in $mat(u_B)$. 
Instead of  performing 
 $|mat(u_A)|\times|mat(u_B)|$ pairwise reachability
queries to check for each node $v\in mat(u_A)$  whether it can reach a node $v'\in
mat(u_B)$, (1) we
  merge all index lists of $mat(u_B)$ to a single list of the minimum  size
 that preserves all the reachability information saved in the original
lists; and (2) for each $v\in mat(u_A)$,  use  the list of $v$ and the merged
list rather than individual lists for $mat(u_B)$  to holistically determine whether
$v$ reaches some node in $mat(u_B)$. Intuitively, we can regard the set $mat(u_B)$ as a single dummy node
 which   is reachable from all nodes that are ancestors of nodes in
$mat(u_B)$.  
%  For a node $v$,  $v'$ does not necessarily to be exactly identified.   Recall the processing paradigm of typical reachability schemes. 
% We can merge all lists of $mat(u_B)$ to a single list of minimum  size while 
% preserving all the reachability information that can be derived from original
% lists.  Now,   Because the merged list is expected to  be much smaller than the
% total size of lists before merged,  it can efficiently prune  redundant nodes
% before doing more  costly operations for exactly   determining which two nodes
% in $mat(u_A)$ and $mat(u_B)$  form a match of  the query.

% Such an approach can be readily applied to process TPQs with disjunction and
% negation. For queries with only a portion of query
% nodes as output nodes,  it also allows us to avoid  generating matches for every
% query edge,  thereby saving substantial
% time and space cost.  

In this paper, we use 3-hop \cite{th} as the underlying reachability index scheme, 
as 3-hop has both a very compact index size and
reasonable query processing time. 
As different labeling schemes are often preferable to different graph structures, 
it is also very flexible for our framework to use other 
labeling schemes to efficiently process different types of graphs.
 
We restrict our attention to in-memory processing and do not address the issues
relating to disk-based access methods and physical representation of graph data.
   
\begin{mdef}[Algorithm outline]
Our \underline{GT}PQ \underline{e}valuation \underline{a}lgorithm (referred to
as GTEA) is outlined as follows. First, it prunes candidate matching nodes that do
not satisfy downward structural constraints (i$.$e$.$ not satisfy the subtree
pattern query rooted at the corresponding query node). Second, it performs the
second round pruning process on a carefully selected subtree pattern, called
prime subtree, to remove nodes not satisfying upward structural constraints
(i.e. not reachable from any candidate nodes of the root). Third, the prime subtree is
further shrunk if possible, and GTEA generates the matches of the
shrunk prime subtree while representing the intermediate results as a graph,
from which the final results can be efficiently obtained. We begin with  focusing on evaluating GTPQs
with AD edges only and show how to extend  the algorithm to process PC edges in Section \ref{pc-edge}.
 
% In this paper, we   
\end{mdef}

%Overall, our algorithm for evaluating GTPQs is outlined as follows. First, we
%filter candidate matching nodes that do not satisfy downward structural
%constraints. Second, we further filter nodes not satisfying
%upward structural constraints  for a portion of query nodes chosen  based on
%output nodes and the results of the first step. Third, we extract a subtree from
%the tree pattern such that it imposes enough constraints to induce the final
%matches. Finally, we
%generate the matches of the subtree and represent the results as a graph,  from
%which the final results can be obtained. We begin with  focusing
%on evaluating GTPQs with AD edges only and show how to extend the algorithm to process PC
%edges in Section 4.4. 
% In this paper, we  
\end{mdef}

\subsection{Pruning  Candidate Matching Nodes}\label{prune}
We use 
%[[]]
a
two-round pruning process to filter unqualified data nodes. The first
round selects data nodes that satisfy downward
structural constraints of the query pattern for each query node. At the second
round, we then obtain a minimum subtree that contains all output nodes having
more than one
%[[]]
candidate matching node, and select necessary edges from this subtree to
find nodes satisfying upward structural constraints.

\begin{figure}[t]
\centering
\includegraphics[height=1.4in]{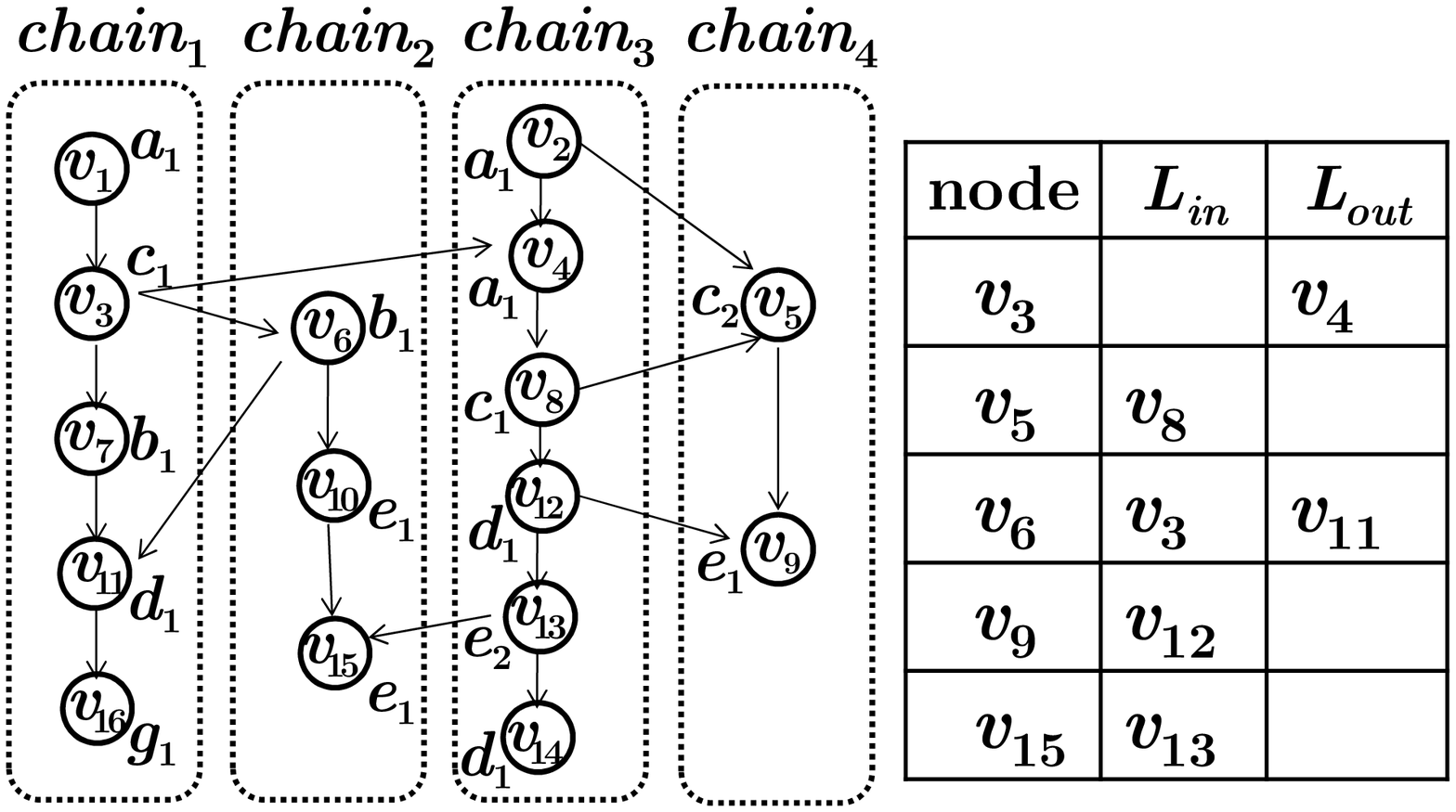}
\caption{Chain decomposition and 3-hop index}
\label{chains} 
\end{figure}  

\subsubsection{Preliminary: Merging  3-hop index}
\label{secion41}
3-hop  is a recent graph reachability indexing scheme well-known for
its compact index size and reasonable query time. It follows the 
indexing paradigm mentioned in Section \ref{frame}. It uses the chain-cover
which consists of a set of disjoint chains covering all nodes in the graph. 
Each node in the graph is assigned a chain ID $cid$ and its sequence number
$sid$ on its chain. For two nodes $v$ and $v'$ on the same chain (i.e$.$,
$v.cid=v'.cid$), $v\leq_c v'$, if $v.sid\leq v'.sid$. In particular, if
$v.sid<v'.sid$, we  say $v$ is \emph{smaller} than $v'$. Obviously, reachability
on the chain-cover can be answered using chain IDs and sequence numbers. To encode the  
remaining reachability information outside chain-cover, 3-hop records a
successor list $L_{out}(v)$ $\big($resp$.$ predecessor list $L_{in}(v)$$\big)$
of ``entry'' (resp$.$ ``exit'') nodes to (resp$.$ from) other chains for each node $v$.  The
entry (resp$.$ exit) node to (resp$.$ from) a chain is the smallest (resp$.$
largest) one on that chain that $v$ reaches (resp$.$ reaches $v$). See \cite{th} for details of 3-hop
index construction. For answering the reachability between two nodes $v_1$ and $v_2$ on different 
chains, 3-hop takes the following steps. (1) Collect the smallest nodes on any
other chain that $v_1$ can reach through exit nodes of chain $v_1.cid$. That
is, we get a set of nodes $X_{v_1}=\{x|x\in
\bigcup_{v_1\leq_c v'} L_{out}(v') $$\textrm{ and } \forall v'$$\geq_c$$v_1,
x\leq_c L^{x.cid}_{out}(v')\}$ $\cup\{v_1\}$, where $L^{x.cid}_{out}(v')$ is the
entry node of $v'$ on chain $x.cid$. We call $X_{v_1}$ the
\emph{complete successor list} of $v_1$. (2) Collect the largest nodes on any chain that can reach $v_2$ through entry nodes of
chain $v_2.cid$. In this step, we get a set of nodes $Y_{v_2}=\{y|y\in
\bigcup_{v'\leq_c v_2} L_{in}(v') \textrm{ and } \forall v'$$\leq_c$$v_2$, $
$$L^{y.cid}_{in}(v')\leq_c y\}\cup\{v_2\}$, where $L^{y.cid}_{in}(v')$ is the
exit node of $v'$ on chain $y.cid$. We
call $Y_{v_2}$ the \emph{complete predecessor list} of $v_2$. (3) If there is a
pair $(x, y) (x\in X_{v_1}, y\in Y_{v_2})$ such that $x\leq_c y$, then we can
conclude that $v_1$ can reach $v_2$.

\begin{exmp}
Fig$.$ \ref{chains} gives a chain decomposition of $G$ of Fig$.$
\ref{datagraph} and the corresponding 3-hop index. Chain IDs and sequence
 numbers are omitted. As an example,
$v_3.cid=v_{11}.cid=1$, $v_{11}.sid=4$ and $v_3.sid=2$.  Because 
$v_3.sid<v_{11}.sid$, $v_3\leq_c v_{11}$ and $v_{11}$ is reachable from $v_3$. 
To answer whether $v_3$ can reach $v_{9}$, we collect the entry nodes in 
$L_{out}(v_i) (i=3, 7, 11, 16)$ into $X_{v_3}=\{v_3, v_4\}$.  Then we look up
the exit nodes in $L_{in}(v_j) (j=9, 5)$ and get $Y_{v_{9}}=\{v_9, v_{12}\}$. 
Since there is a pair $(v_4, v_{12})$ such that $v_4\in X_{v_3}, v_{12}\in Y_{v_{9}}$, and
$v_4\leq_c v_{12}$, we say $v_3$ can reach $v_{9}$.
\qed 
\end{exmp}

Note that to obtain the complete predecessor (resp$.$ successor) lists, the
original 3-hop needs to visit all larger (resp$.$ smaller) nodes. We can assign a 
forward (and backward) tracing pointer to each node which points to the smallest
larger (resp$.$ largest smaller) node whose $L_{out}$ (resp$.$  $L_{in}$) list is
nonempty so that nodes with empty lists can be skipped.  We define two
operations \emph{next}($v$) and \emph{pre}($v$) on each node $v$, which return the node that the forward and the backward tracing
pointer points  to respectively. For example, since
$v_6$ is the largest smaller node that has a non-empty $L_{in}$ w.r.t. $v_{15}$,
prev$(v_{15})=v_6$.
% \subsubsection{Merging process}
% As  introduced in our framework, we merge the
% predecessor and successor lists of the candidate matching nodes for each query
% node to facilitate the pruning process.

A basic operation of the pruning
process is merging the complete predecessor/successor  lists for a given set of
data nodes (denoted by $S$). For the 3-hop case, it picks the largest (resp$.$
smallest) nodes on each chain from the complete predecessor (resp$.$ successor)
list and we call the resultant list  \emph{predecessor contour} $C^p$ (resp$.$
\emph{successor contour} $C^s$). A node $v$ is said to reach (resp$.$ be
reachable from) $S$ if $v$ reaches (resp$.$ is reachable from) at least one node in $S$. 
We have the following proposition.

\begin{prop}
\label{prop-reach}
A data node $v$ reaches $mat(u)$ iff there is a pair $\textup{(}x,
y\textup{)}\in X_v \times C^p$ such that $x\leq_c y$, while $mat(u)$ reaches
$v$ iff there exists a pair $\textup{(}x,
y\textup{)}\in C^s \times Y_v$ such that $x\leq_c y$. \qed
\end{prop} 

% Clearly, \emph{
% a node $v$ can reach  $S$, iff there is a pair ($x, y$)($x\in X,
% y\in C^p$) such that $x\leq_c y$} (for simplicity, we say
% $v$ can reach a set of nodes  if $v$ can reach at least one node in
% the set).   Similarly, when checking whether a node
% is reachable from one in $S$, we only need to look up nodes in $C^s$ instead of
% the complete successor list for every node in $S$.

% Procedure MergePredLists (Fig$.$ \ref{merge}) sketches the process for
% calculating the predecessor contour. For each chain $i$, we use
% $visited_i$ to record the largest node  whose predecessor list has been looked up in the chain $i$. We assume nodes in $S$ follow the
% descending order of sequence numbers, which can be easily maintained in the
% process  of evaluation without additional cost. For each node $v\in S$,
% procedure mergePredLists processes $v$, and every node $v'$ that is larger
% than $v$ and not processed,  as follows.  In each iteration,  it compares the
% sequence number of the entry node  in the predecessor list with nodes
% recorded in $C^p$,  and then update $C^p$ if a larger node is found (line 4--9).

\SetAlgorithmName{Procedure}{Procedure}  
  
% \begin{algorithm}[tb] 
% \small
% \DontPrintSemicolon
% \SetAlFnt{\small}
% \SetKwInput{KwData}{Input}\SetKwInput{KwResult}{Output}
% \KwData{A set of nodes $S$. The nodes on the same chain follow the descending
% order of sequence numbers.}
% \KwResult{The predecessor contour $C^p$} 
% %\BlankLine 
% \ForEach{node $v\in S$}{
% 	\lIf{$C^p[v.cid]<v.sid$}{
%     	$C^p[v.cid]:=v.sid$\;	
%     }
%     $v':=v$\;
%     \Repeat{$v' = null$ or $visited_{v'.cid}\geq v'.sid$}{
% 		\ForEach{index node $v''\in L_{in}(v')$}{ \label{merge:update}
% 			\If{$C^p[v''.cid]<v''.sid$}{
% 				$C^p[v''.cid]:=v''.sid$\;\label{merge:end}
% 			}
% 		}
% 		$v':=\textrm{prev}(v')$
% 	}
% 	\lIf{$visited_{v.cid} < v.sid$}{
% 		$visited_{v.cid}:=v.sid$\;	
% 	}
% % 	\ForEach{node $v'$ $s.t.$ $v'\leq_c v$}{
% % 		\lIf{$v'.visited={\rm true}$}{ \label{merge:check}
% % 			\textbf{break} \;
% % 		}
% % 		 $v'.visited:={\rm true}$ \; \label{merge:mark}
% % 		\ForEach{index node $v''\in L_{in}(v')$}{ \label{merge:update}
% % 			\If{$C^p[v''.cid]<v''.sid$}{
% % 				$C^p[v''.cid]:=v'.sid$\;\label{merge:end}
% % 			}
% % 		}
% % 	}	
% } 
% \Return{$C^p$}
% \caption{MergePredLists\label{merge}}
% \end{algorithm} 

\begin{algorithm}[tb] 
\small
\DontPrintSemicolon
\SetAlFnt{\small}
\SetKwInput{KwData}{Input}\SetKwInput{KwResult}{Output}
\KwData{A set of nodes $S$.}
\KwResult{The predecessor contour $C^p$ of $S$.} 
%\BlankLine 
\ForEach{node $v\in S$}{
	\lIf{$C^p[v.cid]<v.sid$}{
    	$C^p[v.cid]:=v.sid$\;	
    }
    $v':=v$\;
    \Repeat{$v' = null$ or $visited_{v'.cid}\geq v'.sid$}{
		\ForEach{index node $v''\in L_{in}(v')$}{ \label{merge:update}
			\If{$C^p[v''.cid]<v''.sid$}{
				$C^p[v''.cid]:=v''.sid$\;\label{merge:end}
			}
		}
		$v':=\textrm{prev}(v')$
	}
	\lIf{$visited_{v.cid} < v.sid$}{
		$visited_{v.cid}:=v.sid$\;	
	}
} 
\Return{$C^p$}
\caption{MergePredLists\label{merge}}
\end{algorithm} 

Procedure \ref{merge} sketches the process of
calculating the predecessor contour $C^p$, where $visited_i$ records
  the largest node on chain $i$ whose predecessor list
has been looked up. 
For each node $v\in S$, MergePredLists processes $v$ and
those smaller nodes whose predecessor lists have not been looked up as
follows. For each node $v'$ to be processed and each exit node $v''$ in 
$L_{in}(v')$, it compares $v''$ with the nodes in $C^p$ on the
same chain of $v''$, and update $C^p$ if $v''$ is larger (line 4--9).
 To retrieve nodes from $C^p$ efficiently, $C^p$ can be implemented as a map 
 that uses chain IDs as keys and the sequence numbers as values.

\begin{exmp}
We show how to compute the predecessor contour of 
$mat(u_{10})$ for the query $Q$ of
Fig$.$ \ref{exmp:des}.  Example \ref{exmp:gtpq} have given that
$mat(u_{10})=\{ v_9, v_{10}, v_{13}, v_{15}\}$.  The procedure collects the complete predecessor
lists for each of $mat(u_{10})$ one by one,  but no predecessor list is
repeatedly visited. For example,  assume that $v_{10}$ is read before 
$v_{15}$. When collecting $Y_{v_{15}}$,  although prev($v_{15}$) points to
$v_6$, MergePredLists needs not look up $L_{in}(v_{6})$, because the list has
 been looked up when collecting $Y_{v_{10}}$. The predecessor contour of
 $mat(u_{10})$ is $\{v_3, v_9, v_{13}, v_{15}\}$.  It can be easily verified that the
size of this predecessor contour is  a half of the total size of the four
individual complete lists of $v_9,
v_{10}, v_{13}$ and $v_{15}$. Note that the size of a predecessor contour is
bounded by  the number of chains. This example actually gives the worst
case but still has a high  compression rate (50\%).
\qed 
\end{exmp}

\begin{algorithm}[tb]  
\small
\DontPrintSemicolon
\SetAlFnt{\small}
\SetKwInput{KwData}{Input}\SetKwInput{KwResult}{Output}
\KwData{3-hop index $L_{out}$, a GTPQ $Q$.}
\KwResult{Updated candidate matching nodes satisfying downward structural
constraints.}
%\BlankLine
\lForEach{node $u\in V_q$}{
	$mat(u):=\{x|x\in V, x\sim u\}$\;
}
\lForEach{leaf node $u'$ in $V_q$}{
	$C^p_{u'}:=\textrm{MergePredLists}\big(mat(u')\big)$\;
}
$V'_q=V_q\backslash\{u'|u'\textrm{ is a leaf node}\}$ \;
\ForEach{ $u\in V'_q$ in bottom-up order}{
	\lForEach{ $v\in mat(u)$}{
		$chain_{v.cid}:=chain_{v.cid}\cup \{v\}$\;
	}	
	\ForEach{$chain_i$ that is not empty}{
		\lForEach{child $u'$ of $u$}{
			$val[p_{u'}]:=0$\;
		}
		
		\ForEach{node $v_{i}\in chain_i$}{		
			\ForEach{child $u'$ of $u$ \emph{s.t.}  $val[p_{u'}]=0$ }{
				\If(\tcp*[f]{using Proposition \ref{prop-reach}}){$v_i$ reaches $mat(u')$ }{
					$val[p_{u'}]:=1$
				}
			}
			\If{$f_s(u)$ evaluates to false with the valuation $val$}{
				$mat(u):=mat(u)\backslash\{v_i\}$\;
			}
%			$visited_i:=v_i.sid$\;
		}		
% 		\ForEach{node $v_{i}\in chain_i$}{
% 			\ForEach{node $v'_i$ such that $v_i\leq_c v'_i$}{
% 				\lIf{$v'_i.visited=\textrm{true}$}{
% 					{\bf break}\;
% 				}\lElse{				
% 					$v'_i.visited:=\textrm{true}$ \;
% 				}
% 				\ForEach{index node $v''_i\in L_{out}(v'_i)$}{
% 					\ForEach{child $u'$ of $u$ \emph{s.t.}  $val[p_{u'}]=\textrm{false}$ }{
% 						\If{$C^p_{u'}[v''_i.cid]\neq null$ and $C^p_{u'}[v''_i.cid]\geq
% 						v''_i.sid$}{ $val[p_{u'}]:=\textrm{true}$\;
% 						}
% 					}
% 				}
% 			}
% 			\If{$f_b(u)$ evaluate false with the valuation $val$}{ 
% 				$mat(u):=mat(u)\backslash\{v_i\}$\;
% 			}
% 		}
	}
	$C^p_{u}:=\textrm{MergePredLists}\big(mat(u)\big)$\;
}
\caption{PruneDownward\label{alg:prunedownward}}
\end{algorithm}

\begin{mdef}[Time complexity]
The  time complexity of the procedure is $O(|S|+|L_{in}|)$, where
$|L_{in}|$ is the total size of all predecessor lists in 3-hop index. It can
be observed from the fact that no index node in a predecessor list has been ever
repeatedly visited. 
\end{mdef}
\vspace{1pt}

Following the same line of MergePredLists, we develop 
MergeSuccLists that calculates the successor contour of a node set with time complexity of
$O(|S| + |L_{out}|)$, where $|L_{out}|$ is the total size of all successor
lists in 3-hop index.

\subsubsection{Pruning process for downward structural constraints} 
Procedure  \ref{alg:prunedownward} describes the
first round of the pruning process. In the procedure, $val$ refers to a
valuation for variables associated with query nodes. PruneDownward first
collects   $mat(\cdot)$ sorted in the descending order of sequence numbers for
each query node and calculates the predecessor contours for leaf nodes (line
1--2). Then, it processes each non-leaf query node $u$ following a bottom-up  fashion (line 4--14). For each node $u$, it first groups nodes  $mat(u)$  by chain ID (line 5). 
%Note that the descending 
%order of sequence numbers for nodes in $chain_i$ can be
%naturally maintained without sort operations.  
Then for each candidate matching node $v_i$ of $u$ on each chain $i$,
PruneDownward checks whether $v_i$ satisfies downward structural
constraints (line 8--13). To do this, (1) it first assigns a valuation to
$p_{u'}$ for each child node $u'$ of $u$  according to
the reachability from $v_i$ to $mat(u')$ (line 9--11) %(line 10 and line 15--16)
, (2) and then remove $v_i$ from $mat(u)$ if
the structural predicate $f_s(u)$ of $u$  evaluates to false under the
valuation (line 12--13). Note that when processing the next node  on the same
chain,  the valuation for  the previous node is inherited due to  the transitive
property of transitive closure in a chain. Therefore, no predecessor
list is repeatedly looked up.  After all candidate matching  nodes for $u$ have
been processed, the remaining data nodes in $mat(u)$ must satisfy the  downward
structural constraints. Then  the predecessor contour for $u$ is  computed
(line 14), and used in the  pruning process of the parent node of $u$. The 
procedure terminates after the root is processed. 
 
\begin{exmp} 
We first  show how procedure PruneDownward prunes\\ $mat(u_3)$ of Fig$.$
\ref{exmp:des}. In a bottom-up fashion, before pruning $mat(u_3)$, PruneDownward first processes its
non-leaf child $u_7$. No nodes in $mat(u_7) ($i$.$e$.$ $\{v_6, v_7\})$ are
removed, because $v_6$ can reach both $mat(u_9)$ and $mat(u_{10})$ while $v_{7}$ can reach $mat(u_{10})$. The predecessor contour for
$mat(u_7)$ is then computed and $C^p_{u_7}=\{v_6, v_7\}$. For determining
whether $v_5$ should be removed from $mat(u_3)$, PruneDownward checks the reachability between 
$v_5$ and $mat(u_6)$, $mat(u_7)$, $mat(u_8)$ respectively by using the
predecessor contours. One can verify that $v_5$ cannot reach $mat(u_6)$, which
means $val[p_{u_6}]\\ =0$  and  the structural predicate $f^{v_5}_s(u_3)$
evaluates to true. Thus, $v_5$ remains in  $mat(u_3)$. Because the other two
nodes $v_3$ and $v_8$ are in different chains, they do not inherit the valuation determined by
$v_5$ and PruneDownward needs to check pairwise reachability  between
$\{v_3, v_8\}$ and $\{mat(u_6)$, $mat(u_7)$, $mat(u_8)\}$. Only $v_8$ is subsequently removed, because $p_{u_8}=1, p_{u_6}=p_{u_7}=0$  and
$f^{v_8}_{ext}(u_3)$  evaluates to false. Finally, after this pruning round,
$mat(u_3)=\{v_3,v_5\}$.
  
When PruneDownward  refines $mat(u_1)$ and reads $v_2$, the assignments
of $p_{u_2}$ and $p_{u_3}$ are directly inherited from the result computed in the previous
step of processing $v_4$ and $f^{v_2}_{ext}(u_1)$ immediately evaluates to true
without any index lookups.

PruneDownward gets the following refined candidate
matching nodes which satisfy the downward structural constraints:
 $mat(u_2)\\=\{v_3, v_8\}, mat(u_3)=\{v_3,
v_5\}.$\qed
\end{exmp} 
 
\begin{mdef}[Time complexity]
Since no successor list is repeatedly checked, the 3-hop index is looked up for
at most $|E_q||L_{out}|$ times, where $|E_q|$ is the number of edges in the tree pattern. 
MergePredLists is invoked ($|V_q|-1$) times to compute predecessor
contours for each non-root query node, and the total time cost is $O(|V_{mat}|+|V_q||L_{in}|)$, where $|V_q|$ is the
number of query nodes and $|V_{mat}|$ is the total
size of initial candidate matching nodes (i$.$e$.$ $|V_{mat}|=\Sigma_i|mat(u_i)|$).
Therefore, PruneDownward is in
$O(|V_q|(|L_{in}|+|L_{out}|)+|V_{mat}|)$ time.
\end{mdef}

\begin{algorithm}[tb] 
\small
\SetAlFnt{\small\sf}
\DontPrintSemicolon
\SetKwInput{KwData}{Input}\SetKwInput{KwResult}{Output}
\KwData{3-hop index $L_{in}$, the prime subtree $(V_t, E_t)$.}
\KwResult{Updated candidate matching nodes satisfying upward structural
constraints.}
%\BlankLine
$C^s_{u_{root}}:=\textrm{MergeSuccLists}\big(mat(u_{root})\big)$\;
$V_t:=V_t\backslash\{u_{root}\}$ \;
\ForEach{node $u\in V_t$ in top-down order such that $|mat(u)|>1$}{
	\ForEach{child $u'$ of $u$ \label{upward:groupb} such that $|mat(u')|>1$}{
		\ForEach{node $v\in mat(u')$}{
% 			\If{$|Group_v|=0$}{$chain_{v.cid}=chain_{v.cid}\cup\{v\}$\;}
			$chain_{v.cid}:=chain_{v.cid}\cup\{v\}$\;
			$Group_v:=Group_v \cup \{u'\}$ \;\label{upward:groupe} 
		}
	}
	%\ForEach{$chain_i$ that is nonempty}{  
		%\ForEach{node $v_{i}\in chain_i$}{
	\ForEach{node $v_i$ in a nonempty $chain_i$}{
% 			initialize $mark_j$ for each chain $j$ \;      
%%			\lIf{$C^s_{u}[i]\leq v_i.sid$}{     
%%				$reach:=\textrm{true}$; \textbf{break}\;		  		
%%			}
%%			$v'_i:=v_i$\;
%%			\Repeat{$v'_i$ = null or $visited_i\geq v'_i.sid$}{
% 				\lIf{$v'_i.visited=\textrm{true}$}{\textbf{break}\;}
% 				\lElse{$v'_i.visited:=\textrm{true}$\;}
%%				\ForEach{index node $v''\in L_{in}(v'_i)$}{
%%					\If{$C^s_{u}[v''.cid]\leq v''.sid$}{
%%						$reach:=\textrm{true}$; \textbf{break}\;						
%%					} 
%%				}
%%				\lIf{$reach=\textrm{true}$}{\textbf{break}\;} 
%%				$v'_i:=\textrm{prev}(v'_i)$ \;
%%			}
			
% 			\ForEach{node $v'_i$ such that $v'_i\leq_c v_i$}{
% % 				\lIf{$v'_i.visited=\textrm{true}$}{\textbf{break}\;}
% % 				\lElse{$v'_i.visited:=\textrm{true}$\;}
% 				\ForEach{index node $v''\in L_{in}(v'_i)$}{
% 					\If{$C^s_{u}[v''.cid]\leq v''.sid$}{
% 						$reach:=\textrm{true}$\;
% 						\textbf{break}\;
% 					}
% 				}
% 				\lIf{$reach=\textrm{true}$}{\textbf{break}\;}
% 			}
%%			\If{$reach=\textrm{false}$}{
			\If(\tcp*[f]{using Proposition \ref{prop-reach}}){$mat(u')$ do not reach
			$v_i$}{ \ForEach{$u'\in Group_{v_i}$}{
%%					$mat[u']:=mat[u']\backslash\{v_i\}$ \;
					$mat(u'):=mat(u')\backslash\{v_i\}$ \;
				}
			}
			\lElse{\textbf{break} \;}
%%			 $visited_i:=v_i.sid$\;
		}
	%}
	\ForEach{non-leaf child $u'$ of $u$}{
		$C^s_{u'}:=\textrm{MergeSuccLists}\big(mat(u')\big)$\;
	}
}
\caption{PruneUpward\label{alg:pruneupward}}
\end{algorithm}

\subsubsection{Pruning process for upward structural constraints}

After the fist-round pruning process,  for each
backbone node $u$, the remaining nodes in $mat(u)$ satisfy all the structural
constraints imposed by predicates. Because the results of the query should  consist of  matches 
of output nodes only, the matches for predicate nodes are no longer useful and
do not need to be considered.
Moreover, some backbone nodes may not contribute to determining which
candidate matching  output nodes are in the same instance and hence 
can be also discarded. With these two observations, the structural constraints
of a backbone subtree are enough  to derive the relationships among
candidate matching nodes for the output query nodes. Such a subtree, we call the
\textit{prime subtree}, can be induced by the paths from the query root to all
such output nodes that $|mat(\cdot)|>1$. The next pruning step  only
needs to consider this subtree pattern which in essence is reduced to a 
conjunctive GTPQ.
 
% For example, the
% prime  subtree of $Q$ of Fig$.$ \ref{exmp:des} is induced by $u_1$, $u_2$ and
% $u_4$.  The pruning process needs filter $mat(u_2)$ and $mat(u_4)$ based on the
% reachability conditions $u_1$--$u_2$ and $u_3$--$u_4$.  
% \end{mdef}  
%   \textsc{medium weight}
% \begin{mdef}[Procedure PruneDownward]
 In the opposite 
direction  to  PruneDownward, procedure  PruneUpward (Procedure
\ref{alg:pruneupward})  traverses down the prime subtree.
For each query node $u$, it filters the candidate matching nodes
of each child $u'$ of $u$ (line 3--14). 
All the candidate nodes to be processed are first clustered and merged into
duplicate-free sets according to their chain IDs, where the order of
nodes is reversed (line 4--7). As a data node can match multiple query nodes, the
algorithm uses $Group_v$ to record the corresponding query nodes that $v$ matches (line 7) in order to update $mat(\cdot)$ when a reachability condition is determined (line 10--11). Then,
for each node $v_i \in mat(u')$ on a nonempty $chain_i$, $v_i$ should be
removed if $mat(u)$ cannot reach  $v_i$ according to Proposition
\ref{prop-reach}.  Observe that once a node is confirmed to satisfy the
condition of the incoming edge,  all other  larger nodes  do not need to
be checked since they must also satisfy the condition. 
% \begin{comment}
 
\begin{exmp}
In this example, assume that $u_2$ and
$u_3$ are output nodes of $Q$ of Fig$.$ \ref{exmp:des}. The prime subtree is
induced by $u_1$, $u_2$ and $u_3$.  PruneUpward starts from $u_1$ to refine
$mat(u_2)$ and $mat(u_3)$. After grouping distinct data nodes into $chain$, it
gets $chain_1=\{v_3\}$, $chain_3$$=$$\{v_8\}$, and $chain_4=\{v_5\}$. $v_3$ is in both $mat(u_2)$ and $mat(u_3)$, 
but the procedure only stores one copy
in $chain$ to avoid processing it repeatedly when checking reachability with
$mat(u_1)$. After the two query nodes whose matching candidate nodes have the
identical $v_3$ are inserted to $Group_{v_3}$,  $Group_{v_3}=\{u_2, u_3\}$. Because $mat(u_1)$ reaches
$v_3$, $v_3$ is not removed from either $mat(u_2)$ or $mat(u_3)$. Similarly, it
can be verified that $mat(u_1)$ can reach $v_8$ and $v_5$.  In the end, none 
is removed from $mat(u_2)$ and $mat(u_3)$ after this pruning round.\qed
\end{exmp}  

% \end{comment}
%  \end{mdef} 
\begin{mdef}[Time complexity]
The time complexity is
$O(|V'_{mat}|+(|L_{in}|+|L_{out}|)|V'_t|)$, where
$|V'_t|$ is the number of internal nodes in the prime subtree and $|V'_{mat}|$
is the total size of the remaining candidate matching nodes after the first pruning
round.
\end{mdef}

\subsection{Computing Final Results}
\begin{figure}[t]
\centering
\includegraphics[height=1.23in]{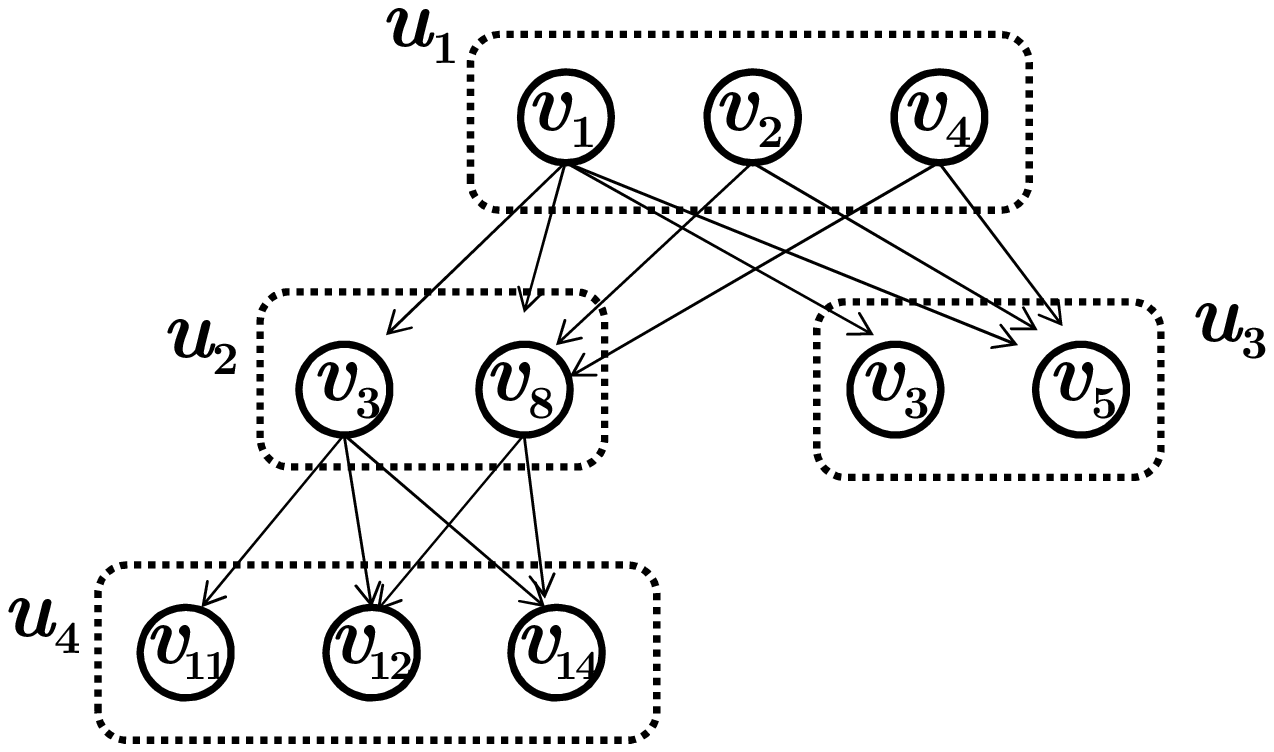}
\caption{Example of the maximal matching graph  for $Q$ over $G$ depicted in
Fig$.$  
\ref{exmp:des} }
\label{match}   
\end{figure}    

\begin{mdef}[Shrunk prime subtree]
As a result of the pruning process, the matching output nodes are
guaranteed to be in the answer. The left to do is to identify how they form
the final results by computing the matches of edges in the prime subtree. Given
a prime subtree, assume that $u$ is the lowest common ancestor of all output
nodes. We can further shrink the  subtree by (1) removing the ancestors of $u$  if $u$ is not the
root, and (2) removing all such nodes $u'$ that  $|mat(u')|=1$.
 If the removing
process leads to disjoint subtrees,  we just compute results for each  subtree, do
 a Cartesian product of  them and add the
candidate matching nodes of removed output nodes to assemble the whole final
results.  From now on, we only need to compute edge matches for the shrunk
prime subtree(s).
\begin{exmp}
The shrunk prime subtree of $Q$ of Fig$.$ \ref{exmp:des}
is induced by  $u_2$ and $u_4$. Even
if we change the query to mark $u_5$ also as an output node,  the  shrunk prime
subtree is \emph{still} the same since $|mat(u_5)|=|\{v_{13}\}|=1$ and $v_{13}$
must be in every answer.\qed
\end{exmp} 
\end{mdef}
\begin{mdef}[Maximal matching graph]
The full matches of the shrunk prime subtree can be represented by a \emph{maximal
matching graph} $Q_g(G)=(V_r, E_r)$, where (1) $V_r\subseteq V$ such that $v\in
V_r$, if  there is a query node $u\in V_q$ such that $v\models u$; (2)
$E_r\subseteq V_r\times V_r$ such that $(v_1, v_2)\in E_r$, if $(v_1, v_2)$ is
a match of an edge $(u_1, u_2)\in E_q$.
  
We group the nodes and edges in
the graph according to what query nodes and edges they match. Specifically, in
an implementation, each node $v$ has several branch lists, each of which
corresponds to the child of the query node that $v$ matches and includes
pointers pointing to nodes matching the child. 

\begin{exmp}
Recall the GTPQ $Q$ and data graph $G$ in Fig$.$ \ref{exmp:des}. Let $u_2$,
$u_3$ and $u_4$ be output nodes. Fig$.$ \ref{match} shows the corresponding
maximal matching graph. As an example, $v_1$ has two branch lists corresponding
to the two incident query edges, denoted by $bch_1$ and $bch_2$ respectively.
$bch_1=\{ptr_{v_3}, ptr_{v_8}\}$, and $bch_2=\{ptr_{v_3}, ptr_{v_5}\}$, where
$ptr_{v_i} (i=3,5,8)$ is pointer to $v_i$.
\label{exmp:maximalgraph}
\qed 
\end{exmp}
\end{mdef}

\begin{algorithm}[t]
\small
\DontPrintSemicolon
\SetKwInput{KwData}{Input}\SetKwInput{KwResult}{Output}
\KwData{The maximal matching graph $MaximalGraph$, a query
node $u$ and one of its candidate matching node $v$.}
\KwResult{the answer to the subGTPQ rooted at $u$ and dominated by $v$.} 
\SetAlFnt{\small} 
%\BlankLine
\lIf{$v$ is a leaf node}{ 
	\Return $\{u:v\}$\; 
}\Else{   
	$results:=\emptyset$\;
	\ForEach{branch list $bch$ of $v$}{ 
		$branchResults:=\emptyset$\;
		\ForEach{node $v'$ that a pointer in $bch$ points to}{
			$branchResults:=branchResults\cup \textrm{ CollectResults}(MaximalGraph,
			v'$)\label{alg:res:merge}\; }  
		$results:=results\times branchResults$ \;		 
	}
	\lIf{$u$ is an output node}{
			$results:=\{u:v\}\times results$\;
	} 
	\Return $results$\;
}  
\caption{CollectResults\label{alg:result}}
\end{algorithm}
\begin{mdef}[Computing the maximal matching graph]
Since the nodes of the maximal matching graph have been obtained after
the pruning process, we only need to compute matches for each query
edge whose head and tail both have more than one
matching node. Given a query edge $(u_1, u_2)$, a straightforward way is to
 check  the reachability between  nodes in $mat(u_1)$ and $mat(u_2)$  using 
 3-hop index. The time complexity is
 $O((|L_{in}+L_{out}|)|E_q||V_{mat}|^2_{max})$, with $|V_{mat}|_{max}$ being the
 maximal size of the candidate matching nodes after the pruning process. Since
 in practice  many queries are highly selective and $|V_{mat}|_{max}$ is
 presumably   pretty small, the straightforward way is expected to be fast and practical.
   
A more sophisticated approach that we choose is to utilize the similar technique
used in procedure PruneUpward. Observe that the loop
from line 9 to 12 in PruneUpward is to determine whether a data node 
matching some child of $u$ is reachable from $mat(u)$. By
replacing $C^s_u$ with the successor list of a node $v$, we can simultaneously
get all edges from $v$ in the maximal matching graph in
$O(|L_{in}|+|L_{out}|+|E_v|)$,  where $|E_v|$ is the out-degree of $v$ in the 
resulting graph. The total  time complexity then is 
$O((|L_{in}|+L_{out})|V^{inter}_{mat}|+|E_{mg}|)$, where $|V^{inter}_{mat}|$ is
the number of candidate matching nodes for internal query nodes and $|E_{mg}|$
is the number of edges in the resulting maximal matching graph. 
\end{mdef} 
\begin{mdef}[Enumerating results]  
We next present procedure \ref{alg:result}, referred to as  CollectResults, 
which derives final results from the maximal matching graph. Each result is in a
tuple format. To avoid ambiguity in
presentation, we explicitly specify in the tuple which query
node a data node matches. Specifically, each element in a tuple is of the form
$u:v$, which means $v$ is an image of $u$ in a match.

Procedure CollectResults traverses
down the maximal graph. For a leaf node, since its corresponding query node
 must be an output node, the procedure returns a tuple with only an element of
 it (line 1). For an internal node, it collects results from each
 child for every branch list, and then does a Cartesian product of them (line
 4-8). If the query node it matches is an output node, it is inserted
 into each  result (line 9). The final answer to the query is the union of the
 results of those nodes matching the query root. When query nodes in the shrunk
 prime subtree are all output nodes, no redundant intermediate results would be produced. Note that no existing
 algorithms for pattern queries on graphs can achieve this. When there are non-output query
 nodes in the shrunk prime subtree, our algorithm is not
 duplicate free. Recall Example \ref{exmp:maximalgraph}. The results
 obtained from $v_1$ are the same as those obtained from $v_3$, since $u_1$ is
 not an output node and $v_1$ can reach $v_3$.  However, the duplicate intermediate tuples
 are a subset of the counterpart of other works, because (1) the prime
 subtree we pick is a minimum subtree of the original query pattern that
 contains all output nodes,  (2) for non-output nodes, the algorithm merges the intermediate partial
 results in advance (line \ref{alg:res:merge}). 
\end{mdef}  
% % \vspace{-2pt}
\begin{mdef}[Remark]
In practical languages, there is also group operation that require grouping the
results. We can also easily  adapt our algorithm to support the
group operator. In CollectResults, when $u$ is a group node, the result
returned for $v$ is a tuple containing $v$ and a special group element which is the set
of matches of the subtree dominated by $v$. That is, in line 9, $result:=\{u:v,
(result)\}$.
\end{mdef}

\subsection{Evaluating Queries with PC Edges}
\label{pc-edge}
In the context of graph database, the research on pattern queries often focuses
on reachability patterns. Indeed, the reachability pattern query is an important building
block for other queries. Adding PC edges to a pattern  significantly
increases the complexity of evaluation. Even for tree-structured data,
\cite{complexity} has theoretically demonstrated the difficulty  of handling
TPQs with arbitrary combination of PC and AD edges. \cite{complexity} has
 proved that no holistic
algorithms can achieve optimality as for queries with  AD edges only.
 For graph-structured data, the evaluation of conjunctive pattern queries whose
 edges all represent PC relationship  is essentially a computationally-hard
 labeled graph isomorphism problem. Nevertheless, we can use the similar idea of our framework to support GTPQs with PC edges.
 
When processing a node $u$ in PruneDownward: (1) if $u$
has only PC outgoing edges, we merge the set of parents of $mat(u')$ for each child $u'$
of $u$ into $P_{u'}$, instead of computing the predecessor contours. Then
we sort $mat(u)$ and each $P_{u'}$, and check for each node $v$ in $mat(u)$
whether it is in some $P_{u'}$ in a multiway merge-sort  style. If yes, then
$val[p_{u'}]:=1$, otherwise $val[p_{u'}]:=0$. (2) If $u$ has both AD and PC
edges, we process these two type of edges separately to refine $mat(u)$.
Similarly, when
performing PruneUpward, we collect sets of children of $mat(u)$
instead of computing the successor contour.

 After the pruning stage, all candidate
matching nodes are guaranteed to be in final results. To  compute the
maximal matching graph, we can either do nested joins to check the  adjacent 
relationships, or perform multiway merge-join  to derive the  adjacent edges
in the resulting graph. Other operations including determining    the prime
subtree and enumerating final results are the same.

Alternatively, we can also use another strategy to deal with PC edges.
Regarding PC edge as a special type of AD edge, we can first process  PC edges in the same way
with AD edges in the process of pruning, except those whose tail's structural
variable is the  operand of a negation operator and which need to be processed
as stated before. The prime subtree becomes a minimum subtree that contains all
output nodes and those  PC edges that are regarded as AD edges when pruning.
After computing the maximal matching graph, we check  whether the two incident nodes of the corresponding
edge in the maximal matching graph are adjacent in  the data graph and remove
them if not. Next, the unsatisfied nodes are removed in a top-down
fashion, followed by enumerating final results. We use this strategy in our
implementation. 
% \vspace{-2pt}   
  
\section{Experimental Evaluation}  
\label{section5}
In this section, we present an experimental study using both real-life and
synthetic data to evaluate (1) the efficiency and scalability of our algorithm,
(2) the effectiveness of representing intermediate results as graphs, and (3)
the efficiency of the pruning process.  

We only give the experimental results for conjunctive TPQs with all query nodes
being output nodes (i$.$e$.$ the traditional TPQs). We found that  our algorithm has 
 better performance than other algorithms even for them.
% , based on the
% following consideration. First, we find it  difficult to fairly choose or
% randomly generate structural predicates  for query nodes, although it
% is common in practice. The experiments may otherwise become ad-hoc.
%  More
% importantly, our algorithm not only supports a broader class of TPQs, 
% i$.$e$.$  
% GTPQs, but 
 Since there has been no other algorithms designed for GTPQs and the
 decomposition-based approach that may be applied on top of them to process
 GTPQs incurs high overhead as analyzed in Related work and empirically
 demonstrated   in prior studies \cite{twigor} and \cite{twignot}, our
 algorithm can do even far better for general GTPQs than those
 algorithms, compared to the results  reported here.  Additional experimental
 results concerning  I/O cost and the results on GTPQs with disjunctive and
 negative predicates can be found in the Appendix.
   
\begin{mdef}[Implementation] We have implemented the algorithm proposed in 
Section \ref{section4} (GTEA), TwigStack \cite{holistic},
Twig$^2$Stack \cite{twig2stack}, TwigStackD \cite{stackd} and HGJoin 
\cite{hjoin}. TwigStack is the classical holistic twig join algorithm.
 Twig$^2$Stack is the latest algorithm for evaluating TPQs on
tree-structured data which has a distinct feature of representing results in
hierarchical stacks. Other algorithms for
tree-structured data that can support disjunction and/or negation,  such as
BTwigMerge \cite{alltwig}  and TwigStackList$\neg$ \cite{twignot}, are in essence the same
as TwigStack with respect to the conjunctive TPQs and hence are not included in our experiments.
 TwigStackD can evaluate conjunctive TPQs over graph-structured
data. In our implementation, we fixed the problems in the original paper \cite{comm}. HGJoin is a hash-based 
structural join algorithm for processing graph pattern queries. We did not
implement the query plan generation in the original algorithm which relies on
selective estimation techniques \cite{est} and takes exponential time in the query size; 
instead, for each query, we generated all valid plans and took evaluation on
each. The  minimum query processing time on the best plan is reported; thus, the time
presented in this paper is always smaller than the real time
of the original HGJoin. This version is denoted by HGJoin+. 
By representing intermediate results as graphs, we have  also implemented
another version denoted by HGJoin*. All experiments are performed on a 2.4GHz
Intel-Core-i3 CPU with 3.7 GB RAM.
%  11.04 
% (with gcc 4.5.2).  
\end{mdef}
\subsection{On XMark Data} 
In this set of experiments, we use large synthetic XMark data \cite{xmark} to
evaluate the efficiency and scalability of various algorithms. As mentioned in
Section \ref{section1}, many graph-structured XML database can be modeled by a
special form of graphs consisting of trees connected by cross edges (ID/IDREF
links).  In this case, we can use existing twig join algorithms to
process conjunctive TPQs by decomposing them into a set of subqueries on separative
trees.  We use TwigStack and Twig$^2$Stack to investigate the efficiency of
applying this approach. 

\begin{mdef}[Datasets] We generated five XMark datasets with the scaling
factors from 0.5 to 4. For each dataset, we generate a graph, where nodes
correspond to XML elements and edges represent the internal links
(parent-child) and ID/IDREF links. The attribute for graph nodes is the tag of
elements except for nodes corresponding to \textsf{person}, \textsf{item}
elements, for each type of which we randomly classify them into ten groups to represent
different properties. A label is assigned to each
node according to the tag or the group it belongs to. Distinct labels indicate
different attribute values. The details of the generated documents and graphs
are presented in Table \ref{tab:xmark}.
\end{mdef} 

\begin{table}[t] 
\caption{Statistics of XMark datasets}
\small
\centering
\begin{tabular}{l|l|l|l|l|l}  
\hline 
Scaling factor & 0.5 & 1 & 1.5 & 2 & 4 \\
\hline
Dataset size (MB) & 55 & 111& 167 & 223 & 447 \\
\hline
Nodes (Million) & 0.64 & 1.29 & 1.94 & 2.52 & 5.17 \\
\hline
Edges (Million) & 0.77 & 1.54 & 2.32 & 3.09 & 6.20 \\
\hline 
\end{tabular} 
\label{tab:xmark} 
\end{table}
\begin{table}[t] 
\caption{The average size of query results on XMark}
\small
\centering 
\begin{tabular}{l|l|l|l|l|l}   
\hline 
Queries & 55M & 111M & 167M & 223M & 447M \\
\hline
$Q_1$ & 368 & 762.8 & 1115.8 & 1496.8 & 2986.8 \\
$Q_2$ & 34.6 & 75.8 & 117.8 & 150.3 & 297.2 \\
$Q_3$ & 1.9 & 4.1 & 5.8 & 6.1 & 17.1 \\
\hline
\end{tabular} 
\label{tab:res}
\end{table}   

\begin{figure}[t] 

\subfigure[$Q_1$]{
\label{exp:q1} 
\begin{minipage}[t]{0.11\textwidth}
\centering
\includegraphics[height=1.1in]{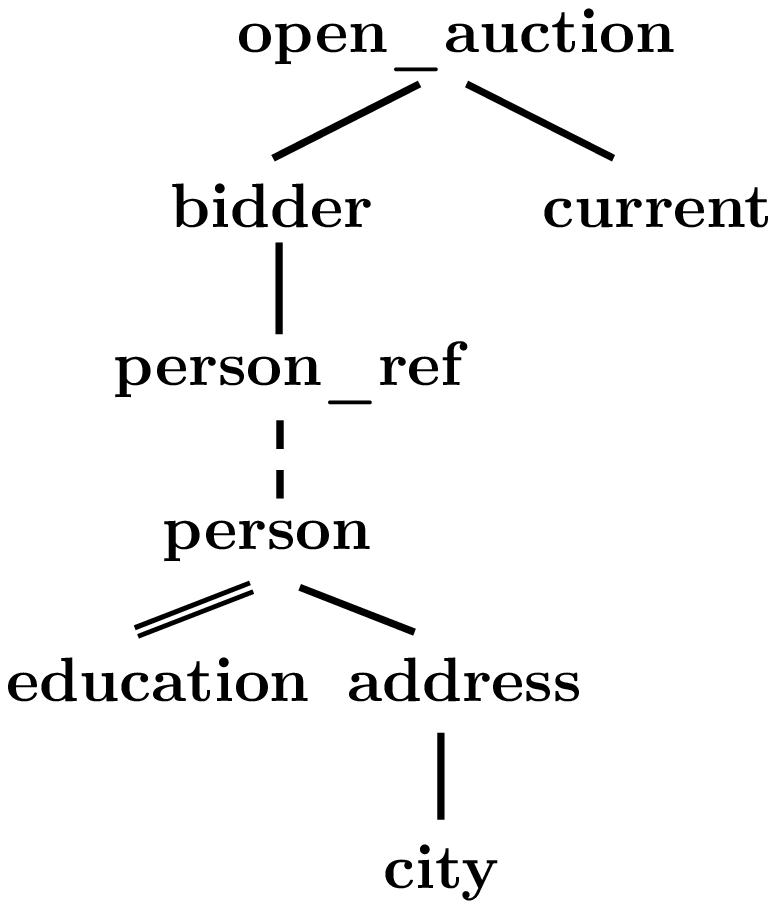}
\end{minipage}} 
\subfigure[$Q_2$]{  
\label{exp:q2}  
\begin{minipage}[t]{0.14\textwidth}
\centering
\includegraphics[height=1.1in]{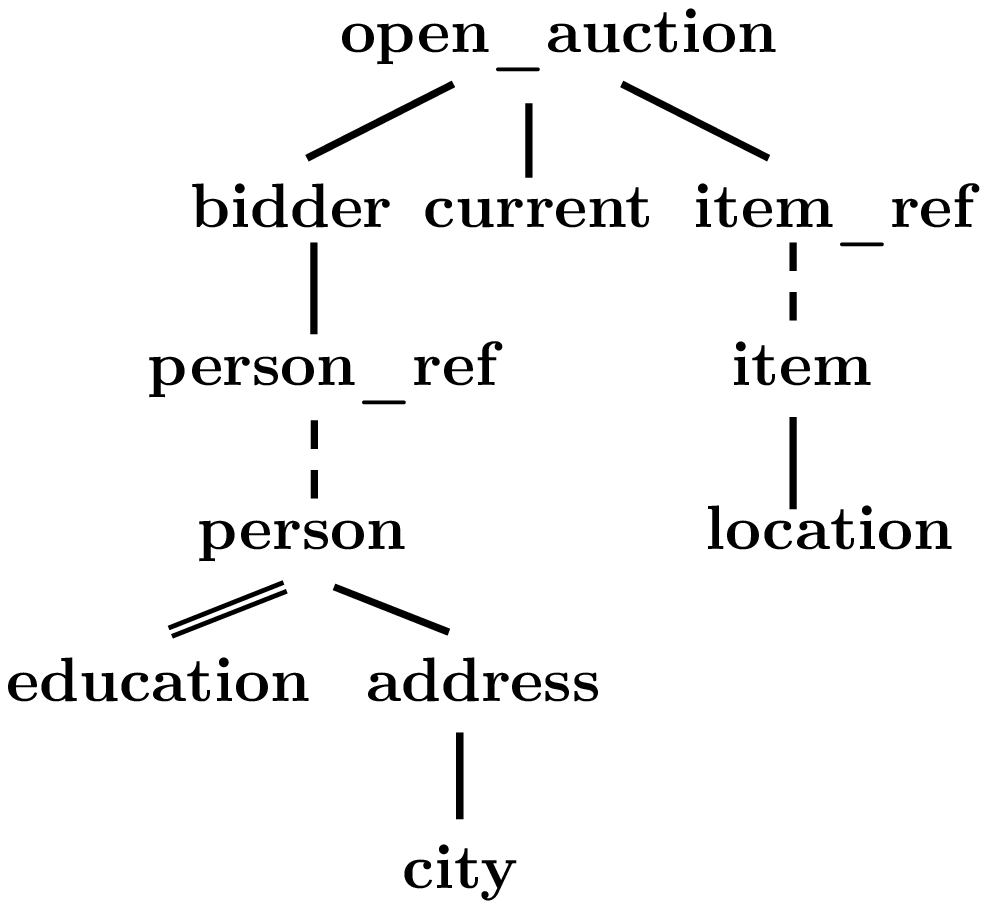}
\end{minipage}} 
\subfigure[$Q_3$]{      
\label{mingtpq} 
\begin{minipage}[t]{0.14\textwidth}
\centering
\includegraphics[height=1.1in]{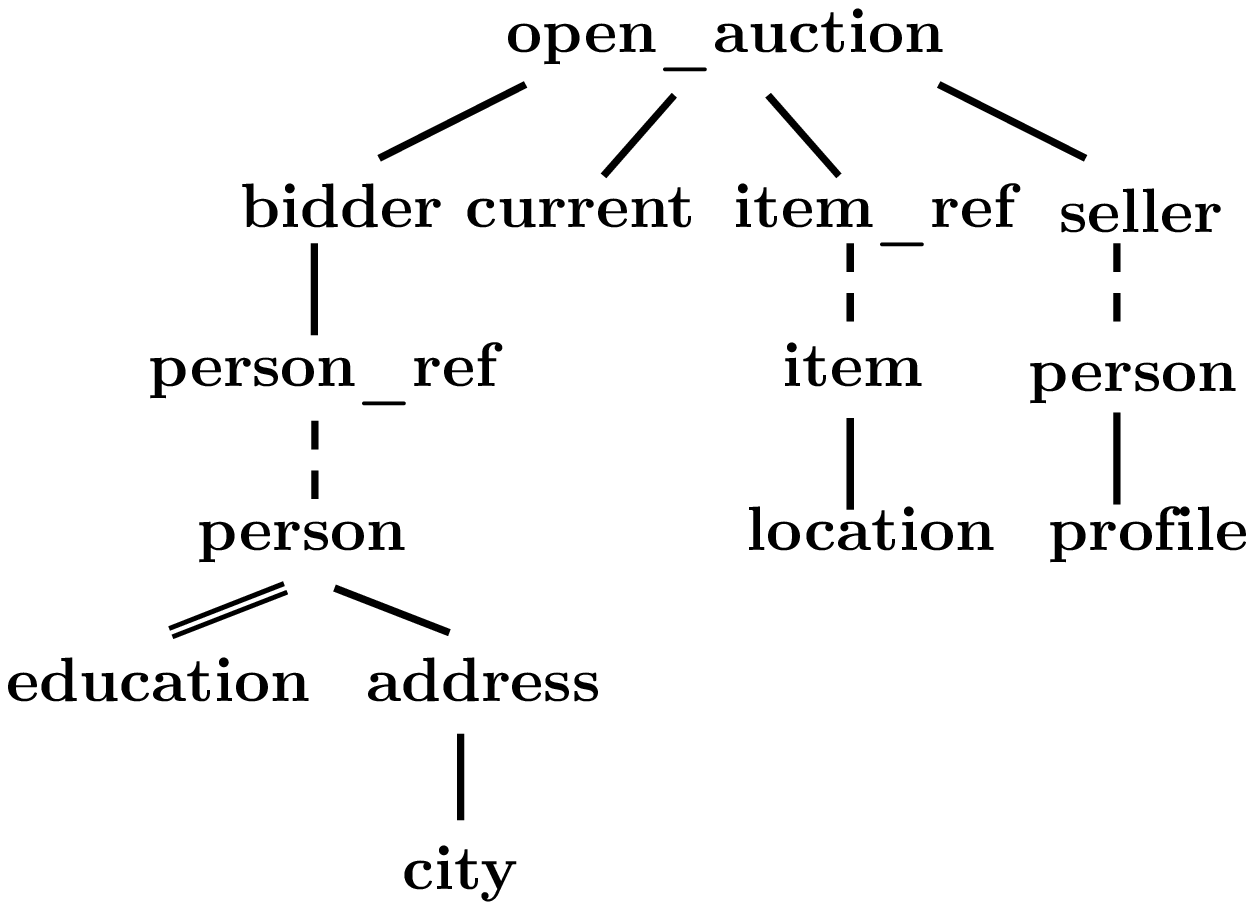}  
\end{minipage}}
\caption{Queries for XMark data}
\label{exp:queries}  

\end{figure} 

\begin{comment} 
\begin{figure*} 
\centering
\subfigure[$Q_1$]{
\label{exp:q1-res} 
\begin{minipage}[c]{0.22\textwidth}
\centering
\includegraphics[height=1in]{Q1-res.eps}
\end{minipage}} 
\subfigure[$Q_2$]{
\label{exp:q2-res}  
\begin{minipage}[c]{0.22\textwidth}
\centering
\includegraphics[height=1in]{Q2-res.eps}
\end{minipage}}
\subfigure[$Q_3$]{    
\begin{minipage}[c]{0.22\textwidth}  
\label{exp:q3-res}   
\centering
\includegraphics[height=1in]{Q3-res.eps}
\end{minipage}}
\subfigure[XMark 55M]{      
\label{exp:mark-res}      
\begin{minipage}[c]{0.22\textwidth}  
\centering
\includegraphics[height=1in]{mark55-res.eps}
\end{minipage}}
\subfigure[varying $|V|$ with small results]{ 
\begin{minipage}[c]{0.3\textwidth} 
\centering
\includegraphics[height=1in]{Graph-small.eps} 
\end{minipage}}
\subfigure[varying $|V|$ with large results]{    
\begin{minipage}[c]{0.3\textwidth}
\centering
\includegraphics[height=1in]{large_test2.eps}
\end{minipage}}
\subfigure[Efficiency of the pruning process]{    
\begin{minipage}[c]{0.3\textwidth}
\centering
\includegraphics[height=1in]{filter-res.eps}
\end{minipage}}
\caption{Performance evaluation} 
\label{exp:results} 
\end{figure*}  
\end{comment}
       
\begin{figure}[t]    
\centering 
\subfigure[Varying data size]{
\label{exp:q1-res} 
\begin{minipage}[c]{0.23\textwidth}
\centering
\includegraphics[width=1.23in, angle=270]{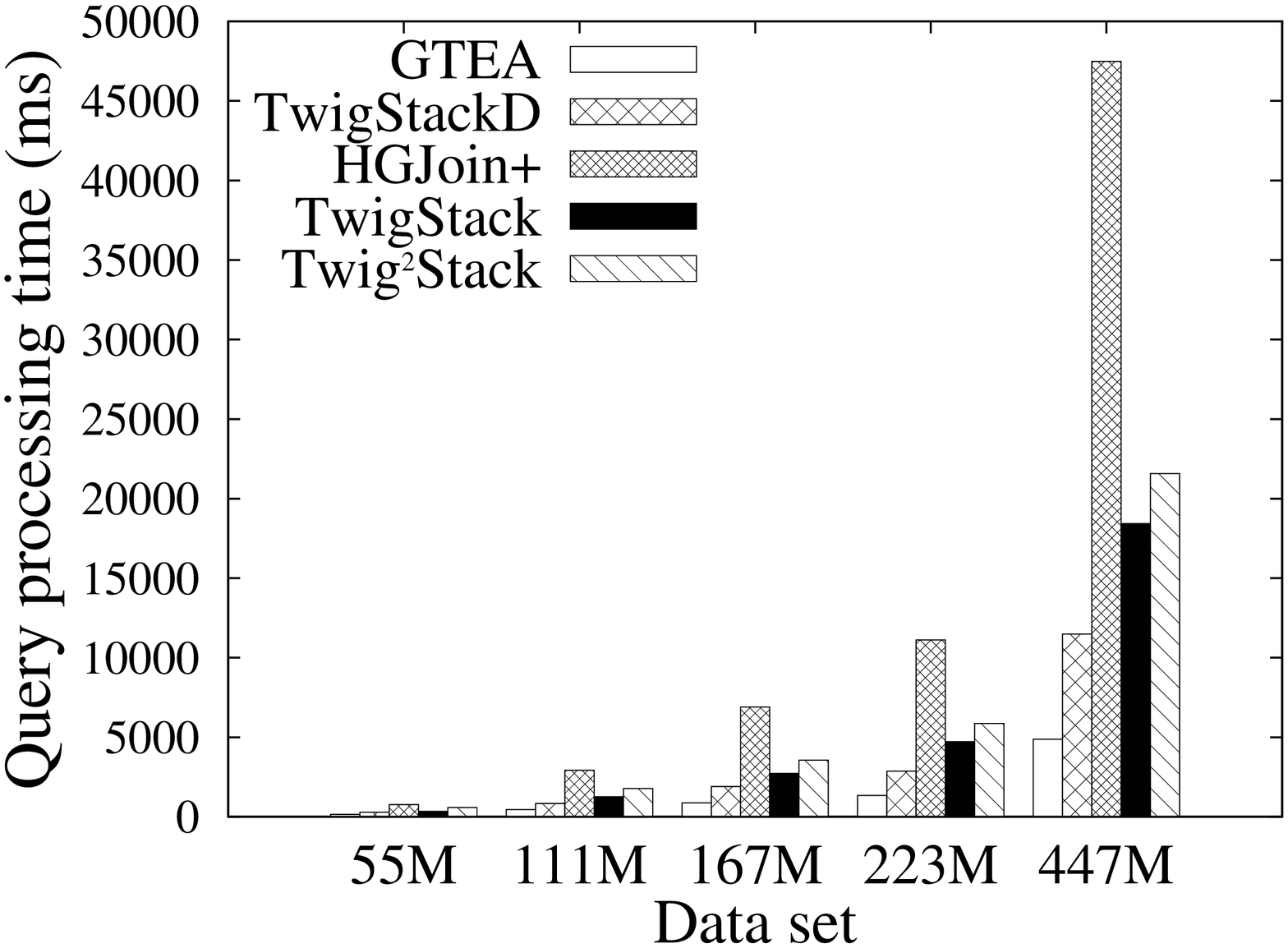}
\end{minipage}} 
\subfigure[Varying query size]{          
\label{exp:mark-res}        
\begin{minipage}[c]{0.22\textwidth}  
\centering
\includegraphics[width=1.23in, angle=270]{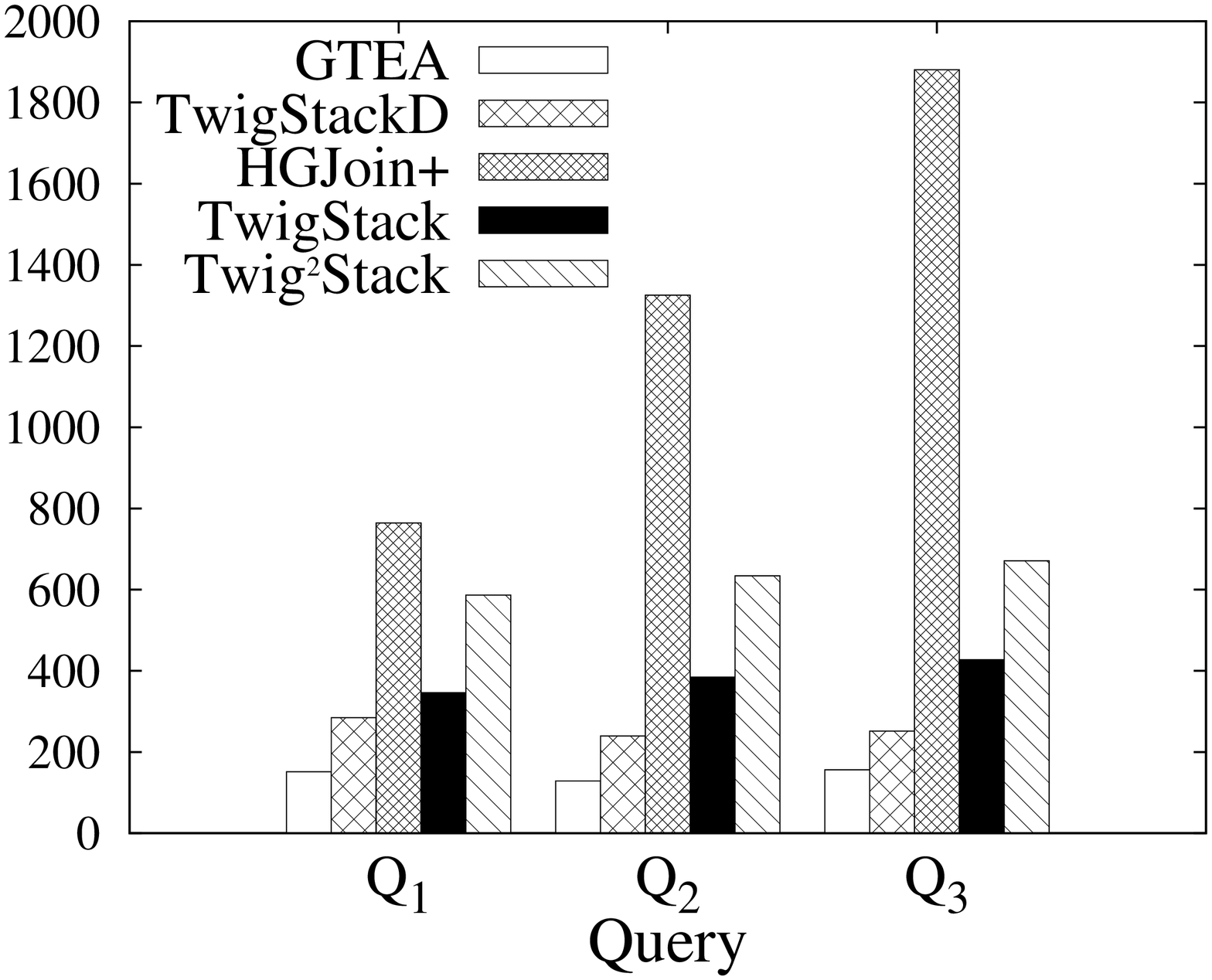}
\end{minipage}}
% \subfigure[]{   
% \begin{minipage}[c]{0.18\textwidth} 
% \centering
% \includegraphics[height=1in]{Graph-small1.eps}  
% \end{minipage}} 
% \subfigure[]{    
% \begin{minipage}[c]{0.18\textwidth} 
% \centering
% \includegraphics[height=1in]{Graph-large1.eps}
% \end{minipage}}  
% \subfigure[]{     
% \begin{minipage}[c]{0.18\textwidth}
% \centering 
% \includegraphics[height=1in]{filter-res1.eps}
% \end{minipage}} 
\caption{Performance results on XMark data}   
\label{exp:results1}   
\end{figure}

% \begin{figure*}  
% \centering
% \subfigure[]{
% \label{exp:q1-res} 
% \begin{minipage}[c]{0.21\textwidth}
% \centering
% \includegraphics[height=1in]{Q1-res1.eps}
% \end{minipage}} 
% \subfigure[]{      
% \label{exp:mark-res}      
% \begin{minipage}[c]{0.2\textwidth}  
% \centering
% \includegraphics[height=1in]{mark55-res1.eps}
% \end{minipage}}
% \subfigure[]{   
% \begin{minipage}[c]{0.18\textwidth} 
% \centering
% \includegraphics[height=1in]{Graph-small1.eps}  
% \end{minipage}} 
% \subfigure[]{    
% \begin{minipage}[c]{0.18\textwidth} 
% \centering
% \includegraphics[height=1in]{Graph-large1.eps}
% \end{minipage}}  
% \subfigure[]{     
% \begin{minipage}[c]{0.18\textwidth}
% \centering 
% \includegraphics[height=1in]{filter-res1.eps}
% \end{minipage}} 
% \caption{Performance evaluation}   
% \label{exp:results}   
% \end{figure*}  

\begin{mdef}[Queries] Three types of queries we used for experiments are
depicted in Fig$.$ \ref{exp:queries}, where dotted edges refer to ID/IDREF links
in the original data. For each query type, we generated ten 
queries by randomly choosing a label for each of \textsf{person} and
\textsf{item} nodes representing a different attribute predicate. The average is
reported.
\end{mdef}
\begin{mdef}[Experimental results]
Fig$.$ \ref{exp:results1}(a)  shows the query evaluation time for
$Q_1$ on  datasets varying the data size. The
results for $Q_2$ and $Q_3$ are quite similar.  The results reveal the
following. (1) GTEA constantly outperforms all other algorithms. Specifically, GTEA is three times to more than
 one order of magnitude faster than TwigStack and
Twig$^2$Stack, five times to more than two orders of magnitude faster than
HGJoin, and in the best cases three times faster than TwigStackD. When data size
becomes larger, the performance gain by GTEA becomes more significant. (2)
TwigStackD also has very good performance in this set of experiments with the following reasons. 
(a) It utilizes SSPI, a reachability index with pretty small size and
good querying time for tree-like graphs.  (b) Its basic idea is extended from the holistic twig join
algorithms, and so TwigStackD also has the advantages taken by
the  stack encoding and the blocking method for path results \cite{holistic}. (c)
Although TwigStackD has to buffer every nodes in pools (a special structure used to store nodes popped
from stacks) and large amounts of the operations of
checking edge conditions with all nodes in pools have to be done (indicated as
reasons of inefficiency in \cite{hjoin} and \cite{jointkde}), the pre-filtering process
 it uses can filter redundant nodes and relieve the cost of
the above operations. Indeed, without the pre-filtering process, TwigStackD is 
slower by orders of magnitude \cite{comm}. (3) It is sort of
surprising that TwigStack has slightly better performance than Twig$^2$Stack. 
The reason is that although   Twig$^2$Stack can avoid generating path matches 
(as a primary reason for the efficiency in \cite{twig2stack}), the 
overhead brought by merging stack trees and maintaining the hierarchical structures overrides the benefits
in the experiments. The fact that the depth of XMark graphs is small (with an
average of 5),   also make the hierarchical stack encoding have not a strong
advantage. Besides, the enumeration of path matches (as a reason for
inefficiency for TwigStack in \cite{twig2stack}) can be done fast 
using the blocking technique. (4) HGJoin has the worst performance, mainly
because (a) the structural-join way has to generate a large number of
(largely redundant) intermediate results for small substructures and (b)
non-trivial merge-join operations on them have to be done even with the best
plan. The query processing time increases significantly when the size of data graphs
increases.   
% (5)The performance of TwigStackD is mostly close to our algorithm.

Fig$.$ \ref{exp:results1}(b) shows the results on the XMark dataset of scale 0.5
 for different queries. (1) The query processing time of GTEA
nearly maintains the same as the query size increases. In particular, the time
cost for evaluating $Q_2$ is smaller than that for $Q_1$. It is because the size
of the results of $Q_2$ is much smaller than that for $Q_1$ as presented in
Table \ref{tab:res}, resulting in smaller cost for enumerating the final results. (2)
The processing time of TwigStack and
Twig$^2$Stack does not increase  significantly over $Q_1$, $Q_2$ and $Q_3$,
although they have to evaluate a increasing number of subqueries and perform
a growing number of merge operations. Indeed, as shown in Table
\ref{tab:res}, the sizes of the results of $Q_1$ and $Q_2$, which are
a subquery of $Q_2$ and $Q_3$ respectively, are small and thus the extra cost
for evaluating $Q_2$ and $Q_3$ is very limited. (3) However,  HGJoin is  
much more sensitive to the increase of the query size, which is  due to the
impact of the redundant intermediate results and expensive sort operations
involved in performing multi-structural joins. 
The results for HGJoin  highlight the crucial importance of using a pruning
process to reduce the size of intermediate results not contributing to the
answer.
\end{mdef}

% \begin{figure}
% \centering
% \includegraphics[height=1in]{filter-res.eps}
% \caption{Efficiency of the pruning algorithm}
% \label{prune}
% \end{figure}

 \begin{figure*}   
 \centering
% \begin{minipage}[t]{0.24\textwidth}
% \centering
% \includegraphics[height=1.2in]{distribution.eps}
% \caption{The result distributions of the tested queries}
% \end{minipage}
% \subfigure[]{
% \begin{minipage}[t]{0.22\textwidth}
% \centering
% \includegraphics[height=1.2in]{distribution.eps}
% \end{minipage}} 

\subfigure[]{   
\begin{minipage}[t]{0.24\textwidth}
\centering
\includegraphics[width=1.23in, angle=270]{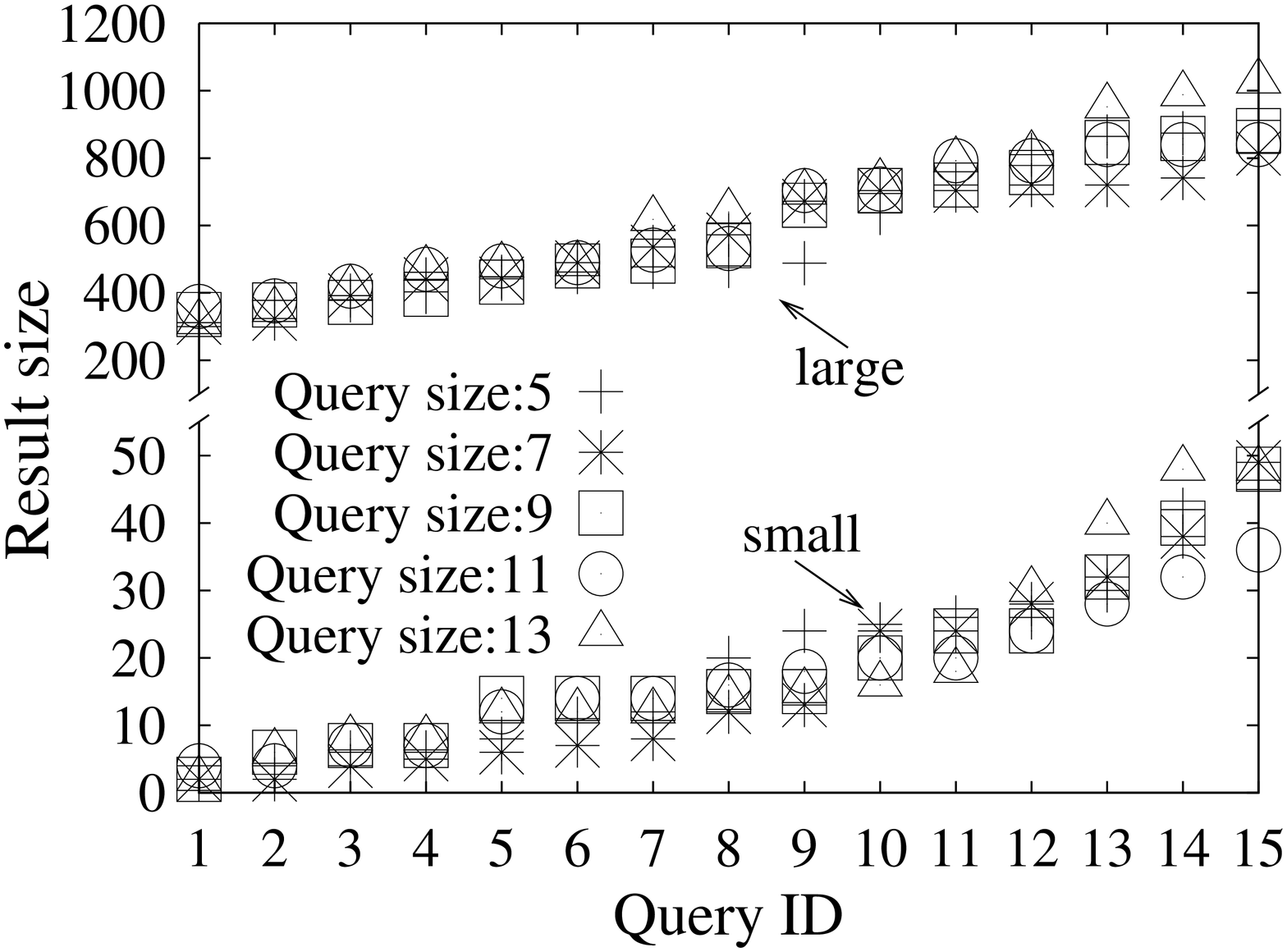}
\end{minipage}} 
\subfigure[]{       
\begin{minipage}[t]{0.24\textwidth}
\centering
\includegraphics[width=1.23in, angle=270]{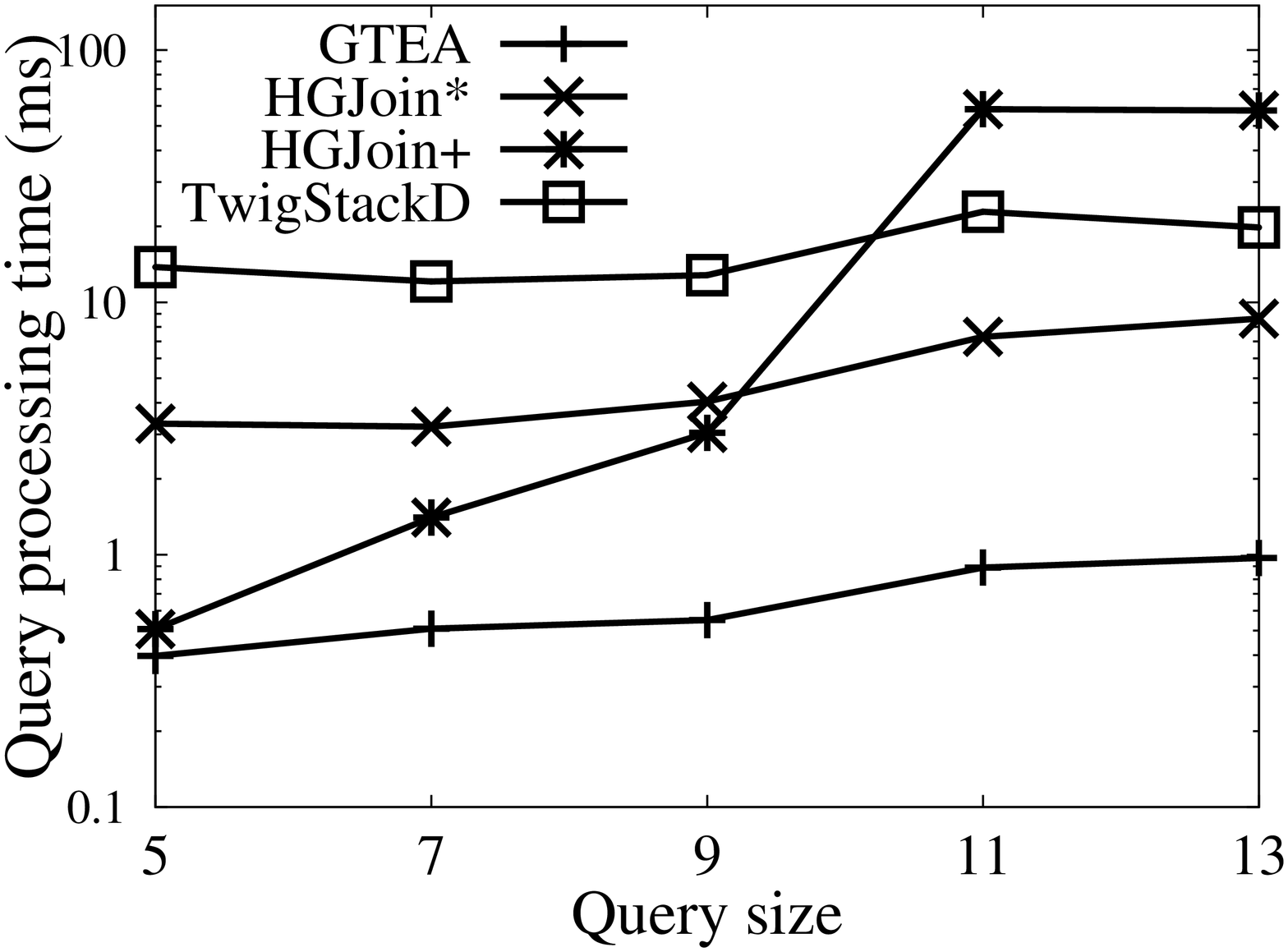}
\end{minipage}}   
\subfigure[]{   
\begin{minipage}[t]{0.24\textwidth} 
\centering
\includegraphics[width=1.23in, angle=270]{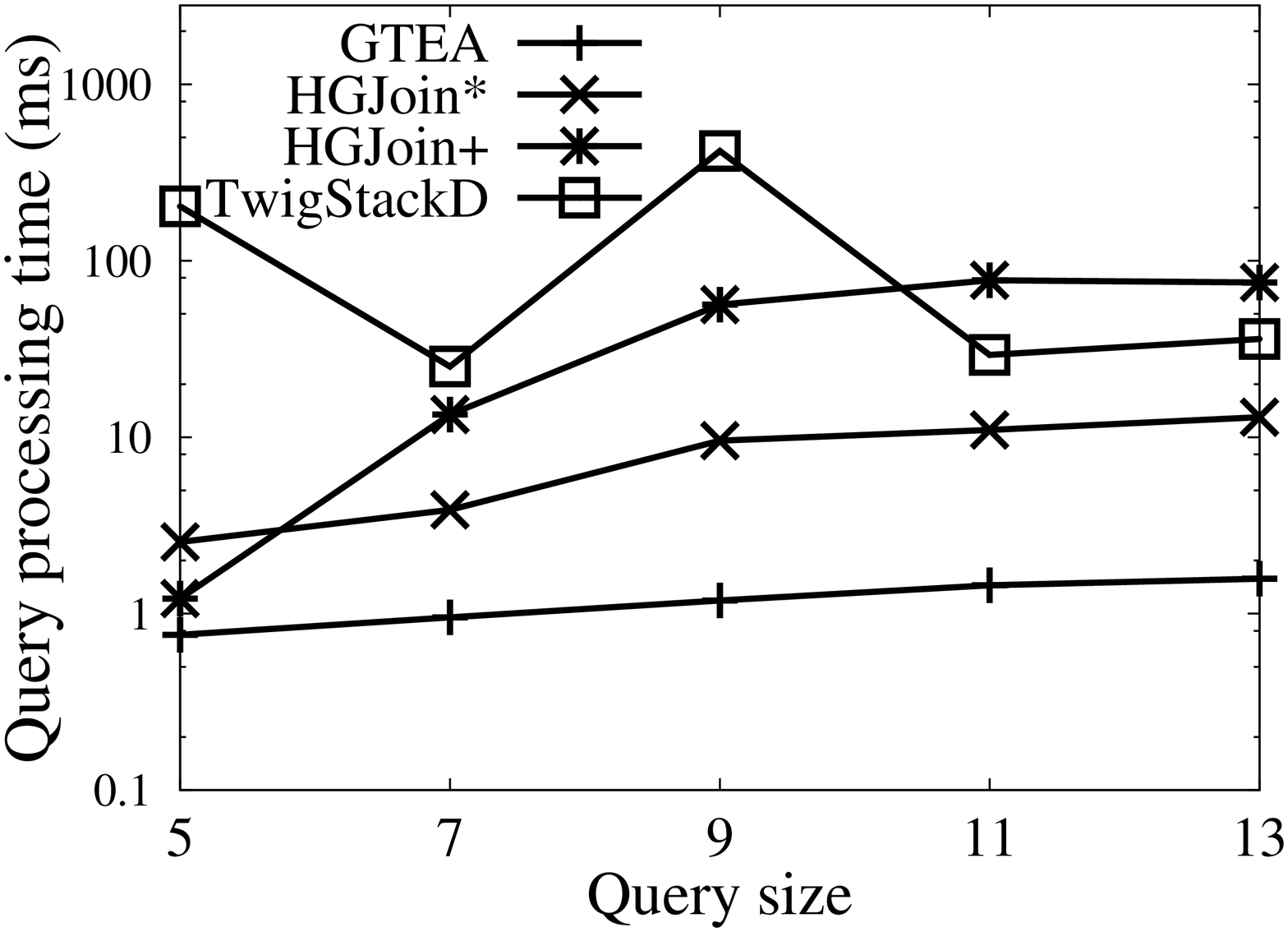}
\end{minipage}} 
\subfigure[]{
\begin{minipage}[t]{0.24\textwidth} 
\centering 
\includegraphics[width=1.23in, angle=270]{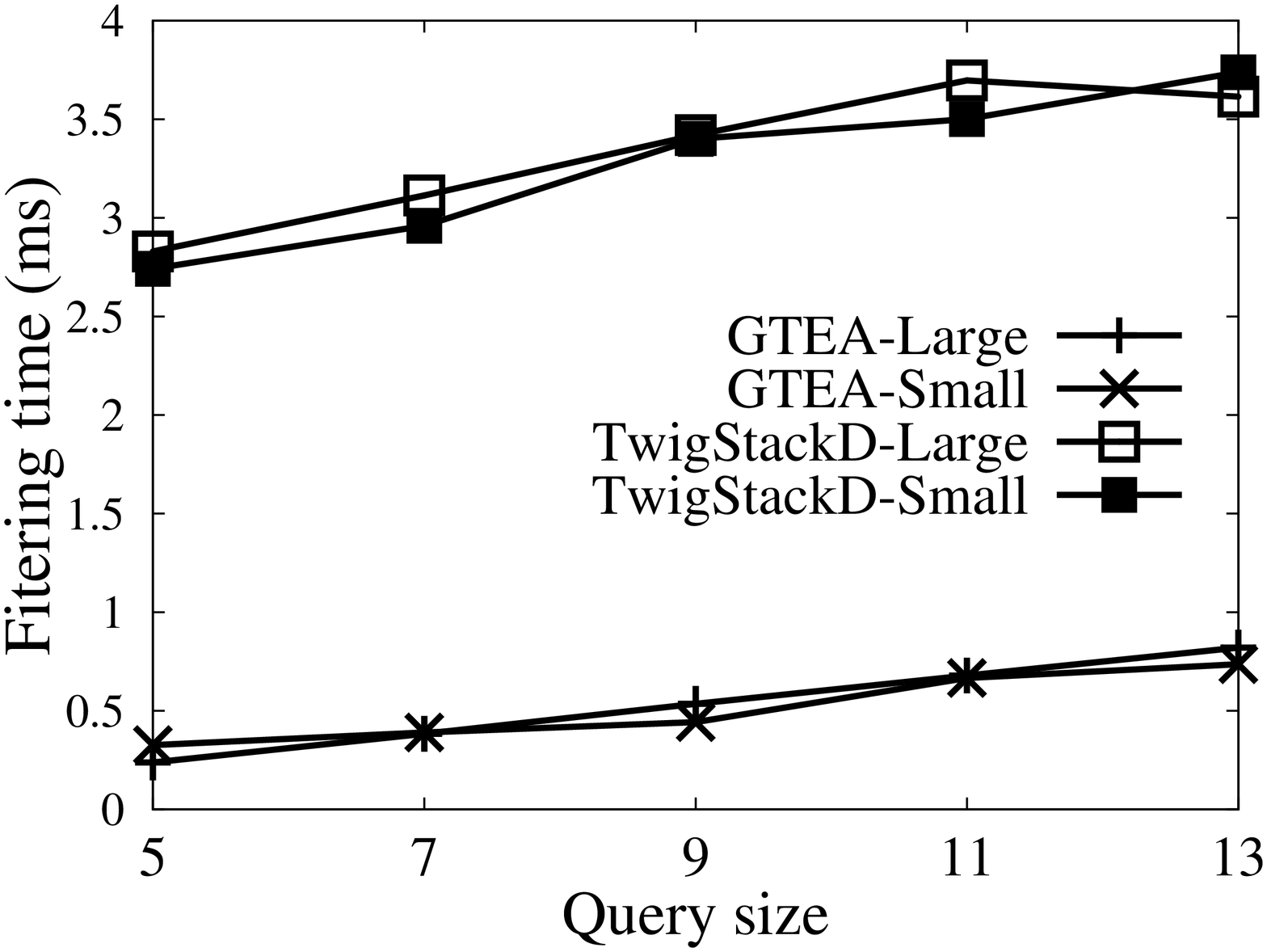}
\end{minipage}}
\caption{Performance results on arXiv data. (a) Distribution of the result
sizes. (b) Query processing time on the queries with small sizes of results.
(c)  Query processing time on the queries with small sizes of results. (d)
Comparison with the pre-filtering process. }
\label{exp:results2}
\end{figure*}

\subsection{On arXiv Data}
In this set of experiments, we used a real-life graph to evaluate the
performance of GTEA, TwigStackD and HGJoin for general graph data, verify the effectiveness
of graph representation of intermediate results and the efficiency of the
pruning process. 
\begin{mdef}[Dataset]
We generated a graph from the HEP-Th
database\footnote{http://kdl.cs.umass.edu/data/hepth/hepth-info.html},
originally derived from the arXiv\footnote{http://arxiv.org/}. There are paper
nodes and author nodes, each associated with multiple properties. For simplicity, we assigned a label to
each author node according to the email domain, and a label to each paper node
based on its area and journal it is published in, to represent the attributes.
The edges of the graph represent author or citation relationships. The graph has
9562 nodes, 28120 edges, and 1132 distinct labels.
\end{mdef}
\begin{mdef}[Query generator]
We designed a query generator to randomly produce meaningful queries. Each query
node is associated with a label randomly chosen from  the data graph to
represent attribute predicates. Two groups of queries are generated: one has a
small size of results between 2 and 50, the other has a large size
between 200 and 1200. For each group, five sets of queries were generated with
query size varying from 5 to 13. We generated fifteen
different queries for each size scale and report the average. 
The average time can 
 reflect the average case performance of each algorithm, 
 since the queries are generated in a random way.
 The results for queries of distinct sizes in the same group are
 comparable, because the differences of the result sizes  of
the queries have little impact on the query processing time and  the number
of query results for each size scale follow a close distribution as illustrated
in Fig$.$  \ref{exp:results2}(a).
\end{mdef}

\begin{mdef}[Experimental results]
Fig$.$ \ref{exp:results2}(b) and (c) report the results for the two groups of
queries. They tell us the
following. (1) GTEA has the best query processing time, significantly smaller
than the processing time of other algorithms (more than one order of magnitude
in most cases). It also has the best scalability in both two groups
of experiments.  (2) TwigStackD no longer has  good performance as   on XMark
data. In fact, it has the longest querying time for queries with size of 5 to 9.
The arXiv graph is much denser and deeper than XMark data, causing the
inefficiency of the pool structure as well as SSPI. The problem of
TwigStackD is highlighted by Fig$.$ \ref{exp:results2}(c) where it fluctuates 
sharply for queries with large results. The results reflect that TwigStackD has
rather poor performance for particular queries. In contrast, GTEA is most robust
since it always maintains good performance for all experiments.  (3) HGJoin+ is not
quite scalable similar to the performance on the XMark data. Yet it now has
better performance than TwigStackD when the query size is smaller than 11. (4) 
The revised HGJoin (i$.$e$.$ HGJoin*)  has better scalability than HGJoin+.
For the group of queries with large results, the query processing time of
HGJoin* is smaller than that of HGJoin+ when the query size is larger than 7, compared to
11 for the group of queries with small results. This observation demonstrates
that graph representation of intermediate results can improve the performance and achieve better scalability
especially when there are many intermediate/final results and when the query
size is large. The reason why the revised one takes more time 
than the original one for processing the queries of small sizes  is that
HGJoin* incurs costs for dynamically and recursively deleting  unqualified nodes (not exist
in our algorithm though), which offset the benefits taken by avoiding merge-join
operations on tuples.  
 
Fig$.$ \ref{exp:results2}(d) evaluates the efficiency of our pruning process and
the pre-filtering algorithm in TwigStackD, which clearly
shows that our pruning method greatly outperforms the counterpart and also
has better scalability with the query size. It is because the pre-fltering
algorithm in TwigStackD requires two  traversals of the data graph.
\end{mdef} 
 
% In summary,  our algorithm is more efficient,
% scalable, and robust; the graph representation of intermediate results  can
% effectively improve the performance of query answering; the pruning algorithm
% also has good performance.
% \vspace{-2pt}
\section{Conclusions}     
We have proposed the GTPQ, a new class of tree pattern queries on
graph-structured data, which incorporates structural predicates defined in terms of propositional
logic to specify  structural conditions. We studied
several fundamental problems, and 
 established a general framework for evaluating GTPQs using a graph
representation of graphs and a pruning approach.  An
algorithm has been developed for evaluating GTPQs, which can achieve a small
size of intermediate results due to the effective pruning process and largely
avoid generating redundant matches by dynamically shrinking the tree pattern
during pruning and enumerating processes.   
   
\noindent{\bfseries Acknowledgement.} This work is supported  by the National Science
Foundation of China (61075074).

% \vspace{-2mm} 

\bibliographystyle{abbrv} 
\bibliography{gtpq}

\begin{appendix}
\section{XQuery Example}
$Q_1$ in Example 1 can be expressed in XQuery:

{
\small
\begin{tabbing}
let\qquad \=$\textdollar$dblp $:=$ doc(dblp.xml)\\
for \>$\textdollar$paper in $\textdollar$dblp//inproceedings,
\\\>$\textdollar$conf in $\textdollar$dblp//proceedings\\
where \>$\textdollar$paper/author $=$ ``Alice'' and
$\textdollar$paper/author $=$ ``Bob'' and\\
\>$\textdollar$paper/crossref $=$ $\textdollar$conf/@key and
 data($\textdollar$conf/year) $\geq$ 2000 and\\ \>data($\textdollar$conf/year)
 $\leq$ 2010\\ return\\
\>if (exists($\textdollar$paper/year) and exists($\textdollar$conf/title))\\
\>then
\=$<$paper$>$\\
\>\>$<$title$>\{\textdollar$paper/title$\}$$<$/title$>$\\
\>\>$<$year$>\{\textdollar$paper/year$\}$$<$/year$>$\\
\>\>$<$conf$>\{\textdollar$conf/title$\}$$<$/conf$>$\\
\>\>$<$/paper$>$
\end{tabbing}
}

\section{Proofs}
\begin{proof}[Proof Sketch  of Theorem \ref{thm:sat}]
Given a GTPQ $Q$, we can safely remove two kinds of nodes as well as their
descendants without changing the satisfiability: the nodes whose attribute predicates are
unsatisfiable and those non-independently constraint nodes. We  next only
consider the case where there does not exist such two kinds of nodes. We prove
that $Q$ is satisfiable, iff for the root node $u_r$, $f_{cs}(u_r)$ is
satisfiable.

\vspace{4pt}
(1) $\rightarrow:$
Suppose $G$ is such a data graph that $Q(G)$ is non-empty. Let $C$ be a
certificate and $T$ be the corresponding truth assignment $T$ on variables in
structural predicates: For a query node $u$, if there exists a data node $v$
such that $\in C$ and $v\models  u$, $p^T_{v}:=1$, otherwise $p^T_{v}:=0$.

By the definition of semantics, if $p^T_{v}=1$, $f^T_{ext}=1$; thus,
$f^T_{tr}=1$. For each clause $(\neg p_{u_1}\vee (p_{u_2}\wedge
f_{ext}(p_{u_2}))$ in $f_{cs}(u_r)$, because $u_2\trianglelefteq u_1$,
$p_{u_1}\rightarrow p_{u_2}$ and $f_{ext}(u_1)\rightarrow f_{ext}(u_2)$ hold;
thus,  $(\neg p_{u_1}\vee (p_{u_2}\wedge
f_{ext}(p_{u_2}))$ is true. Therefore, $f^T_{cs}(u_r)=1$.

\vspace{4pt}
(2)$\leftarrow:$
Suppose $T$ is a satisfying truth assignment of $f_{cs}(u_r)$. We initialize a
data graph $G=(V, E, f)$ as follows.
\begin{enumerate}[(a)]
  \item For each variable $p_{u_i}$ in $f_{cs}(u_r)$ such that $(p_{u_i}\wedge
  f_{ext}(u_i))^T=1$, add a node $v_i$ to $G$.
  \item Add an edge $(v_i, v_j)$ to $G$, iff $(u_i, u_j)$ is an edge in $Q$.
  \item For each node $v_i$, choose $f(v_i)$ such that $f(v_i)$  satisfies
  $f_a(u_i)$.
\end{enumerate}

We simulate the  process of evaluating $Q$ on $G$
 and denote the truth assignment in the evaluation by
$T'$. We assign a truth value to each node variable in  a bottom-up process
according to the semantics of GTPQ and at the same time modify $G$ if necessary
to make $V$ as a certificate. 

For any query node $u_i$, if $p^T_{u_i}=0$ and
$p^{T'}_{u_i}=1$, it can be inferred that there exists $v_j\in G$ such that
$f(v_j)$ satisfies $f_a(u_i)$. If  $f_a(u_j)\not\rightarrow f_a(u_i)$, we change $f(v_j)$ so that $f(v_j)$ satisfies $f_a(u_j)$, but does not satisfy $f_a(u_i)$, leading to $p^{T'}_{u_i}=0$. 

We next prove by contradiction that after the above processing, if $p^{T}_{u}=1$, 
$p^{T'}_{u}=1$.  Assume one  node  at the largest depth, for which 
$p^{T}_{u}=1$ and $p^{T'}_{u}=0$,  is $u$. By assumption, for any descendant
$u_d$ of $u$, if $p^{T}_{u_d}=1$, 
$p^{T'}_{u_d}=1$. So there must be a child $u'$ of $u$ for which 
$p^{T}_{u'}=0$ and $p^{T'}_{u'}=1$. From the way $G$ is constructed, there is
a mapping from $u'$ and its descendants to another descendant $u''$ of $u$ and
its descendant such that  $u'\trianglelefteq u''$ and $p^T(u'')=1$. However,
since $p^T(f_{cs}(u_r))=1$, if $p^T(u'')=1$, $p^T(u')=1$, which is contradictory
to our assumption.

For each backbone node $u$, $(p_{u}\wedge
f_{ext}(u))^{T'}=(p_{u}\wedge f_{ext}(u))^{T}=1$. So each output node has
a non-empty image in $V$ and those images constitute an answer to $Q$.
\end{proof}
 
\begin{proof}[Proof Sketch  of Theorem \ref{thm:sat:com}]
Since attribute predicates are conjunctive, the satisfiability of them can be
determined in linear time. We assume in the following that all attribute
predicates are satisfiable.

\vspace{6pt}
(1) A union-conjunctive GTPQ where all attribute predicates are satisfiable is
always satisfiable.
\vspace{4pt}

(2) We prove that the satisfiability problem of a general GTPQ is NP-Complete by
a reduction from SAT.

Given any instance $\phi$ of SAT, we suppose 
$\phi$ has $n$ variables and  construct a GTPQ $Q$ with $n+1$ nodes
as follows. (a) First, choose the first $n$ nodes, each $v_i$ corresponding to a
distinct variable $x_i$ in $\phi$. Then, construct an edge from the $(n+1)$-th
node to each of them. (b) Each node is associated with a satisfiable attribute
predicate with a distinct attribute variable. The structural predicate of the
root is $\phi$, with $p_{v_i}$ replacing $x_i$ for each non-leaf node $v_i$. (c)
The root node, denoted by $u_r$, is the only output node.

Since $f_{cs}(u_r)=f_{s}(u_r)=\phi$, $f_{cs}(u_r)$ is satisfiable iff $\phi$ is
satisfiable. By Theorem \ref{thm:sat}, the conclusion that $Q$ is satisfiable
iff $\phi$ is
satisfiable immediately follows.

It is easy to check that the reduction takes linear time and the satisfiability
is in NP.
\end{proof}

\begin{proof}[Proof Sketch of Theorem \ref{thm:homo}]
\mbox{}

(1) $\rightarrow.$ According to the truth table of the complete structural
predicate of the root node of $Q_1$, we can enumerate all (potentially
exponential) combinations of query nodes of $Q_1$ such that for each
combination, there exists a bijection $\lambda$ from a certificate to it.
Informally, for each combination as a GTPQ, we  can construct a data
graph $G$ from a satisfying truth assignment in the
way we use in the proof of Theorem \ref{thm:sat}, so that the data nodes
constitute a certificate.  By assumption, $G$ is also a certificate with respect
to $Q_2$, and there is a mapping $\lambda'$ from $Q_2$ to the certificate. Further, $\lambda'\circ\lambda^{-1}$
 is a mapping from $Q_2$ to $Q_1$ satisfying the first three conditions in the definition of homomorphism. 
Finally, a homomorphism can be derived from
all such mappings with respect to the combinations.

(2) $\leftarrow.$ For the opposite direction, suppose there is a homomorphism
$\lambda$ from $Q_2$ to $Q_1$. Let $G$ be a data graph, on which the answer of
$Q_1$ is not empty. Suppose $res\in Q_1(G)$ and $C$ is a corresponding
certificate with the truth assignment  denoted by  $T$. It is clear that $C$ is a certificate of $Q_2$ with a truth assignment $T'$ such
that (a) $p^{T'}_{u}=1$, iff $p^{T}_{\lambda(u)}=1$; (2) $f^{T'}_{cs}(u_r)=1$
for the root $u_r$.
\end{proof}

\begin{proof}[Proof Sketch of Theorem \ref{thm:homo:time}]
The proof is based on a reduction from the tautology checking problem (TCP)
of propositional formulas to the containment problem of GTPQs by constructing a
GTPQ from an instance of TCP using the same technique  in the proof of
Theorem \ref{thm:sat:com}.
\end{proof}

\begin{proof}[Proof Sketch of Theorem \ref{thm:minnp}]
The proof is based on a reduction from  the variable minimal equivalence 
problem (VME) \cite{vme} in propositional logic 
of propositional formulas to the decision version of the minimization problem of
GTPQs by constructing a GTPQ from an instance of VME using the same technique  in the proof of
Theorem \ref{thm:sat:com}.
\end{proof}
   
\setlength{\textfloatsep}{2pt plus 0pt minus 0pt}
\setlength{\floatsep}{2pt plus 0pt minus 0pt}
\setlength{\dbltextfloatsep}{2pt plus 0pt minus 0pt}
\setlength{\dblfloatsep}{2pt plus 0pt minus 0pt}

\section{Additional Experimental \\ Results}
\subsection{Measuring I/O cost}
We measure the I/O cost of each algorithm in  terms of three metrics, namely
the number of data nodes accessed (\#input), the number of index
elements looked up (\#index), and the size of intermediate results
(\#intermediate\_results). 

Regarding the number of index lookups, the value for GTEA is the total number of
elements retrieved   from successor and predecessor lists in 3-hop index; the
value for HGJoin is the total number of ids and interval lables in tag lists
(called Alist and Dlist in \cite{hjoin}); the value for TwigStackD is the total
number of surrogate and surplus predecessors visited in SSPI. Since TwigStack and
Twig$^2$Stack do not use a graph reachability index, they have no such cost.

The cost of intermediate results for each algorithm is computed as follows.
(1) The value for GTEA is twice  the total number of the nodes and edges of the
maximal matching graph. (2) The values for HGJoin, TwigStack and Twig$^2$Stack include
the cost of intermediate results for subqueries in the form of tuples. (3) In
addition, TwigStack and Twig$^2$Stack also involves the space cost of stack
encoding. (4) Apart from the cost of stack encoding, TwigStackD introduces the space cost
of pool encoding. It is necessary to clarify that in our experiments, all 
intermediate results are maintained in  main memory and not stored on disk.
This metric is to evaluate the
worst-case I/O cost caused by the intermediate results. When measuring this cost, we assume that any intermediate result is written to
disk and read back to main memory when needed. 
   
\begin{figure}[htb]
\begin{center}
  \includegraphics[width=2in,angle=270]{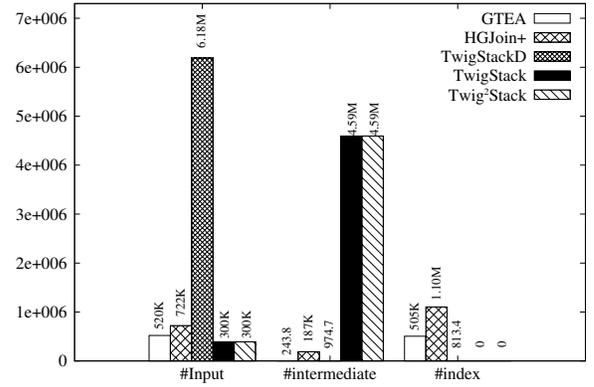}
  \caption{I/O cost}
  \label{exp-io}   
\end{center}
\end{figure}

Fig$.$ \ref{exp-io} depicts the experimental results for processing $Q_3$ on the
XMark dataset with scale factor 1.5. The detailed costs are reported
above columns.
Note that TwigStack and Twig$^2$Stack involve exactly the same I/O cost.  

From the results, TwigStack and Twig$^2$Stack read the smallest number of
data nodes. They only need to scan those data nodes corresponding to all query
nodes for once. In comparison, GTEA accesses more, because it needs to perform a
two-round pruning process (bottom-up and top-down). The value, however, is
bounded by two times of that of TwigStack. As HGJoin splits a query to subtree queries and
 the different subqueries have identical query
 nodes, HGJoin also accesses some data nodes for more than once,  with a
 bound of the maximum number of children of any node in the tree pattern.
 TwigStackD reads far more data nodes than others in the experiments, resulting from the two traversals of
the data graph in the pre-filtering process.

The results clearly show that GTEA creates much fewer intermediate results
than all other four algorithms. TwigStack and Twig$^2$S-tack have more 
intermediate results than GTEA by four orders of
magnitude. The huge gap results from the fact that
TwigStack and Twig$^2$Stack need to output a large number of intermediate path and twig solutions to each subtree 
query which is far less selective 
than the whole query. The structural joins adopted by HGJoin also introduce many
partial solutions and lead to a large size of intermediate results as shown in
the figure. For TwigStackD, its pre-filtering process selects  nodes
potentially in the final answers and considerably saves the space cost of stacks
and pools. GTEA shows the best performance, as it can prune non-answer nodes as
TwigStackD and represent the intermediate results as a maximal matching graph.

Fig$.$ \ref{exp-io} shows that  GTEA again outperforms
HGJoin, due to the compact 3-hop index and the effectiveness of the merging
operations in the pruning process. Yet GTEA incurs  more cost for looking up
indexes than TwigStackD. GTEA uses the 3-hop index in the two-round pruning process and
when constructing the maximal matching graph, while TwigStackD looks up the
reachability index only when expanding the partial solutions in pools. However,
the small cost achieved by TwigStackD is at the expense of the large I/O cost
for scanning data nodes in the pre-filtering process which significantly
reducing the number of nodes to be processed in the stacks and pools. Moreover,
since indexes are often  kept mostly in main memory, the difference in the
number of disk I/O's needed for GTEA and TwigStackD to support the index lookup is
supposed to be actually  small.

Overall, GTEA achieves good performance gain over other competitors in terms of
I/O cost. The results indicate that the pruning process does not incur high I/O
cost as TwigStackD and the graph representation can keep the space cost of
intermediate results pretty small.

\begin{figure}[tb]
\begin{center}
  \includegraphics[height=1.1in]{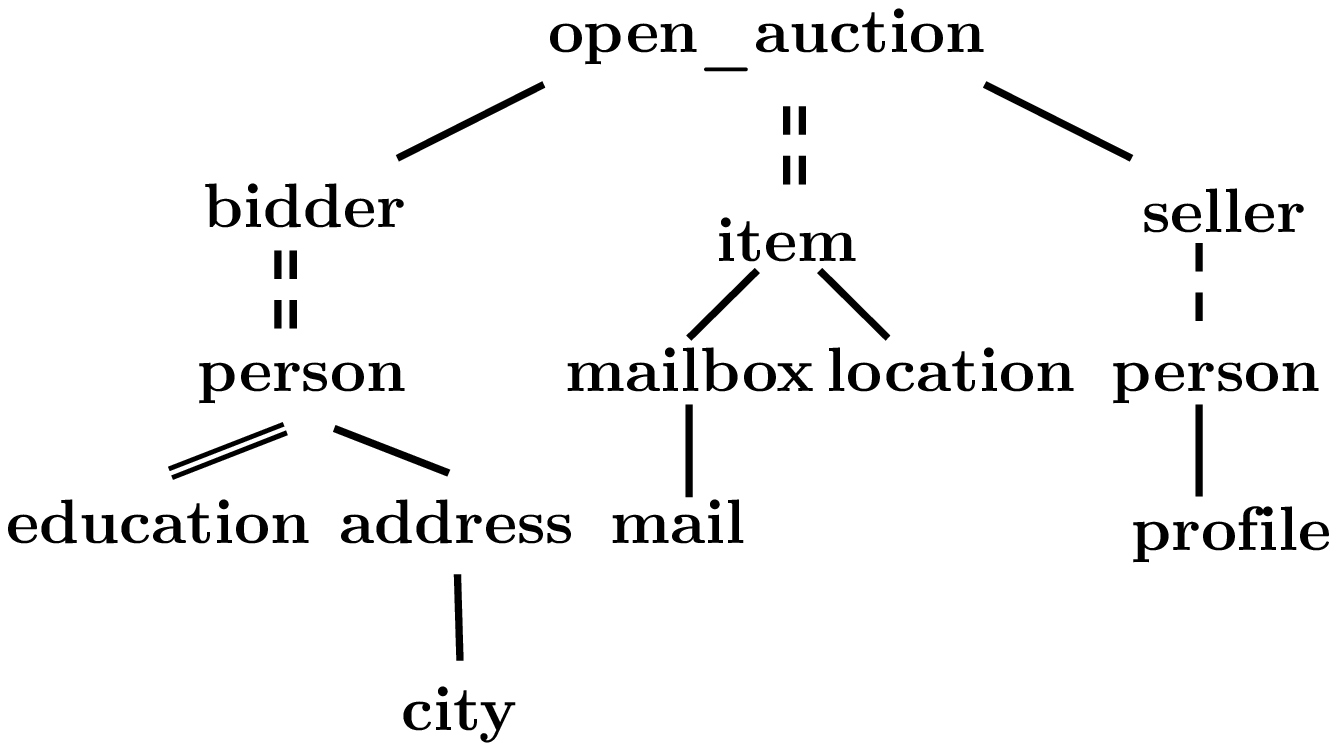}
  \caption{The tree structure of tested queries}
  \label{exp-tree} 
\end{center}
\end{figure} 

\subsection{GTPQ Processing}
  
\begin{table}[t] 
\caption{The output nodes of the queries in Exp-1}
\small
\centering 
\begin{tabular}{l|l}  
\hline
$Q_4$ & open\_auction \\
$Q_5$ & open\_auction, bidder, seller \\
$Q_6$ & open\_auction, bidder, seller, city, profile \\
$Q_7$ & open\_auction, item, location\\
$Q_8$ & all query nodes\\
\hline
\end{tabular} 
\label{tab:output}
\end{table}
\begin{table}[t] 
\caption{The structural predicates of the queries in Exp-2}
\small
\centering
\begin{tabular}{l|l}   
\hline
DIS$_1$ & $f_s(\textrm{open\_auction})= \textrm{bidder}\vee \textrm{seller}$
\\\hline 
\multirow{2}{*}{DIS$_2$} & $f_s(\textrm{open\_auction})= \textrm{bidder}\vee
\textrm{seller}$\\
& $f_s(\textrm{item})= \textrm{mailbox}\vee \textrm{location}$
\\ \hline 
 DIS$_3$ & $f_s(\textrm{open\_auction})= \textrm{bidder}\vee \textrm{seller}\vee
\textrm{item}$ \\ 
\hline
\hline
NEG$_1$ & $f_s(\textrm{person})=\neg \textrm{education}$ \\\hline 
NEG$_2$ & $f_s(\textrm{open\_auction})=\neg  \textrm{bidder},
f_s(\textrm{person})=\neg \textrm{education}$ \\\hline 
\multirow{2}{*}{NEG$_3$} & $f_s(\textrm{open\_auction})=\neg 
\textrm{bidder}\wedge \neg \textrm{seller}$\\
& $ f_s(\textrm{person})=\neg
\textrm{education}$ \\
\hline
\hline
\multirow{2}{*}{DIS\_NEG$_1$} & $f_s(\textrm{open\_auction})=\neg 
\textrm{bidder}\vee \textrm{seller}$\\ 
& $f_s(\textrm{person})=\neg  
\textrm{education}$ \\\hline
 DIS\_NEG$_2$ & $f_s(\textrm{open\_auction})=(\neg 
\textrm{bidder}\wedge \textrm{seller})\vee (\textrm{bidder}\wedge 
\neg \textrm{seller})$ \\ \hline 
\multirow{2}{*}{DIS\_NEG$_3$} & $f_s(\textrm{open\_auction})=(\neg 
\textrm{bidder}\wedge \textrm{seller})\vee (\textrm{bidder}\wedge 
\neg \textrm{seller})$\\
& $f_s(\textrm{person})=\neg \textrm{education}$ \\ \hline 
\multirow{3}{*}{DIS\_NEG$_4$} & $f_s(\textrm{open\_auction})=$\\
& $(\neg 
\textrm{bidder}\wedge \textrm{seller}\wedge\textrm{item})\vee (\textrm{bidder}\wedge 
\neg \textrm{seller}\wedge \neg \textrm{item}),$\\
& $f_s(\textrm{person})=\neg
\textrm{education}$ \\
\hline
\end{tabular} 
\label{tab:predicates}
\end{table}

 \begin{table}[t] 
\caption{Numbers of query results}
\small
\centering 
\begin{tabular}{c|c|c|c|c}  
\hline
$Q_4$ & $Q_5$ & $Q_6$  & $Q_7$ & $Q_8$ \\
\hline
 88 & 98 & 98 & 88 & 151\\
\hline \hline
DIS$_1$ & DIS$_2$ & DIS$_3$ & NEG$_1$ & NEG$_1$ \\
\hline
1236 & 26352 & 2052 & 456 & 1938 \\
\hline \hline
NEG$_3$ & DIS\_NEG$_1$ & DIS\_NEG$_2$ & DIS\_NEG$_3$ & DIS\_NEG$_4$ \\
\hline
1240 &  4156 & 2328 & 2300 & 5643 \\
\hline
\end{tabular} 
\label{tab:result_size}
\end{table}

\begin{figure*}[tb] 
\centering
\subfigure[Varying the number of non-output nodes]{
\begin{minipage}[c]{0.4\textwidth}
\centering
\includegraphics[width=1.45in, angle=270]{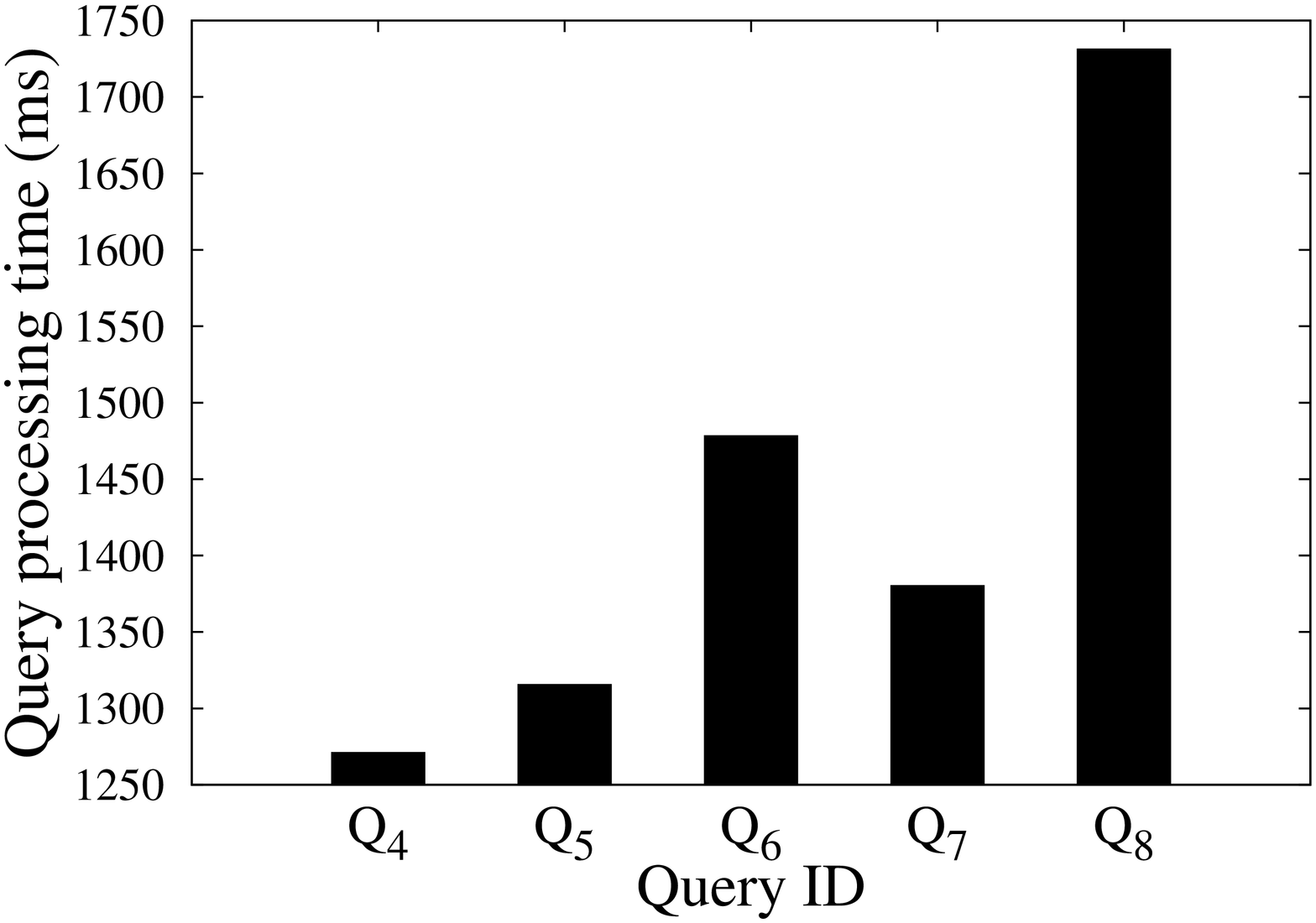}
\end{minipage}}
\subfigure[Union-conjunctive queries]{
\label{exp:q1-res} 
\begin{minipage}[c]{0.4\textwidth}
\centering
\includegraphics[width=1.45in, angle=270]{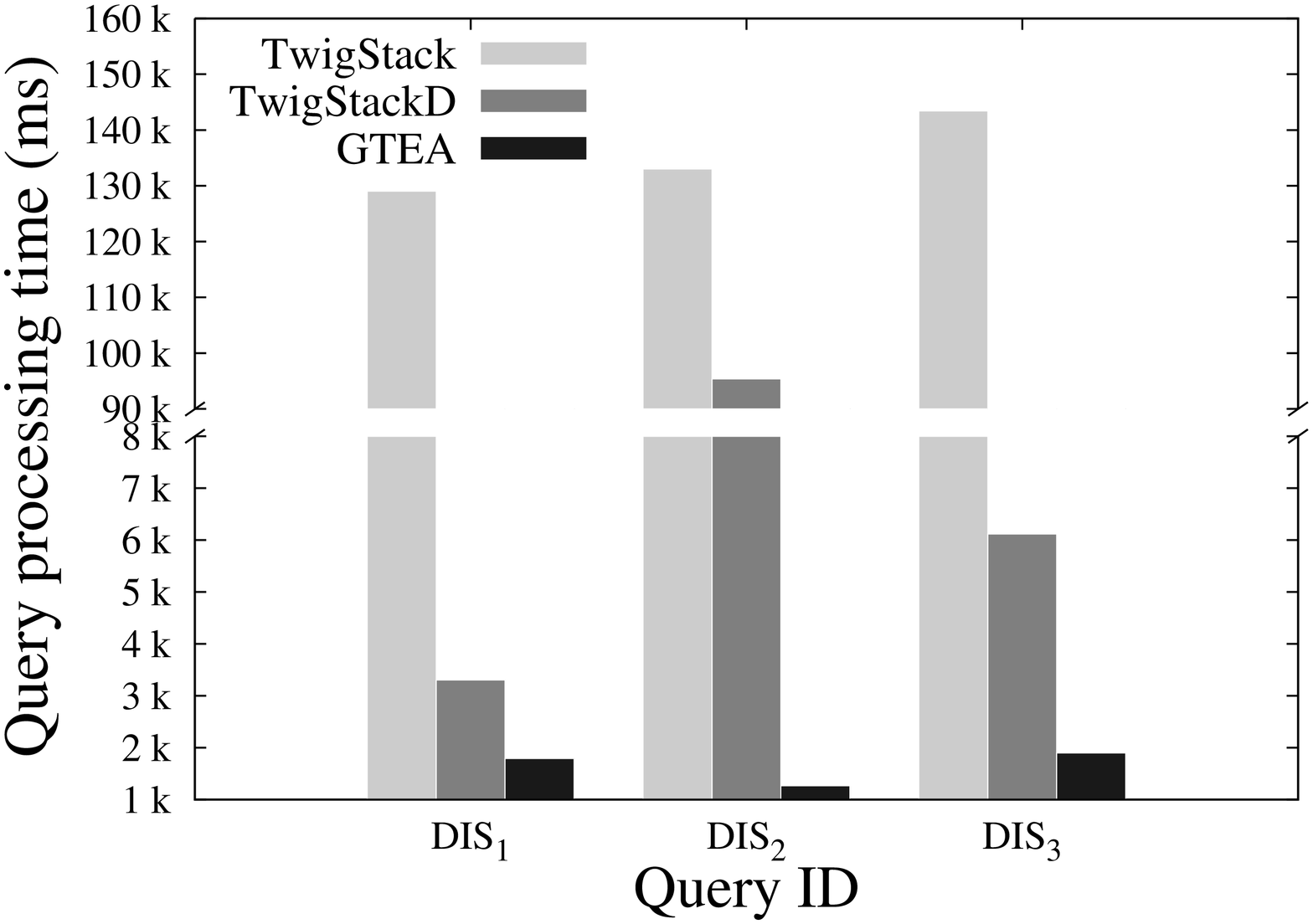}
\end{minipage}} \\
\subfigure[Queries with negation]{      
\label{exp:mark-res}      
\begin{minipage}[c]{0.4\textwidth}  
\centering
\includegraphics[width=1.45in, angle=270]{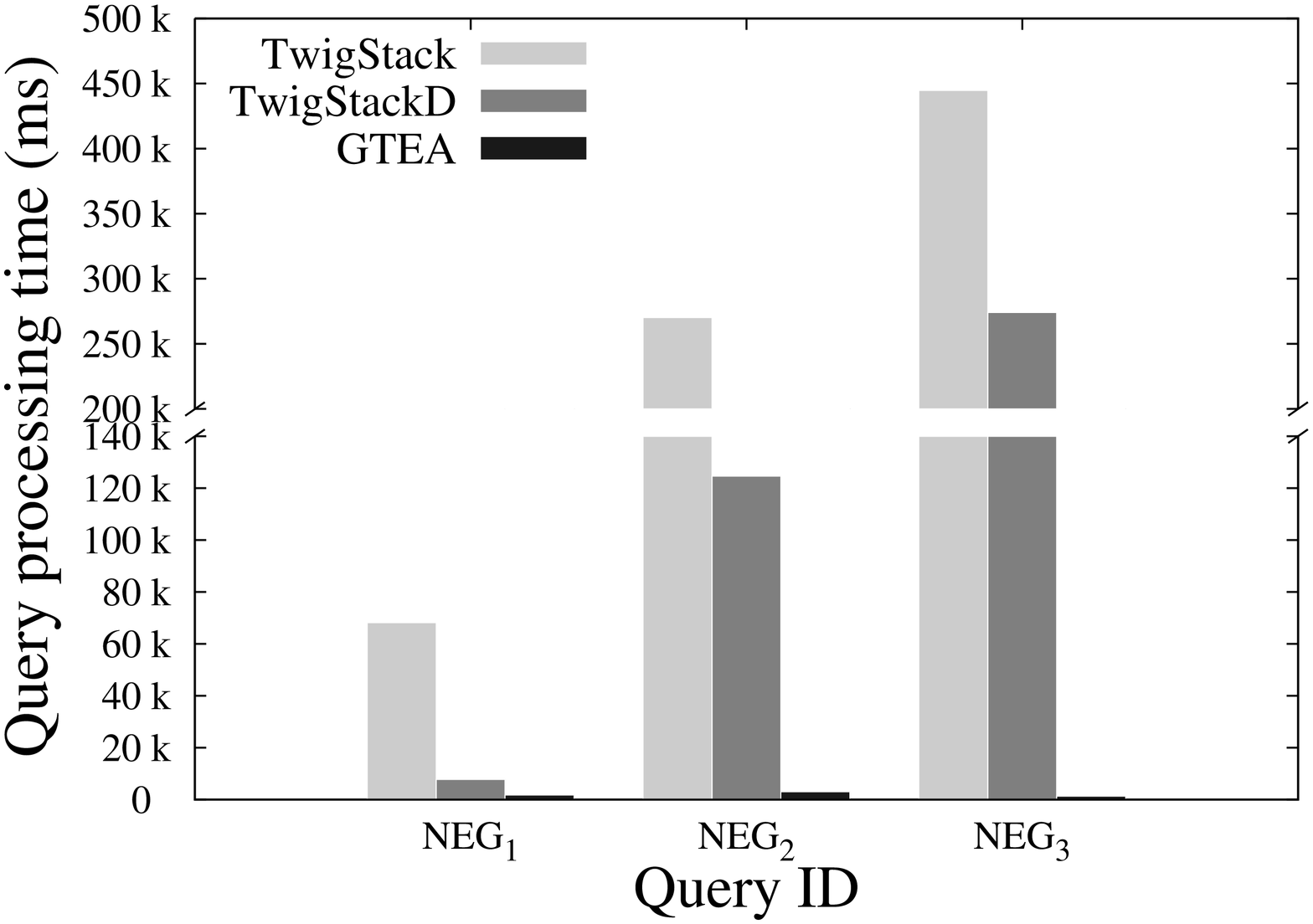}
\end{minipage}}
\subfigure[Queries with disjunction and negation]{   
\begin{minipage}[c]{0.4\textwidth} 
\centering
\includegraphics[width=1.45in, angle=270]{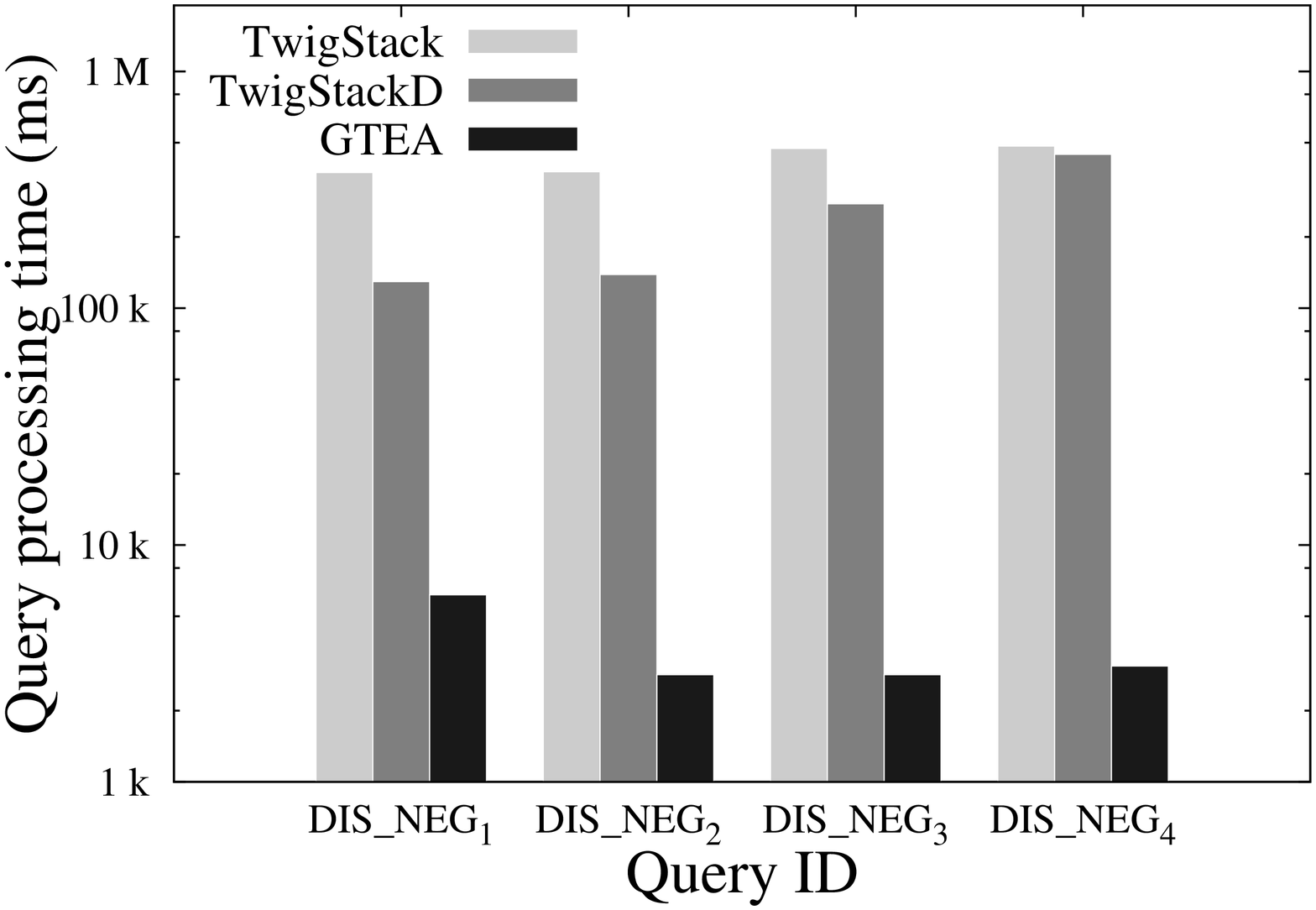}  
\end{minipage}} 
\caption{GTPQ Processing}    
\label{exp:gtpq}   
\end{figure*}

In this section, we present the experimental results for GTPQs with the same
structure (Fig$.$ \ref{exp-tree}) on the XMark data set with scale factor 4.  Since
HGJoin and TwigStackD need to do the same deco-mpose-and-merge operations to
process GTPQs and our experiments for conjunctive queries have shown that TwigStackD significantly outperforms HGJoin, 
 we did not include
HGJoin in this set of experiments. Twig$^2$Stack was also not included as it has
 comparable performance to TwigStack and the post-process on  
 top of the two algorithms for processing GTPQs is also the same.

 \begin{mdef}[Exp-1 Optimization for non-output nodes]
 We first compare GT-EA, TwigStack and TwigStackD for processing  conjunctive
 queries with varying the  size of output nodes. The output nodes for each
 tested query are given in Table \ref{tab:output}. The result sizes of those
 queries are presented in Table \ref{tab:result_size}. Because TwigStack and TwigStackD are not 
 optimized for queries with non-output nodes and the
 differences in the result sizes of the tested queries are small, the 
 processing time on different queries is close to each other for both  
 algorithms. Fig$.$ \ref{exp:gtpq}(a) depicts the results of GTEA only. Recall
 that GTEA uses a prime subtree obtained based on the output nodes and the specific
  matching nodes in procedure PruneDownward  and constructing the maximal
   matching graph for avoiding creating useless matches to non-output nodes. Hence,
  the processing time of GTEA  depends on the structure of the
 prime subtree and the size of the final answers. 
 The results show that the less the number of output nodes is,
  the less processing time the evaluation generally takes.
 \end{mdef}
 
  \begin{mdef}[Exp-2 GTPQ processing]
 We next show the experimental results for queries that may contain negation and
 disjunction. Three classes of  tested queries, namely the queries with
 disjunction only (DIS), those with negation only (NEG) and with both
 disjunction and negation (DIS\_NEG), are shown in Table \ref{tab:predicates}.
All potentially valid backbone nodes are set as output nodes for all queries.
Fig$.$ \ref{exp:gtpq}(b), (c) and (d)  depict the results for the tested GTPQs.
All of them consistently verify the significant performance gain of GTEA (from
several times to three orders of magnitude). Indeed, as mentioned in the related
work, TwigStack and TwigStackD need to process a number of small subqueries 
and do expensive post merge-join operations for processing GTPQs.
 It is non-trivial to fine tune the two algorithms for GTPQs. 
 It may be possible to derive an efficient mechanism that makes the intermediate results output by TwigStack 
 and TwigStackD in sorted order so that the 
 merge-join operations take less cost. However, it is difficult to reduce the
 large size of intermediate results which considerably impairs the efficiency
 of TwigStack and TwigStackD, so they are unlikely to outperform our algorithm
 anyway.
  \end{mdef}
\section{Processing queries with multiple output structures}
GTEA can be straightforwardly extended to process queries not restricted to
backbone nodes. The only modification is in procedure CollectResults. For an internal node in the maximal matching graph, instead of doing one Cartesian product of the results of branches, the procedure may perform several Cartesian products of
the results of different branches depending on the specified result structures.
Take the query DIS$_1$ (the query structure is shown in Fig$.$ \ref{exp-tree} and
the predicates are defined in Table \ref{tab:predicates}) for example, and
suppose that the results of the query should be of the form
(\textsf{open\_auction}, \textsf{bidder}, \textsf{item}) or (\textsf{open\_auction},
\textsf{seller}, \textsf{item}). The (shrunk) prime subtree is constructed by considering \textsf{bidder} and
\textsf{seller} as the originally defined backbone nodes. In the maximal
matching graph, for each matching nodes of \textsf{open\_auction}, the
CollectResults procedure performs two Cartesian products to derive the
answers: one product of the two branch results corresponding to \textsf{bidder}
and \textsf{item}, and the other product of the two branch results corresponding
to \textsf{seller} and \textsf{item}.

\section{Expanded Algorithms}
We show procedure PruneDownward and PruneUpward in more details in Procedure 6
and Procedure 7.

\begin{algorithm}[htbp]
\small
\DontPrintSemicolon
\SetAlFnt{\small}
\SetKwInput{KwData}{Input}\SetKwInput{KwResult}{Output}
\KwData{3-Hop index $L_{out}$, a GTPQ $Q$}
\KwResult{Candidate matching nodes satisfying downward structural constraints.} 
%\BlankLine
\lForEach{node $u\in V_q$}{
	$mat(u):=\{x|x\in V, x\sim u\}$\;
}
\lForEach{leaf node $u'$ in $V_q$}{
	$C^p_{u'}:=\textrm{MergePredLists}(mat(u'))$\;
}
$V'_q=V_q\backslash\{u'|u'\textrm{ is a leaf node}\}$ \;
\ForEach{ $u\in V'_q$ in bottom-up order}{
	\lForEach{ $v\in mat(u)$}{
		$chain_{v.cid}:=chain_{v.cid}\cup \{v\}$\;
	}	
	\ForEach{$chain_i$ that is not empty}{
		\lForEach{child $u'$ of $u$}{
			$val[p_{u'}]:=0$\;
		}
		
		\ForEach{node $v_{i}\in chain_i$}{
			
			\ForEach{child $u'$ of $u$ \emph{s.t.}  $val[p_{u'}]=0$ }{
				\lIf{$C^p_{u'}[i]\geq
					v_i.sid$}{ $val[p_{u'}]:=1$\;
				}
			} 
			$v'_i:=v_i$\;
			\Repeat{$v'_i$ = null or $visited_i\leq v'_i.sid$}{				
				\ForEach{index node $v''_i\in L_{out}(v'_i)$}{
					\ForEach{child $u'$ of $u$ \emph{s.t.}  $val[p_{u'}]=0$ }{
						\If{ $C^p_{u'}[v''_i.cid]\geq
						v''_i.sid$}{ $val[p_{u'}]:=1$\;
						}
					}
				}
				$v'_i:=\textrm{next}(v'_i)$\;
			}
			\If{$f_s(u)$ evaluates to false with the valuation $val$}{
				$mat(u):=mat(u)\backslash\{v_i\}$\;
			}
			
			$visited_i:=v_i.sid$\;
 
		}		
% 		\ForEach{node $v_{i}\in chain_i$}{
% 			\ForEach{node $v'_i$ such that $v_i\leq_c v'_i$}{
% 				\lIf{$v'_i.visited=\textrm{true}$}{
% 					{\bf break}\;
% 				}\lElse{				
% 					$v'_i.visited:=\textrm{true}$ \;
% 				}
% 				\ForEach{index node $v''_i\in L_{out}(v'_i)$}{
% 					\ForEach{child $u'$ of $u$ \emph{s.t.}  $val[p_{u'}]=\textrm{false}$ }{
% 						\If{$C^p_{u'}[v''_i.cid]\neq null$ and $C^p_{u'}[v''_i.cid]\geq
% 						v''_i.sid$}{ $val[p_{u'}]:=\textrm{true}$\;
% 						}
% 					}
% 				}
% 			}
% 			\If{$f_b(u)$ evaluate false with the valuation $val$}{ 
% 				$mat(u):=mat(u)\backslash\{v_i\}$\;
% 			}
% 		}
	}
	$C^p_{u}:=\textrm{MergePredLists}(mat(u))$\;
}
\caption{PruneDownward\label{alg:prunedownward}}
\end{algorithm}

\begin{algorithm}[htbp] 
\small
\SetAlFnt{\small\sf}
\DontPrintSemicolon
\SetKwInput{KwData}{Input}\SetKwInput{KwResult}{Output}
\KwData{3-hop index $L_{in}$, the prime subtree $(V'_t, E'_t)$ of a GTPQ}
\KwResult{Candidate matching nodes satisfying upward structural constraints} 
%\BlankLine
$C^s_{u_{root}}:=\textrm{MergeSuccLists}(mat(u_{root}))$\;
$V'_t:=V'_t\backslash\{u'|u' \textrm{ is a leaf node}\}$ \;
\ForEach{node $u\in V'_t$ from top to bottom such that $|mat(u)|>1$}{
	\ForEach{child $u'$ of $u$ \label{upward:groupb} such that $|mat(u')|>1$}{
		\ForEach{node $v\in mat(u')$}{
% 			\If{$|Group_v|=0$}{$chain_{v.cid}=chain_{v.cid}\cup\{v\}$\;}
			$chain^{u'}_{v.cid}:=chain^{u'}_{v.cid}\cup\{v\}$\;
			$Group_v:=Group_v\cup\{u'\}$ \;\label{upward:groupe}
		}
	}
	merge all lists $chain^{u'}_{i}$($u'$ is a child of $u$) into $chain_{i}$
	for each chain $i$\;
	\ForEach{$chain_i$ that is nonempty}{  
		\ForEach{node $v_{i}\in chain_i$}{
% 			initialize $mark_j$ for each chain $j$ \;      
			\lIf{$C^s_{u}[i]\leq v_i.sid$}{     
				$reach:=\textrm{true}$; \textbf{break}\;		  		
			}
			$v'_i:=v_i$\;
			\Repeat{$v'_i$ = null or $visited_i\geq v'_i.sid$}{
% 				\lIf{$v'_i.visited=\textrm{true}$}{\textbf{break}\;}
% 				\lElse{$v'_i.visited:=\textrm{true}$\;}
				\ForEach{index node $v''\in L_{in}(v'_i)$}{
					\If{$C^s_{u}[v''.cid]\leq v''.sid$}{
						$reach:=\textrm{true}$; \textbf{break}\;						
					} 
				}
				\lIf{$reach=\textrm{true}$}{\textbf{break}\;} 
				$v'_i:=\textrm{prev}(v'_i)$ \;
			}
			
% 			\ForEach{node $v'_i$ such that $v'_i\leq_c v_i$}{
% % 				\lIf{$v'_i.visited=\textrm{true}$}{\textbf{break}\;}
% % 				\lElse{$v'_i.visited:=\textrm{true}$\;}
% 				\ForEach{index node $v''\in L_{in}(v'_i)$}{
% 					\If{$C^s_{u}[v''.cid]\leq v''.sid$}{
% 						$reach:=\textrm{true}$\;
% 						\textbf{break}\;
% 					}
% 				}
% 				\lIf{$reach=\textrm{true}$}{\textbf{break}\;}
% 			}
			\If{$reach=\textrm{false}$}{
				\ForEach{$u'\in Group_{v_i}$}{
					$mat[u']:=mat[u']\backslash\{v_i\}$ \;
				}
			}
			\lElse{\textbf{break} \;}
			 $visited_i:=v_i.sid$\;
		}
	}
	\ForEach{non-leaf child $u'$ of $u$}{
		$C^s_{u'}:=\textrm{MergeSuccLists}(mat(u'))$\;
	}
}
\caption{PruneUpward\label{alg:pruneupward}}
\end{algorithm}
\end{appendix}

\end{document}